\documentclass[a4paper,11pt]{article}
\pdfoutput=1 
\usepackage{jcappub} 
\usepackage{amssymb,amsthm,bbold,bm,mathtools,slashed}
\usepackage{comment,enumerate,footnote,graphicx,subfloat,relsize}
\usepackage{array,tabularx,tabu,multirow,framed,xcolor}
\usepackage[font=small]{caption}
\usepackage{subcaption}
\allowdisplaybreaks
\usepackage{tikz}
\usetikzlibrary{arrows,shapes}
\usetikzlibrary{trees,patterns}
\usetikzlibrary{matrix,arrows} 				
\usetikzlibrary{positioning}				  
\usetikzlibrary{calc,through}				  
\usetikzlibrary{decorations.pathreplacing}  
\usepackage{pgffor}							

\usetikzlibrary{decorations.pathmorphing}	
\usetikzlibrary{decorations.markings}
\tikzset{
	>=stealth', 
    vector/.style={decorate, decoration={snake}, draw},
	provector/.style={decorate, decoration={snake,amplitude=2.5pt}, draw},
	antivector/.style={decorate, decoration={snake,amplitude=-2.5pt}, draw},
	bigvector/.style={decorate, decoration={snake,amplitude=4pt}, draw},
    fermion/.style={draw=black, postaction={decorate},
        decoration={markings,mark=at position .55 with {\arrow[draw=black]{>}}}},
    fermionbar/.style={draw=black, postaction={decorate},
        decoration={markings,mark=at position .55 with {\arrow[draw=black]{<}}}},
    fermionnoarrow/.style={draw=black},
    gluon/.style={decorate, draw=black,
        decoration={coil,amplitude=4pt, segment length=5pt}},
    scalar/.style={dashed,draw=black, postaction={decorate},
        decoration={markings,mark=at position .55 with {\arrow[draw=black]{>}}}},
    scalarbar/.style={dashed,draw=black, postaction={decorate},
        decoration={markings,mark=at position .55 with {\arrow[draw=black]{<}}}},
    scalarnoarrow/.style={dashed,draw=black},
    momentum/.style={draw=black, postaction={decorate},
        decoration={markings,mark=at position 1 with {\arrow[draw=black]{>}}}},
    antimomentum/.style={draw=black, postaction={decorate},
        decoration={markings,mark=at position 0.1 with {\arrow[draw=black]{<}}}}
}

\tikzstyle{block} = [draw, rectangle, 
    minimum height=3em, minimum width=6em]

\usepackage{cleveref}
\crefformat{pluralequation}{#2\color{black}{eqs.~(}#1\color{black}{)}#3}
\Crefformat{pluralequation}{#2\color{black}{Equations~(}#1\color{black}{)}#3}

\newcommand\supsetsim{\mathrel{%
  \ooalign{\raise0.2ex\hbox{$\supset$}\cr\hidewidth\raise-0.8ex\hbox{\scalebox{0.9}{$\sim$}}\hidewidth\cr}}}

\title{ 
The role of unitarisation on dark-matter freeze-out via metastable bound states
}

\author{Kalliopi Petraki, Anna Socha, Christiana Vasilaki}
\affiliation{Laboratoire de Physique de l'École Normale Supérieure, ENS, Université PSL, CNRS, Sorbonne Université, Université Paris Cité, F-75005 Paris, France}

\emailAdd{kalliopi.petraki@phys.ens.fr}
\emailAdd{anna.socha@phys.ens.fr}
\emailAdd{christiana.vasilaki@phys.ens.fr}

\abstract{ 
In many Abelian and non-Abelian theories, standard calculations of radiative bound-state formation violate partial-wave unitarity -- even at arbitrarily small couplings -- when capture into excited states is considered. Recent work demonstrated that unitarity can be restored by the proper resummation of squared inelastic processes in the self-energy of the incoming state. We examine how unitarisation affects dark-matter thermal decoupling, given that the formation and decay of metastable dark-matter bound states are critical in determining the relic abundance, especially for multi-TeV dark matter. We consider an Abelian model featuring bound-state formation via emission of a light scalar that carries a conserved charge, whose dynamics also emulates relevant aspects of non-Abelian theories. Incorporating capture into excited states, we show that, without proper treatment, unitarity violation is so severe as to prevent freeze-out. Resumming the squared bound-state formation processes restores unitarity and ensures freeze-out, while capture into excited levels still significantly depletes dark matter. We further discuss the impact of higher partial waves, both within and beyond the present model. Finally, we point out the intriguing possibility of late dark-matter decoupling that can affect structure formation.
}

\keywords{dark matter, unitarity, bound states, excited states, freeze out.}

\begin{document}
\maketitle
\flushbottom

\section{Introduction \label{Sec:Introduction}}

If dark matter (DM) couples to force mediators that are much lighter than itself, its interactions become effectively long-ranged. This behaviour arises generically in viable thermal-relic DM scenarios with particle masses at or above the TeV range, as has been shown by explicit computations in a variety of models and is supported by model-independent unitarity arguments~\cite{Flores:2024sfy,Baldes:2017gzw}. Long-range interactions imply the existence of bound states. The formation and decay of metastable bound states introduces additional annihilation channels for DM, thereby modifying the expected mass-coupling relations necessary to reproduce the observed DM density~\cite{vonHarling:2014kha}. The role of bound-state formation (BSF) in DM thermal decoupling has been extensively studied across various scenarios~\cite{vonHarling:2014kha, Ellis:2015vaa, Petraki:2015hla, Petraki:2016cnz, Baldes:2017gzw, Kim:2016zyy, Biondini:2017ufr, Biondini:2019int, Binder:2018znk, Harz:2018csl, Harz:2019rro, Oncala:2018bvl, Oncala:2019yvj, Oncala:2021tkz, Oncala:2021swy, Ko:2019wxq, Binder:2019erp, Binder:2020efn, Bottaro:2021snn, Binder:2021vfo, Bollig:2021psb, Garny:2021qsr, Binder:2023ckj, Vasilaki:2024fph}, and has been shown to impact the DM density even up to orders of magnitude.

Recent studies have shown that standard BSF computations may severely violate the partial-wave unitarity limits on inelastic cross sections. This issue arises when BSF occurs via the emission of a boson charged under a symmetry, including a charged scalar~\cite{Oncala:2019yvj}, such as the Higgs doublet~\cite{Oncala:2021tkz, Oncala:2021swy}, or a non-Abelian gauge boson~\cite{Binder:2023ckj}, and persists even when the emitted charge is global~\cite{Oncala:2019yvj}. The emission of a conserved charge implies that the incoming and outgoing states are governed by different potentials. Depending on the relative strength of these potentials, this can lead to large overlaps of the corresponding wavefunctions~\cite{Oncala:2019yvj}. In such cases, when capture into excited states is taken into account, unitarity appears to be violated even for arbitrarily small couplings~\cite{Beneke:2024nxh, Petraki:2025}.

Reference~\cite{Flores:2024sfy} demonstrated that unitarity can be restored by properly accounting for the contribution of inelastic processes to the self-energy of the incoming state. It employed a generalised optical theorem, analysed in partial waves and extended to off-shell amplitudes, to derive the imaginary potential generated by squared inelastic processes. While inelastic processes may naively seem to contribute subdominantly to the self-energy, this must be reassessed when their amplitudes approach the unitarity limit. The inclusion of the inelastic contributions to the self-energy is mandated by the continuity equation~\cite{Blum:2016nrz}, and accounts for the reduction of the incoming flux due to inelasticity. This, in turn, suppresses the inelastic scattering rate. Reference~\cite{Flores:2024sfy} provided a simple prescription for computing the unitarised cross sections from those that do not include the resummation of squared inelastic processes.

In this work, we explore for the first time the impact of the regularisation procedure proposed in Ref.~\cite{Flores:2024sfy} on the DM thermal decoupling. To this end, we employ the Abelian model introduced in Ref.~\cite{Oncala:2019yvj}, which features BSF via rapid monopole transitions due to the emission of a charged scalar. This model has the advantage of a simple selection rule, $\Delta \ell = 0$, Abelian dynamics, and fast transitions due to the scalar vertices involved. At the same time, the dynamics we investigate resembles that of BSF via gauge boson emission in non-Abelian theories, providing insight into these more complex scenarios. 

We extend the analysis of Ref.~\cite{Oncala:2019yvj} to include capture into excited bound states, and regularise the cross sections according to~\cite{Flores:2024sfy}, before computing the DM density. We show that summing over all bound levels without regularisation not only leads to unitarity violation for arbitrarily small couplings~\cite{Petraki:2025}, but also prevents DM freeze-out. Indeed, the thermally-averaged BSF cross sections exhibit \textit{super-critical} scaling~\cite{Binder:2023ckj}, causing DM to deplete indefinitely. In contrast, when the inelastic processes are unitarised, DM freezes out. Even upon unitarisation, the capture into excited bound levels significantly alters the predicted relic abundance compared to considering only direct DM annihilations and capture into the ground state.  Moreover, we investigate the role of higher partial waves in BSF processes. In the absence of $\ell$-changing bound-to-bound transitions, their impact on the DM relic density depends largely on the sign of the potential governing the scattering state. Although we do not delve into this aspect here, transitions between bound states with different angular momenta may significantly affect this dynamics, as suggested by the results of Ref.~\cite{Binder:2021vfo}.

The paper is structured as follows. In \cref{Sec:MonopoleTransitions}, we introduce the model, review the unregulated BSF cross sections, show the unitarity violation and present the unitarisation prescription. In \cref{Sec:Decoupling}, we study the DM thermal decoupling, with and without regularisation, compute the relic density, and discuss our results. We conclude in \cref{Sec:Conclusions}.

\clearpage
\section{Monopole transitions \label{Sec:MonopoleTransitions}}

\subsection{Model\label{Sec:MonopoleTransitions_Model}}

We consider the model proposed in Ref. \cite{Oncala:2019yvj}, which involves a complex scalar field $X$ coupled to a dark Abelian gauge force $U(1)$ mediated by the gauge boson $V_\mu$, and a light complex scalar $\Phi$
that is doubly charged under the same force. The interaction Lagrangian reads
\begin{equation}
\begin{aligned}
\mathcal{L}= & -\frac{1}{4} F_{\mu \nu} F^{\mu \nu}+\left(D_\mu X\right)^{\dagger}\left(D^\mu X\right)+\left(D_\mu \Phi\right)^{\dagger}\left(D^\mu \Phi\right)-m_X^2|X|^2-m_{\Phi}^2|\Phi|^2 \\
& -\frac{y m_X}{2}\left(X^2 \Phi^{\dagger}+X^{\dagger^2} \Phi\right)-\frac{\lambda_X}{4}|X|^4-\frac{\lambda_{\Phi}}{4}|\Phi|^4-\lambda_{X \Phi}|X|^2|\Phi|^2,
\end{aligned}
\label{eq:Lagrangian}
\end{equation}
where $F_{\mu \nu} \equiv \partial_\mu V_\nu - \partial_\nu V_\mu$,  $D^\mu_j \equiv \partial^\mu + i q_j g V^\mu$, with $g$ denoting the gauge coupling constant, and $q_X = 1$ and $q_\Phi = 2$ being the charges of the $X$ and $\Phi$ fields, respectively.  The quartic couplings $\lambda_X$, $\lambda_\Phi$ and $\lambda_{X\Phi}$ are real and positive. For later convenience, we define the following structure constants:
\begin{equation}
\alpha_V \equiv \frac{g^2}{4 \pi} 
\quad \mathrm{and} \quad 
\alpha_\Phi\equiv \frac{y^2}{16 \pi}.
\end{equation}

\begin{figure}[h!]
\centering
\begin{tikzpicture}[line width=1pt, scale=1]
%
%
%
\begin{scope}[shift={(0,1)}]
\begin{scope}[shift={(-2,0)}]
\node at (-1.3,1){$X$};
\node at (-1.3,0){$X^{\dagger}$};
\node at (-0.7, 1.3){$p_1$};
\node at (0.7, 1.3){$p_1^\prime$};
\draw (-1,1) -- (1,1);\draw[fermion] 	(-1,1) -- (-0.4,1);\draw[fermion] 	(0.7,1) -- (1,1);
\draw (-1,0) -- (1,0);\draw[fermionbar] (-1,0) -- (-0.4,0);\draw[fermionbar](0.7,0) -- (1,0);
\node at (1.3,1){$X$};
\node at (1.3,0){ $X^\dagger$};
\node at (-0.7, -0.3){$p_2$};
\node at (0.7, -0.3){$p_2^\prime$};
 \draw[shift={(0,0.5)}, fill=blue!30, draw=black] (-0.5,-0.5) rectangle (0.5,0.5);
\node at (0,0.5){${\cal K}_{\mathsmaller{XX^{\dagger}}}^{2\text{PI}}$};
\end{scope}
\node at (0,0.5){$=$};
\begin{scope}[shift={(2,0)}]
\node at (-1.3,1){$X$};
\node at (-1.3,0){$X^\dagger$};
\node at (-0.7, 1.3){$p_1$};
\node at (0.7, 1.3){$p_1^\prime$};
\draw[fermion] 		(-1,1) -- (0,1);
\draw[fermionbar] 	(-1,0) -- (0,0);	
\draw[vector]	(0,0) -- (0,1);
\node at (0.5,0.5){$V$};
\draw[fermion] 		(0,1) -- (1,1);
\draw[fermionbar] 	(0,0) -- (1,0);	
\node at (1.3,1){$X$};
\node at (1.3,0){$X^{\dagger} $};
\node at (-0.7, -0.3){$p_2$};
\node at (0.7, -0.3){$p_2^\prime$};
\end{scope}
\node at (4,0.5){$+$};
\begin{scope}[shift={(6,0)}]
\node at (-1.3,1){$X$} ;
\node at (-1.3,0){$X^{\dagger}$};
\node at (-0.7, 1.3){$p_1$};
\node at (0.7, 1.3){$p_2^\prime$};
\draw[fermion] 		(-1,1) -- (0,1);
\draw[fermionbar] 	(-1,0) -- (0,0);	
\draw[scalar]	(0,1) -- (0,0);
\node at (0.5,0.5){$\Phi$};
\draw[fermionbar] 	(0,1) -- (1,1);
\draw[fermion] 		(0,0) -- (1,0);	
\node at (1.3,1){$X^\dagger$};
\node at (1.3,0){$X$};
\node at (-0.7, -0.3){$p_2$};
\node at (0.7, -0.3){$p_1^\prime$};
\end{scope}

\end{scope}
%
%
%
%

\begin{scope}[shift={(0,-1.5)}]
\begin{scope}[shift={(-2,0)}]
\node at (-1.3,1){$X$};
\node at (-1.3,0){$X$};
\node at (-0.7, 1.3){$p_1$};
\node at (0.7, 1.3){$p_1^\prime$};
\draw (-1,1) -- (1,1);\draw[fermion] (-1,1) -- (-0.4,1);\draw[fermion] (0.7,1) -- (1,1);
\draw (-1,0) -- (1,0);\draw[fermion] (-1,0) -- (-0.4,0);\draw[fermion] (0.7,0) -- (1,0);
\node at (1.3,1){$X$};
\node at (1.3,0){$X$};
\node at (-0.7, -0.3){$p_2$};
\node at (0.7, -0.3){$p_2^\prime$};
 \draw[shift={(0,0.5)}, fill=red!30, draw=black] (-0.5,-0.5) rectangle (0.5,0.5);
\node at (0,0.5){${\cal K}_{\mathsmaller{XX}}^{\mathrm{2PI}}$};
\end{scope}
\node at (-0.3,0.5){$= $};
\node at (0.2,0.5){$\mathlarger{\dfrac{1}{2}} \Bigg($};

\begin{scope}[shift={(2,0)}]
\node at (-1.3,1){$X$};
\node at (-1.3,0){$X$};
\node at (-0.7, 1.3){$p_1$};
\node at (0.7, 1.3){$p_1^\prime$};
\draw[fermion]	(-1,1) -- (0,1);
\draw[fermion]	(-1,0) -- (0,0);	
\draw[vector]	(0,1) -- (0,0);
\node at (0.5,0.5){$V$};
\draw[fermion] 	(0,1) -- (1,1);
\draw[fermion] 	(0,0) -- (1,0);	
\node at (1.3,1){$X$};
\node at (1.3,0){$X$};
\node at (-0.7, -0.3){$p_2$};
\node at (0.7, -0.3){$p_2^\prime$};
\end{scope}
\node at (4,0.5){$+$};
\begin{scope}[shift={(6,0)}]
\node at (-1.3,1){$X$} ;
\node at (-1.3,0){$X$};
\node at (-0.7, 1.3){$p_1$};
\node at (0.7, 1.3){$p_2^\prime$};
\draw[fermion] 		(-1,1) -- (0,1);
\draw[fermion] 	(-1,0) -- (0,0);	
\draw[vector]	(0,1) -- (0,0);
\node at (0.5,0.5){$V$};
\draw[fermion] 	(0,1) -- (1,1);
\draw[fermion] 		(0,0) -- (1,0);	
\node at (1.3,1){$X $};
\node at (1.3,0){$X$};
\node at (-0.7, -0.3){$p_2$};
\node at (0.7, -0.3){$p_1^\prime$};
\end{scope}

\node at (7.8,0.5){$ \Bigg)$};
\end{scope}

\end{tikzpicture}
\caption{The 2-particle irreducible (2PI) kernel generating long-range static potentials for $XX^\dagger$ (upper) and $XX$ pairs (lower). For pairs of identical particles, the factor 1/2 ensures that the resummation of the kernel does not result in double-counting of the loops~\cite[appendix A]{Oncala:2021tkz}. (Figure adapted from Ref.~\cite{Oncala:2019yvj}.)}
\label{fig:2PIdiagrams}
\end{figure}
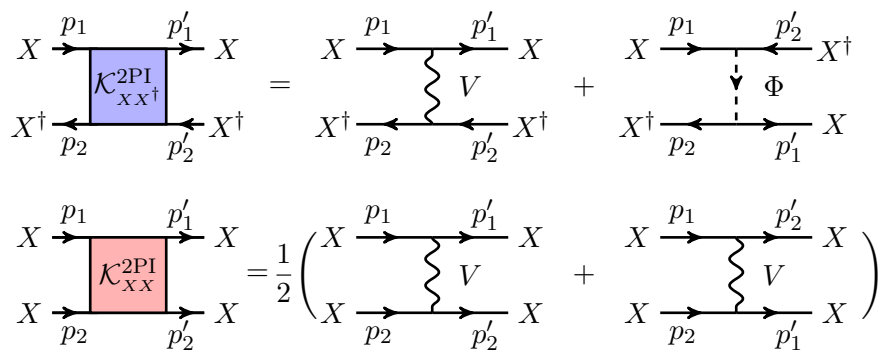 

In the following, we neglect the $\Phi$ mass. The range of validity of this approximation is discussed in \cite[section 3.4]{Oncala:2019yvj}. 
This choice leads to the most pronounced apparent violation of unitarity, which is the focus of our study. In the context of early universe cosmology, it is further motivated by $\Phi$ being a symmetry-breaking scalar at low energies, which acquires only a small, positive, thermal mass-squared at high temperatures that restores the symmetry ~\cite{Flores:2025fvd}. Neglecting the $\Phi$ mass also most accurately emulates the dynamics of BSF via gauge-boson emission in non-Abelian theories, where the same unitarity issues arise.

The long-range potentials of $XX$, $X^\dagger X^\dagger$, and $XX^\dagger$ pairs are generated by the one-boson-exchange diagrams shown in \cref{fig:2PIdiagrams}. In the Coulomb limit, they are~\cite{Oncala:2019yvj}
\begin{subequations}
\label{eq:Potential_V}
\label[pluralequation]{eqs:Potential_V}
\begin{align}
V_{X X}(r) = V_{X^{\dagger} X^{\dagger}}(r) & = + \delta_{\ell,\text{even}} \, \frac{\alpha_V}{r}, 
\label{eq:V_XX}
\\
V_{X X^{\dagger}}(r) 
&\simeq -\frac{\alpha_V + (-1)^{\ell} \alpha_{\Phi}}{r}.
\label{eq:V_XXdagger}
\end{align}
\end{subequations}
Note that in the limit $\alpha_V \rightarrow0$, there is no long-range potential in the $XX$ and $X^\dagger X^\dagger$ pairs, and that the $u$-channel diagrams introduce dependence on the angular momentum mode $\ell$ of the eigenstate~\cite[appendix A]{Oncala:2019yvj}.

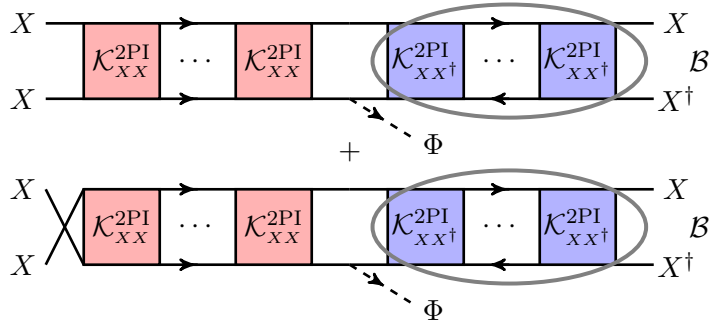
\begin{figure}[h!]
\centering
\begin{tikzpicture}[line width=1pt, scale=1]
\begin{scope}[shift={(0,0.7)}]
\node at (-4.3,1){$X$};
\node at (-4.3,0){$X$};
\draw[fermion]	(-2.1,1) -- (-2,1);
\draw[fermion]	(-2.1,0) -- (-2,0);
\draw (-4,1) -- (0,1);
\draw (-4,0) -- (0,0);
%
\draw[shift={(-3,0.5)}, fill=red!30,draw=black] (-0.5,-0.5) rectangle (0.5,0.5);

\draw[shift={(-1,0.5)},fill=red!30,draw=black] (-0.5,-0.5) rectangle (0.5,0.5);
\node at (-3,0.5){${\cal K}_{\mathsmaller{XX}}^{\mathrm{2PI}}$};
\node at (-1,0.5){${\cal K}_{\mathsmaller{XX}}^{\mathrm{2PI}}$};
\node at (-2,0.5){$\cdots$};
\draw[scalar]	(0,0) -- (0.8,-0.5);
\node at (1.1,-0.6){$\Phi$};
\node at (4.3,1){$X$};
\node at (4.3,0){$X^{\dagger}$};
\draw[fermion]		(2,1) -- (2.1,1);
\draw[fermionbar]	(2,0) -- (2.1,0);
\draw (0,1) -- (4,1);
\draw (0,0) -- (4,0);
%
\draw[shift={(3,0.5)},fill=blue!30,draw=black] (-0.5,-0.5) rectangle (0.5,0.5);
\draw[shift={(1,0.5)},fill=blue!30,draw=black] (-0.5,-0.5) rectangle (0.5,0.5);
\node at (3,0.5){${\cal K}_{\mathsmaller{XX^{\dagger}}}^{\mathrm{2PI}}$};
\node at (1,0.5){${\cal K}_{\mathsmaller{XX^{\dagger}}}^{\mathrm{2PI}}$};
\node at (2,0.5){$\cdots$};
\draw[fill=none,gray,line width=1.5pt] (2.1,0.5) ellipse (1.8 and 0.75);
\node at (4.6,0.5){${\cal B}$};
\end{scope}
\node at (0,0.0){$+$};
\begin{scope}[shift={(0,-1.5)}]
\node at (-4.3,1){$X$};
\node at (-4.3,0){$X$};
\draw[fermion]	(-2.1,1) -- (-2,1);
\draw[fermion]	(-2.1,0) -- (-2,0);
\draw (-4,0) -- (-3.5,1);\draw (-3.5,1) -- (0,1);
\draw (-4,1) -- (-3.5,0);\draw (-3.5,0) -- (0,0);
%
\draw[shift={(-3,0.5)},fill=red!30,draw=black] (-0.5,-0.5) rectangle (0.5,0.5);
\draw[shift={(-1,0.5)},fill=red!30,draw=black] (-0.5,-0.5) rectangle (0.5,0.5);
\node at (-3,0.5){$\mathcal{K}_{\mathsmaller{XX}}^{\mathrm{2PI}}$};
\node at (-1,0.5){${\cal K}_{\mathsmaller{XX}}^{\mathrm{2PI}}$};
\node at (-2,0.5){$\cdots$};
\draw[scalar]	(0,0) -- (0.8,-0.5);
\node at (1.1,-0.6){$\Phi$};
\node at (4.3,1){$X$};
\node at (4.3,0){$X^{\dagger}$};
\draw[fermion]		(2,1) -- (2.1,1);
\draw[fermionbar]	(2,0) -- (2.1,0);
\draw (0,1) -- (4,1);
\draw (0,0) -- (4,0);
%
\draw[shift={(3,0.5)},fill=blue!30,draw=black] (-0.5,-0.5) rectangle (0.5,0.5);
\draw[shift={(1,0.5)},fill=blue!30,draw=black] (-0.5,-0.5) rectangle (0.5,0.5);
\node at (3,0.5){$\mathcal{K}_{\mathsmaller{XX^{\dagger}}}^{\mathrm{2PI}}$};
\node at (1,0.5){${\cal K}_{\mathsmaller{XX^{\dagger}}}^{\mathrm{2PI}}$};
\node at (2,0.5){$\cdots$};
\draw[fill=none,gray,line width=1.5pt] (2.1,0.5) ellipse (1.8 and 0.75);
\node at (4.6,0.5){${\cal B}$};
\end{scope}
\end{tikzpicture}
\caption{ BSF via emission of a charged scalar, $ X + X \rightarrow {\cal B} (X X^\dagger) + \Phi$. (Figure adapted from Ref.~\cite{Oncala:2019yvj}.)
\label{fig:BSFdiagrams}}
\end{figure}
The long-range interactions imply the emergence of non-perturbative phenomena,  including the Sommerfeld effect~\cite{Sommerfeld:1931qaf}, and the existence of bound states~\cite{Salpeter:1951sz}. Bound states exist only between $XX^\dagger$ for an attractive $V_{X X^\dagger}$ potential, corresponding to
$\alpha_{\cal B} = \alpha_V +(-1)^\ell \alpha_\Phi >0$. For the Coulomb potential \eqref{eq:V_XXdagger}, they are characterised by the standard principal and angular momentum quantum numbers, $n\ell m$, and have binding energy $E_n = -\mu \alpha_{\cal B}^2/(2n^2)$, with $\mu = m_X/2$ being the reduced mass of the $XX^\dagger$ pair. $XX^\dagger$ bound states may form radiatively from scattering states, in two ways~\cite{Oncala:2019yvj}, 
\begin{subequations}
\label{eq:BSF}
\label[pluralequation]{eqs:BSF}
\begin{align}
X + X^\dagger &\rightarrow {\cal B} (X X^\dagger) + V^\mu,
\label{eq:BSF_VEmission}
\\
X + X &\rightarrow {\cal B} (X X^\dagger) + \Phi.
\label{eq:BSF_PhiEmission}
\end{align}
\end{subequations}

The emission of a vector boson \eqref{eq:BSF_VEmission} gives rise to dipole transitions ($\Delta\ell =\pm 1$). Since $V^\mu$ is neutral, the scattering and bound state potentials are the same.\footnote{A small difference in the coupling strengths could arise due to the running of the couplings, since the typical momentum exchange differs between scattering and bound states (see e.g.~\cite{Harz:2018csl,Harz:2019rro}).} These QED-like BSF processes are significant inelastic channels, dominating, for example, over $XX^\dagger$ annihilation at low velocities~\cite{vonHarling:2014kha,Petraki:2015hla,Petraki:2016cnz}. Capture into the lowest-lying bound states, $n=1,2$, dominates over capture into excited states, partly due to being more exothermic~\cite{Petraki:2016cnz}. The BSF cross sections are Sommerfeld enhanced due to the attractive potential in the scattering state, and may saturate the unitarity limit in a continuum of non-relativistic momenta for large couplings, $\alpha_V,\alpha_\Phi \sim {\cal O}(1)$~\cite{vonHarling:2014kha,Petraki:2016cnz}. 

The processes \eqref{eq:BSF_PhiEmission} differ from  \eqref{eq:BSF_VEmission} in several important ways. The emission of a charged boson modifies the potential between scattering and bound states, which are here characterised by the coupling strengths $\alpha_{\cal S} = -\alpha_V$ and $\alpha_{\cal B} = \alpha_V + (-1)^\ell \alpha_\Phi$, respectively.  Being eigenstates of different potentials, the scattering and bound states participating in \eqref{eq:BSF_PhiEmission} are \emph{not} orthogonal. 
Moreover, the scalar emission vertex does not introduce any momentum suppression, in contrast to the vector emission vertex of \eqref{eq:BSF_VEmission}. Consequently, the processes \eqref{eq:BSF_PhiEmission} are dominated by monopole transitions ($\Delta \ell = 0$), and, as demonstrated in Refs.~\cite{Oncala:2019yvj, Ko:2019wxq, Oncala:2021swy, Oncala:2021tkz}, can be highly efficient. We emphasise that the reason is two-fold: (i) the change in the potential between the incoming and outgoing states, which leads to large wavefunction overlap integrals, and (ii) the absence of momentum suppression at the radiative vertex. The first feature is shared with BSF via gauge boson emission in non-Abelian theories, and is the essential cause of the unitarity violation at arbitrarily small couplings~\cite{Beneke:2024nxh,Petraki:2025}. The second feature exacerbates the issue in the case at hand.

In the following, we shall focus on BSF via charged scalar emission.

\subsection{Bound-state-formation cross sections \label{Sec:MonopoleTransitions_BSF}} 

The monopole BSF cross sections have been computed in Ref.~\cite{Oncala:2019yvj}. (For monopole bound-to-bound transitions, see Ref.~\cite{Oncala:2021tkz}.) The cross section for capture into the $n\ell$ bound states, $\sigma_{n \ell}^{\rm BSF}$, can be expressed as
\begin{equation}
\label{eq:BSFCrossSection-to-UnitarityBound}
\frac{\sigma_{n \ell}^{{\rm BSF} }}{\sigma^{\rm uni}_\ell} 
= b_\ell R_{n \ell} (\zeta_{\cal B},\zeta_{\cal S}),
\end{equation}
where $\sigma^{\rm uni}_\ell$ is the partial-wave unitarity bound cross section~\cite{Flores:2024sfy}
\begin{equation}
\label{eq:Unitarity-cross section}
\sigma^{\text{uni}}_\ell
\equiv \frac{2^{\delta}4 \pi (2 \ell + 1 ) }{{\bf k}_{\rm CM}^2}
\simeq \frac{2^{\delta}4 \pi (2 \ell + 1 ) }{\mu^2 v_{\rm rel}^2},
\end{equation}
with $\delta =0$ or 1 if the incoming particles are distinguishable or identical, respectively. ${\bf k}_{\rm CM}$ is the momentum of the incoming particles in the centre-of-momentum (CM) frame, approximated in the non-relativistic regime by $|{\bf k}_{\rm CM}| \simeq \mu v_{\rm rel}$, with $\mu$ and $v_{\rm rel}$ being their reduced mass and relative velocity. Note that for monopole BSF, the scattering and bound states are characterised by the same angular mode $\ell$.

The pre-factor $b_\ell$ in \cref{eq:BSFCrossSection-to-UnitarityBound} collects the following:
\begin{equation}\label{eq:bl}
b_\ell \equiv \alpha_\Phi
\left( 1- \frac{\alpha_{\cal S}}{\alpha_{\cal B}}\right)^2 
s_{\mathrm{ps}} 
f_\ell.
\end{equation}
The factor $\alpha_\Phi$ arises from the $\Phi$ emission vertex (cf.~\cref{fig:BSFdiagrams}). $\alpha_{\cal S}$ and $\alpha_{\cal B}$ are the Coulomb potential coupling strengths in the scattering and bound states, respectively,
\begin{align}
\alpha_{\cal S} = -\alpha_V, \qquad
\alpha_{\cal B} = \alpha_V + (-1)^{\ell} \alpha_\Phi,
\label{eq:alpha_SB}
\end{align}
and the factor $(1- \alpha_{\cal S}/\alpha_{\cal B})^2$ confirms that the monopole transitions vanish when the scattering and bound states are eigenstates of the same potential, therefore orthogonal. $s_{p s} \equiv (1-m_{\Phi}^2 / \omega^2)^{1/2}$ denotes the phase-space suppression factor with $\omega$ being the energy of the emitted radiation~\cite{Petraki:2015hla}, 
\begin{align}
\omega = \dfrac{\mu}{2} 
\left(v_{\rm rel}^2 + \dfrac{\alpha_{\cal B}^2}{n^2} \right) .
\label{eq:omega}
\end{align}
Since we neglect the $\Phi$ mass, $s_{ps} \to 1$. Finally, the factor $f_\ell = 2 \delta_{\ell, \text{even}}/2^\delta$ reflects the symmetry properties of the scattering state. For identical bosons, only even-$\ell$ modes contribute, with their strength enhanced by a factor of two. In this case, $\delta = 1$.

The $R_{n \ell}$ factor accounts for the overlap integral between the non-orthogonal wavefunctions of the scattering and bound states. 
Defining the dimensionless parameters,
\begin{equation}
\zeta_{\cal S} \equiv \alpha_{\cal S}/ v_{\rm rel}  \quad \text{ and} \quad \zeta_{\cal B} \equiv \alpha_{\cal B}/ v_{\rm rel},
\end{equation}
it is found to be~\cite{Oncala:2019yvj}
\begin{equation}
\begin{aligned}
\label{eq:rnlGeneral}
R_{n \ell}\left(\zeta_{\cal B}, \zeta_{\cal S}\right) 
&\equiv 
\left[\frac{2^{2 \ell+3} \ell!}{(2 \ell+1)!}\right]^2 
S_{\ell}\left(\zeta_{\cal S}\right) 
\frac{n(n+\ell)!}{(n-\ell-1)!} 
e^{-4 \zeta_{\cal S} {\rm arccot} (\zeta_{\cal B} / n)} 
\left[
\frac{\left(\zeta_{\cal B} / n\right)^{2 \ell+5}}
{\left(1+\zeta_{\cal B}^2 / n^2\right)^{2 \ell+3}}
\right] 
\\
& \times\left|{}_2 F_1 
\left(-n+\ell+1,
1+\ell+\mathbb{i} \zeta_{\cal S} ; 
2 \ell+2 ; 
\frac{\mathbb{i} 4 \zeta_{\cal B} / n}
{\left(1+\mathbb{i} \zeta_{\cal B} / n\right)^2}\right)
\right|^2,
\end{aligned}
\end{equation}
where ${}_2 F_1(a, b; c; z)$ is the ordinary hypergeometric function, and $S_{\ell}$ is the $\ell-$wave Sommerfeld factor \cite{Cassel:2009wt}
\begin{equation}
\label{eq:SommerfeldFactor-l-wave}
 S_{\ell} (\zeta) = \frac{2 \pi \zeta}{1-e^{-2 \pi \zeta}} \prod_{j=1}^{\ell}\left(1+\frac{\zeta^2}{j^2}\right). 
\end{equation}

We note that while $\alpha_{\cal B} > 0$ is required for the existence of bound states, $\alpha_{\cal S}$ may be positive, negative, or zero. Here, $\alpha_{\cal S} \leqslant 0$, as seen in \cref{eq:alpha_SB}. However, extending the present model of \cref{eq:Lagrangian} by a light neutral scalar that mediates an attractive $XX$ and $XX^\dagger$ interaction~\cite{Petraki:2025}, or considering non-Abelian frameworks with charged scalars~\cite{Oncala:2021tkz,Oncala:2021swy} can give rise to $\alpha_{\cal S} >0$. The different possibilities for the scattering-state potential lead to different phenomenological implications~\cite{Petraki:2025}. In the following, we shall focus on the limit $\alpha_{\cal S} \to 0$.

\subsection{Unitarity violation \label{Sec:MonopoleTransitions_UniViolation}}

Unitarity imposes an upper bound on partial-wave inelastic cross sections~\cite{Flores:2024sfy},
\begin{align}
\sigma_{\ell}^{\rm inel} \leqslant \sigma_\ell^{\rm uni} /4,
\label{eq:UniLimit_Inelastic}
\end{align}
with $\sigma_\ell^{\rm uni}$ given by \cref{eq:Unitarity-cross section}. (A tighter constraint on $\sigma_{\ell}^{\rm inel}$ applies as a function of the elastic cross section~\cite{Flores:2024sfy}.)
To assess the compatibility of the monopole BSF processes with this bound, we consider the total inclusive inelastic cross section per partial wave due to BSF defined as
\begin{align}
\sigma_{\ell}^{\rm BSF} / \sigma_\ell^{\rm uni} \equiv
\sum_{n=1+\ell}^\infty \sigma_{n\ell}^{\rm BSF} / \sigma_\ell^{\rm uni} = b_\ell R_\ell (\zeta_{\cal B},\zeta_{\cal S}),
\label{eq:sigmaBSF_SumOverLevels}
\end{align}
where
\begin{align}
R_\ell (\zeta_{\cal B},\zeta_{\cal S}) \equiv \sum_{n=1+\ell}^\infty 
R_{n\ell} (\zeta_{\cal B},\zeta_{\cal S}) .
\label{eq:R_ell}
\end{align}
For $\zeta_{\cal S}/\zeta_{\cal B} \leqslant 1$, $R_\ell$ grows at low velocities (i.e., with increasing $\zeta_{\cal B}$), thus implying that unitarity is violated at sufficiently low velocities even for arbitrarily small couplings~\cite{Petraki:2025}. Similar conclusions hold for dipole and quadrapole transitions~\cite{Beneke:2024nxh}. 

In this section, we aim to compute $R_\ell$ in order to demonstrate the violation of unitarity, and use it in the next section to regularise the BSF cross sections, according to the unitarisation prescription of Ref.~\cite{Flores:2024sfy}. 


\subsubsection*{Limit $\alpha_{\cal S} \to 0$}

For simplicity, in this initial study of how the unitarisation method of Ref.\cite{Flores:2024sfy} affects DM freeze-out, we focus on $\alpha_{\cal S} = 0$. In our model, this limit arises when the $U(1)$ symmetry becomes global, $\alpha_V \to 0$. (Alternatively, it can be achieved by giving the gauge boson a large mass while keeping the global $U(1)$ symmetry of \cref{eq:Lagrangian} unbroken~\cite{Oncala:2019yvj}.) We defer the exploration of $\alpha_{\cal S} \neq 0$ to future work.

For $\alpha_{\cal S} =0$, using appropriate identities, \cref{eq:R_ell} becomes~\cite{Petraki:2025}
\begin{equation}
\begin{aligned}
\label{eq:rnlExact}
R_{n \ell}\left(\zeta_{\cal B}, 0\right) &= 
\left[\frac{2^{2 \ell+3} \ell!}{(2 \ell+1)!}\right]^2 
\frac{n(n+\ell)!}{(n-\ell-1)!}
\left[\frac{(\zeta_{\cal B}/n)^{2 \ell+5}}{\left(1+(\zeta_{\cal B}/n)^2\right)^{2 \ell+3}}\right] 
\\
&\times\left|{}_2 F_1\left[\frac{-n+\ell+1}{2}, \frac{n+\ell+1}{2} ; \ell+\frac{3}{2} ;\left(\frac{2 (\zeta_{\cal B}/n)}{1+(\zeta_{\cal B}/n)^2}\right)^2\right]\right|^2.
\end{aligned}
\end{equation}
Evaluating \cref{eq:rnlExact} numerically for large $n$ becomes computationally expensive due to the rapid oscillations of the hypergeometric function. We thus employ an analytical approximation valid in the regime $n \gg \ell$; since $R_\ell$ is dominated by contributions from highly excited states, this is sufficient for our purposes \cite{Petraki:2025}. For $n \gg \ell$, using $\lim _{a, b \rightarrow \pm \infty}{}_2 F_1(a, b ; c ; z/(ab))={ }_0 F_1( c ; z)$, allows to recast $R_{n \ell}$ as
\begin{align}
\label{eq:rnlAnalyticalApproximation}
R_{n \ell}\left(\zeta_{\cal B}, 0\right)   
\approx R_{n \ell}^{\rm A} (\zeta_{\cal B}) 
&\equiv 
2^6 \zeta_{\cal B}^2\left[\frac{\zeta_{\cal B}^3 / n^3}{\left(1+\zeta_{\cal B}^2 / n^2\right)^3}\right] \times\left|j_{\ell}\left(\frac{2 \zeta_{\cal B}}{1+\zeta_{\cal B}^2 / n^2}\right)\right|^2,
\end{align}
where $j_\ell$ is the spherical Bessel function of the first kind. This is in fact a good approximation for all velocities except around $\zeta_{\cal B} / n \sim 1$, even for small $n$~\cite{Petraki:2025}. We discuss further the applicability of this approximation in \cref{App:Approximations}.

The asymptotic forms of the spherical Bessel function, allow to further simplify \cref{eq:rnlAnalyticalApproximation} into the following form, enveloped by piecewise power-law scaling~\cite{Petraki:2025}
\begin{equation}
\label{eq:rnlAnalyticalApproximationFull}
R^{\rm A}_{n \ell}(\zeta_{\cal B}) \approx 
\begin{cases}{ 
(2/n)^3 \cdot \zeta_{\cal B}^{2 \ell+5}/ c_{\ell}^{2 \ell+2} ,} 
& \zeta_{\cal B} \leqslant c_{\ell}, 
\\[0.5ex] 
2^4 \cdot \dfrac{\left(\zeta_{\cal B} / n\right)^3}
{1+\left(\zeta_{\cal B} / n\right)^2}
\ \sin^2 \left(\dfrac{2\zeta_{\cal B}}{1+\zeta_{\cal B}^2/n^2}\right), 
& c_{\ell}<\zeta_{\cal B} \leqslant n^2 / c_{\ell}, 
\\[1em] 
2^3 \cdot n^{4 \ell+3}/ (c_{\ell}^{2 \ell+2} \zeta_{\cal B}^{2 \ell+1}), 
& n^2 / c_{\ell} < \zeta_{\cal B}, 
\end{cases}
\end{equation}
where $c_\ell$ has been chosen to ensure continuity, 
\begin{equation}
c_{\ell} \equiv 
\frac{1}{2}\left[\frac{(2 \ell+1)!}{2^{\ell+1/2} \ell!}\right]^{1 /(1+\ell)} 
\approx \frac{1+\ell}{e},
\end{equation}
and in the last step we used the Stirling approximation. Using the above, and summing over bound levels, $R_\ell$ can be approximated as~\cite{Petraki:2025}
\begin{equation}
\label{eq:RlSum}
R^{\rm A}_{\ell} \left(\zeta_{\cal B}\right) \approx 
\begin{cases}
\dfrac{4\zeta_{\cal B}^{2 \ell+5} }{c_{\ell}^{2 \ell+2}(1+\ell)^2}, 
& \zeta_{\cal B} < c_{\ell}, 
\\[1em] 
4\zeta_{\cal B} \cdot \ln \left[1+ \dfrac{\zeta_{\cal B}^2 }{ (1+\ell)^2} \right], 
& c_{\ell} < \zeta_{\cal B} < \dfrac{(1+\ell)^2}{c_{\ell}}, 
\\[1em] 
4\zeta_{\cal B}\left\{ 
\dfrac{1}{2(1+\ell)} \left[1-\left(\dfrac{(1+\ell)^2 / c_{\ell}}{\zeta_{\cal B}}\right)^{2 \ell+2}\right]+\ln \left(\dfrac{\zeta_{\cal B}}{c_{\ell}}\right)
\right\}, 
& \dfrac{(1+\ell)^2}{c_{\ell}} < \zeta_{\cal B},
\end{cases}
\end{equation}
where to achieve precision $\mathtt{p}$ in $R_\ell (\zeta_{\cal B})$, one must sum over $n\in [1+\ell,n_\mathtt{p}(\zeta_{\cal B})]$, with~\cite{Petraki:2025}
\begin{align}
n_\mathtt{p}(\zeta_{\cal B}) \approx  
\zeta_{\cal B} / \sqrt{\mathtt{p}\ln (\zeta_{\cal B}/c_\ell)} .
\label{eq:nmaxforRl}
\end{align}

\begin{figure}[t!]
\centering
\includegraphics[width=0.6\linewidth]{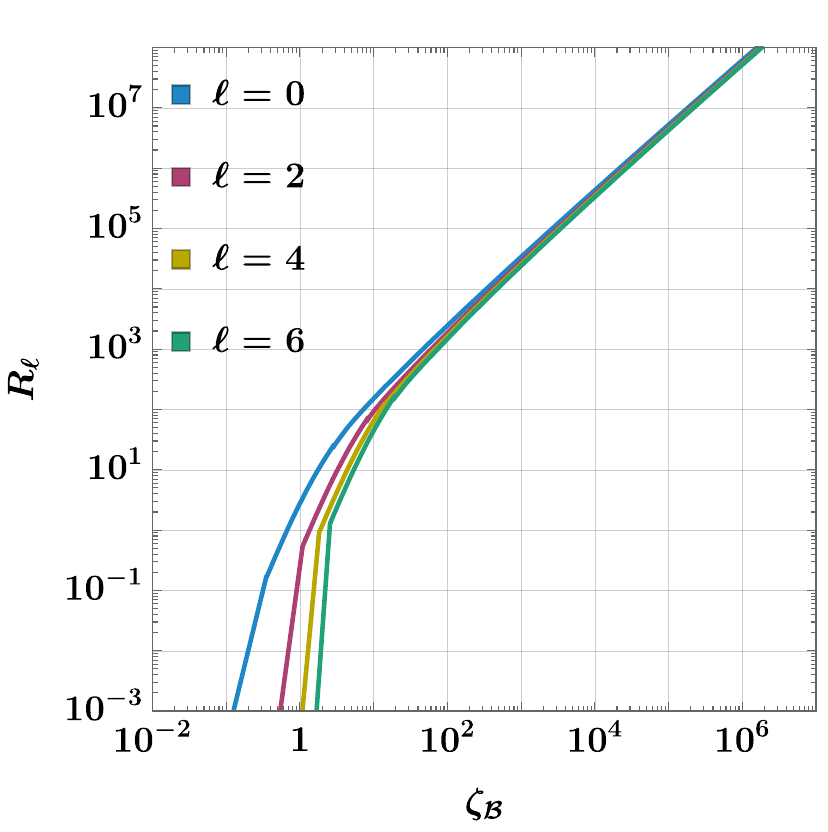}
\caption{The factor $R_\ell$ as a function of $\zeta_{\cal B}$ for different partial waves $\ell$.  The increase of $R_{\ell}$ with $\zeta_{\cal B}$ indicates that unitarity is violated at sufficiently low velocities for arbitrarily small couplings. We have employed the approximate expressions for $R_\ell$ given by \cref{eq:RlSum} \cite{Petraki:2025}.}
\label{fig:rlfigure}
\end{figure}

\Cref{eq:RlSum} is illustrated in \cref{fig:rlfigure}. Evidently, $R_\ell$ increases indefinitely with $\zeta_{\cal B}$. At high velocities ($\zeta_{\cal B} \ll 1+\ell$), the scaling of $R_\ell$ with $\zeta_{\cal B}$ depends on $\ell$, whereas at low velocities ($\zeta_{\cal B} \gg 1+\ell$), $R_\ell$ becomes insensitive to $\ell$ and grows faster than linearly with $\zeta_{\cal B}$ for all partial waves, 
$R_\ell \sim 4\zeta_{\cal B} \ln \zeta_{\cal B}$. 
This behaviour underscores the importance of considering all relevant bound levels for a given partial wave, as well as higher partial waves. Moreover, the growth of $R_\ell$ at low velocities necessitates proper regularisation to maintain consistency with unitarity and ensure accurate predictions for DM thermal decoupling.

In \cref{Sec:Decoupling}, we compute the thermally-averaged cross sections using the analytical approximation for $R_{n\ell}$ given in \cref{eq:rnlAnalyticalApproximation}, and obtain $R_\ell$ by summing over bound states up to $n \leqslant 10 \zeta_B$ as prescribed in \cref{eq:nmaxforRl}, to ensure better than $1\%$ precision. While \cref{eq:rnlAnalyticalApproximationFull,eq:RlSum} provide insight into the underlying scalings and are useful for making making analytical estimates, they can be imprecise in certain regimes, and we do not use them in numerical computations.

\subsection{Unitarity restoration \label{Sec:MonopoleTransitions_UniRestoration}}

\subsubsection*{General discussion}

The cause of unitarity violation --- and ultimately its resolution --- can be traced to the optical theorem. Considering only two-to-two particle scattering for simplicity, and projecting onto partial waves $\ell$, the unitarity of the $\mathbb{S}$ matrix, $\mathbb{S}\mathbb{S}^\dagger = \mathbb{S}^\dagger\mathbb{S} = \mathbb{1}$, implies~\cite{Flores:2024sfy}
\begin{align}
{\rm Im} \, M^{ab}_\ell (k^a,k^b) = 
\sum_c \big |M^{ab}_\ell (k^a,k^c) \big|^2 ,
\label{eq:OpticalTheorem_PartialWaves}
\end{align}
with $M_\ell^{ab} (k^a,k^b) \equiv 
[k^a k^b/ (2^{\delta_a} 2^{\delta_b} s)]^{1/2} \, 
{\cal M}_\ell^{ab} (k^a,k^b)$, where 
${\cal M}_\ell^{ab} (k^a,k^b)$ is the amplitude for the $a\to b$ transition, 
$k^a$ is the magnitude of the three-momentum of a two-particle state $a$ in the CM frame, and $s$ is the first Mandelstam variable. As before, $\delta_a =0$ or 1, if the two particles in the state $a$ are distinguishable or identical respectively. The above can be generalised to include $2 \to N \geqslant 3$ scattering, without affecting the considerations that follow.

\Cref{eq:OpticalTheorem_PartialWaves} leads to (combined) upper bounds on elastic and inelastic cross sections~\cite{Flores:2024sfy}. However, because it is non-linear, any calculation truncated at fixed order in a coupling generally violates \cref{eq:OpticalTheorem_PartialWaves} ---that is, fixed-order calculations violate unitarity. Importantly, this can occur even if the cross sections are consistent with their respective unitarity limits; indeed, the optical theorem \eqref{eq:OpticalTheorem_PartialWaves} constrains not only the modulus but also the phase of amplitudes.

For phenomenological purposes, however, unitarity violation is typically considered problematic when the cross sections exceed their respective unitarity limits. This is expected to occur near resonances or for large couplings. In recent years, the computation of BSF processes in a variety of theories, motivated by DM, has revealed another source of unitarity violation in inelastic processes: large overlap integrals between the initial and final non-relativistic wavefunctions, which arise when the effective Hamiltonians governing the two states are different~\cite{Oncala:2019yvj}. As already shown in \cref{Sec:MonopoleTransitions_UniViolation}, the unitarity violation that ensues can be severe, appearing for arbitrarily small couplings at sufficiently low velocities. 

Irrespective of the source of unitarity violation, restoring unitarity requires resumming the relevant interactions to all orders. Two well-known examples illustrate this point. First, Breit–Wigner resonances are regulated by the imaginary part of the propagator's self-energy, which arises from resumming the one-particle-irreducible (1PI) diagrams that generate its decay width. Second, the elastic scattering of two non-relativistic particles appears to violate unitarity at tree level: while the tree-level amplitude is real, \cref{eq:OpticalTheorem_PartialWaves} implies a positive imaginary part if the amplitude is non-zero. This apparent unitarity violation is resolved by resumming the tree-level elastic diagrams, which form part of the 2PI kernel and generate a real potential in Schr\"odinger’s equation. 

The restoration of unitarity in two-particle inelastic scattering builds on the above two cases. Upon squaring, $2\to N$ inelastic processes generate \emph{imaginary} contributions to the 2PI kernel of the initial state, analogously to how a particle’s decay width generates an imaginary part in its self-energy. Reference~\cite{Flores:2024sfy} argued that the form of the imaginary potential is dictated by the optical theorem \eqref{eq:OpticalTheorem_PartialWaves}, generalised to off-shell amplitudes. 
Such contributions to the potential may naively appear suppressed by higher powers of the couplings and loop factors compared to purely elastic tree-level terms. However, when inelastic cross sections appear to approach or exceed the unitarity limit, these contributions to the 2PI kernel may become significant.  Reference~\cite{Flores:2024sfy} showed that including them in the potential ensures consistency with the combined unitarity bounds on elastic and inelastic cross sections. 

Reference~\cite{Flores:2024sfy} further found that under certain analyticity and convergence conditions for the unregulated inelastic amplitudes (computed using only the real potential), it is possible to express the regulated inelastic cross sections (which include both the real and the imaginary potentials) in terms of the unregulated ones, through a remarkably simple formula,
\begin{align}
y_{\ell, \rm reg}^j  = \dfrac{y_{\ell,\rm unreg}^j}{(1+y_{\ell,\rm unreg})^2},
\label{eq:yreg}
\end{align}
where 
$y_{\ell,\rm (un)reg}^j \equiv \sigma_{\ell, \rm (un)reg}^{{\rm inel}, j} / \sigma_\ell^{\rm uni}$ 
and $y_{\ell,\rm (un)reg} = \sum_j y_{\ell,\rm (un)reg}^j$, 
with $j$ denoting the inelastic channel.  
Evidently, \cref{eq:yreg} is consistent with the unitarity bound \eqref{eq:UniLimit_Inelastic} for any $y_{\ell,\rm unreg}$, with the regulating factor becoming important when $y_{\ell, {\rm unreg}} \gtrsim {\cal O} (1)$. 
The corresponding expression for the elastic cross sections can be found in~\cite{Flores:2024sfy}, but is not needed here. The simplicity of \cref{eq:yreg} stems from the fact that the imaginary part of the 2PI kernel is related to the hard-scattering inelastic amplitudes (i.e.~the inelastic amplitudes without any 2PI resummation) via the optical theorem \eqref{eq:OpticalTheorem_PartialWaves}~\cite{Flores:2024sfy}.
When the required analyticity or convergence conditions are not met, corrections to~\cref{eq:yreg} generally arise at higher orders in the relevant couplings~\cite{Petraki:2025}.

\subsubsection*{Bound-state formation}

The convergence condition required for \cref{eq:yreg} holds for all BSF processes~\cite{Petraki:2025}. However, the unregulated BSF amplitudes exhibit singularities in the complex momentum plane that affect the regularisation prescription, introducing higher-order corrections to \cref{eq:yreg}. These corrections become important only at large couplings~\cite{Petraki:2025}, and we neglect them here. 

We emphasise that $y_{\ell,\rm unreg}$ should include all inelastic processes with the same initial state. In the limit of interest, $\alpha_V \to 0$, the $XX$ state has no 2-to-2 direct annihilation channels. The 2-to-3 annihilations, $XX\to \Phi\Phi\Phi^\dagger$, are subdominant with respect to individual 2-to-2 BSF processes, and entirely negligible compared to the unregulated inclusive BSF cross sections that encompass capture into the tower of $n\in [1+\ell, \infty)$ bound states. We shall thus consider only the BSF processes \eqref{eq:BSF_PhiEmission}.\footnote{
The process $XX \to \Phi\Phi\Phi^\dagger$ is also subdominant compared to $XX^\dagger \to \Phi\Phi^\dagger$, and we therefore neglect it entirely in the freeze-out calculation of \cref{Sec:Decoupling}. We note however, that the $XX \to \Phi\Phi\Phi^\dagger$ cross section is, in principle, very significantly regulated by the unregulated BSF cross sections, according to \cref{eq:yreg}. Since we neglect this process for phenomenological purposes, we do not discuss its regularisation further here.  By contrast, the $XX^\dagger \to \Phi\Phi^\dagger$ cross sections are neither regulated by nor regulate the BSF processes \eqref{eq:BSF_PhiEmission}, as they involve different initial states.} 

\begin{figure}[b!]
\centering
\includegraphics[width=0.95\linewidth]{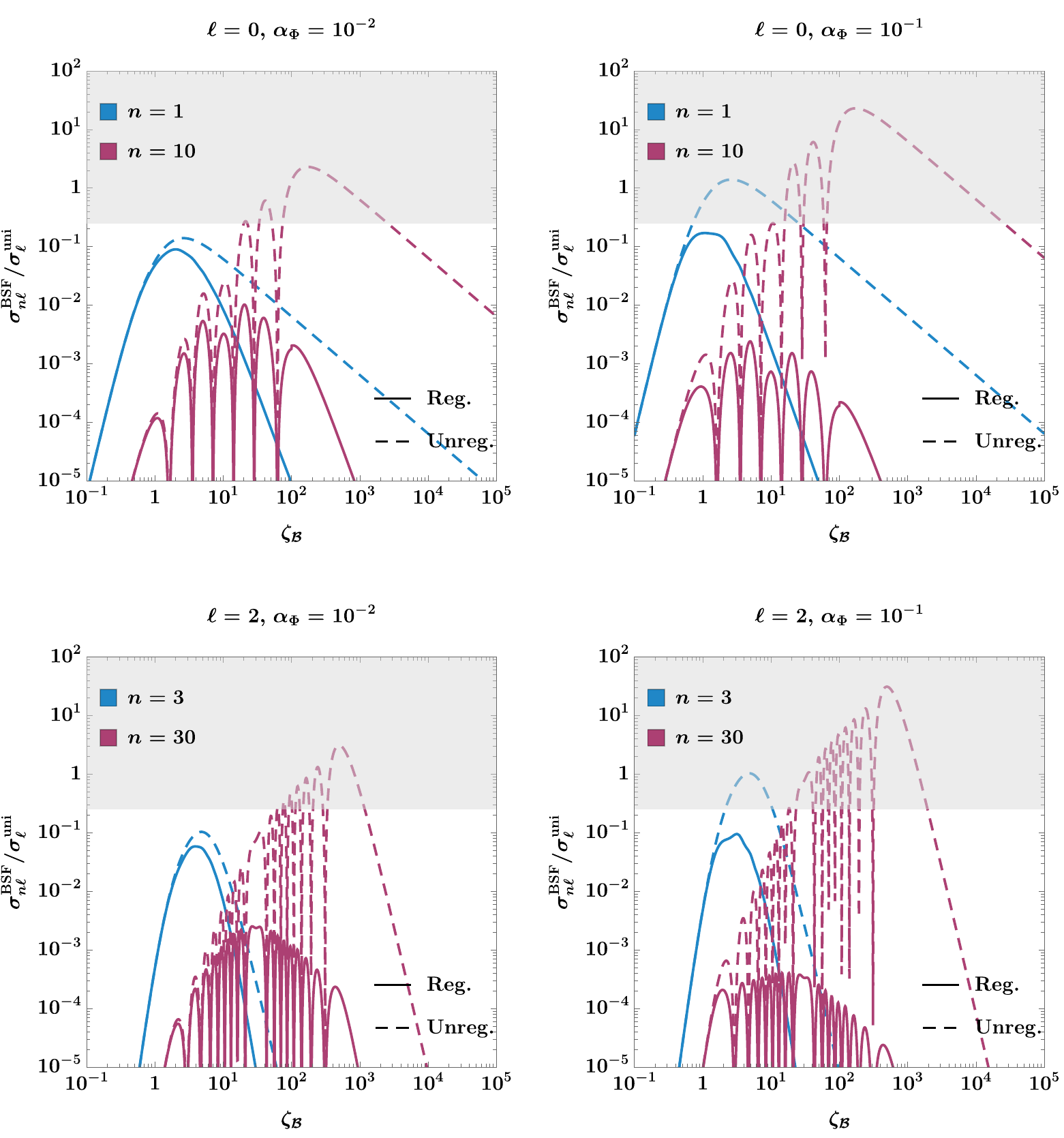}
\caption{The regulated (solid lines) and unregulated (dashed lines) BSF cross sections, normalised to $\sigma_\ell^{\rm uni}$, as functions of $\zeta_{\cal B} = \alpha_{\cal B}/v_{\rm rel}$, for $\alpha_{\cal S}=0$ and $n= (\ell +1), 10 (\ell+1)$ and $\ell \in \{0,2\}$, and two indicative values of the coupling $\alpha_\Phi \in \{ 10^{-2}, 10^{-1} \}$. The cross sections are calculated using the analytical approximation of \cref{eq:rnlAnalyticalApproximation} and the regulating factor $R_\ell$ by summing \cref{eq:rnlAnalyticalApproximation} over $n\in[1+\ell, 10 \zeta_B)$. The shaded region denotes the region forbidden by unitarity.}
\label{fig:UnthermallyAveragedCrossSectionPlots}
\end{figure}

Considering the above, \cref{eq:yreg} adapted to the present case, reads
\begin{equation}
\label{eq:Rnlregulated}
\dfrac{\sigma_{n\ell, {\rm reg}}^{\rm BSF}}{\sigma_\ell^{\rm uni}}
=\dfrac{\sigma_{n\ell, {\rm unreg}}^{\rm BSF} / \sigma_\ell^{\rm uni}}
{(1 + \sigma_{\ell, {\rm unreg}}^{\rm BSF} / \sigma_\ell^{\rm uni})^2}
=\dfrac{b_\ell R_{n \ell}}{\left(1+ b_\ell R_{\ell}\right)^2},
\end{equation}
where $\sigma_{n\ell, {\rm unreg}}^{\rm BSF} / \sigma_\ell^{\rm uni} = b_\ell R_{n\ell}$ as per \cref{eq:BSFCrossSection-to-UnitarityBound}, with $b_\ell = \alpha_{\Phi} \delta_{\ell,\rm even}$ according to \cref{eq:bl} for $\alpha_{\cal S}\to 0$, and $R_{\ell}$ is defined in \cref{eq:R_ell}.  
As noted in~\cite{Flores:2024sfy}, if the momentum scaling of the unregulated inelastic cross sections is different from that of the unitarity limit, the regularisation \eqref{eq:Rnlregulated} modifies their momentum dependence. \Cref{fig:UnthermallyAveragedCrossSectionPlots} shows the scaling of individual regulated and unregulated cross sections with $\zeta_{\cal B} = \alpha_{\cal B}/v_{\rm rel}$, which we now discuss.

The unregulated BSF cross sections exhibit the scaling of \cref{eq:rnlAnalyticalApproximationFull}, and are suppressed both at high and low velocities (at least for $\ell>0$); recasting \cref{eq:rnlAnalyticalApproximationFull} for $\sigma_{n\ell, {\rm unreg}}^{\rm BSF} v_{\rm rel}$, 
\begin{align}
\sigma_{n\ell, {\rm unreg}}^{\rm BSF} v_{\rm rel}  
\approx \dfrac{2^8 \pi (2\ell+1)}{m_X^2}  
\times 
\begin{cases}
\dfrac{1}{c_\ell^{2\ell+2} \ n^3}
\left(\dfrac{\alpha_{\cal B}}{v_{\rm rel}}\right)^{2\ell+6},
&\dfrac{\alpha_{\cal B} }{c_\ell} < v_{\rm rel},
\\[1em]
\dfrac{n^{4\ell+3}}{c_\ell^{2\ell+2}}
\left(\dfrac{v_{\rm rel}}{\alpha_{\cal B}}\right)^{2\ell},
&v_{\rm rel} < \dfrac{\alpha_{\cal B} \, c_\ell }{n^2},
\end{cases}
\label{eq:BSFunreg_HighAndLowVelocities}
\end{align}
where we took into account that $\alpha_{\cal B} = \alpha_{\Phi}$. 
At low velocities, we recognise the standard velocity scaling of perturbative inelastic processes without Sommerfeld effect, since indeed $\alpha_{\cal S}=0$; however, in contrast to annihilation processes, here the $v_{\rm rel}^{2\ell}$ suppression is ameliorated by an $\alpha_{\cal B}^{-2\ell}$ enhancement, manifesting the non-perturbative nature of the bound-state ladder.
For intermediate velocities, $\sigma_{n \ell, \rm unreg}^{\rm BSF}$ oscillates with $\zeta_{\cal B}$, the oscillations increasing with $n$ and reaching their maximum value around $\zeta_{\cal B} \sim n^2/c_\ell$. In this regime, depending on the coupling, even individual unregulated BSF cross sections tend to violate unitarity. 

Regularisation does not affect the BSF cross sections at small $\zeta_{\cal B}$, where $R_\ell$ is small, but alters their intermediate- and low-velocity behaviour, where $R_\ell$ is large. When the unregulated cross sections approach the unitarity limit, an important transition occurs: while the unregulated  cross sections continue to increase with $\zeta_{\cal B}$, the regulated  cross sections begin to decrease~\cite{Flores:2024sfy}.  In the oscillation region, the regulated cross sections experience significant damping, which becomes especially pronounced for higher partial waves. The peak of individual regulated cross sections shifts toward smaller $\zeta_{\cal B}$ values. At low velocities, for $(1+\ell)^2/c_\ell \lesssim \zeta_{\cal B}$, the regulating factor becomes 
$(1+ \alpha_{\Phi} R_\ell)^{-2} \sim 
(4\alpha_{\cal B} \zeta_{\cal B} \ln{\zeta_{\cal B}/c_{\ell}})^{-2}$, 
causing the regulated cross sections to decrease faster with decreasing velocity compared to the unregulated ones. Focusing on $\zeta_{\cal B} \gtrsim (1+\ell)^2/c_\ell$, \cref{eq:rnlAnalyticalApproximationFull,eq:RlSum} imply
\begin{align}
\sigma^{\rm BSF}_{n \ell, \rm reg} v_{\rm rel}  
&\approx  
\dfrac{16\pi (2\ell+1)}
{m_X^2 \alpha_{\cal B}^2 \, \ln^2 (\zeta_{\cal B}/c_\ell)}
\times 
\nonumber \\[1ex]
&\times 
\begin{cases}
\dfrac{2}{n}
\left(\dfrac{\zeta_{\cal B}^2/n^2}{1+\zeta_{\cal B}^2/n^2}\right)
\sin^2 \left(\dfrac{2\zeta_{\cal B}}{1+\zeta_{\cal B}^2/n^2} \right),
&\quad
\dfrac{(1+\ell)^2}{c_\ell} < \zeta_{\cal B} < \dfrac{n^2}{c_\ell},
\\[1em]
\dfrac{n^{4\ell+3}}{c_\ell^{2\ell+2}}
\dfrac{1}{\zeta_{\cal B}^{2\ell+2}},
&\quad
\dfrac{n^2}{c_\ell}<\zeta_{\cal B}.
\end{cases}
\label{eq:BSFreg}
\end{align}
%

\clearpage
\section{Dark matter thermal decoupling \label{Sec:Decoupling}}

We now examine the impact of regularisation on DM thermal decoupling and relic density, considering the monopole capture processes \eqref{eq:BSF_PhiEmission}, and summing over all bound levels that contribute significantly to each partial wave. The DM freeze-out is computed using both (i) the unregulated and (ii) the regulated cross sections. In \cref{Sec:Decoupling_BoltzmannEqs}, we present the Boltzmann equations (BEqs) that govern the dynamics of bound and free states. The relevant interaction rates are introduced in \cref{Sec:Decoupling_Rates}. In \cref{Sec:Decoupling_Stages}, we discuss how regularisation modifies the DM thermal decoupling. Finally, in \cref{Sec:Decoupling_Results}, we examine the impact of higher partial waves and regularisation on the DM relic density.

\subsection{Boltzmann equations\label{Sec:Decoupling_BoltzmannEqs}}

We parametrise time evolution in terms of the variables,
\begin{equation}
x \equiv\frac{m_X}{T},  \qquad 
z\equiv \frac{|E_{n=1}|}{T} = \frac{\alpha_{\cal B}^2x}{4},
\end{equation}
which we will use interchangeably as convenient. We define the yields of the scattering and bound states, $Y_X \equiv  n_X/s$ and $Y_{n\ell} \equiv n_{n\ell}/s$, respectively, with $ s(T) = \left(2 \pi^2 /45\right) \cdot  g_{* S}(T) \, T^3$ being the entropy density of the universe and $g_{* S}(T)$ the corresponding degrees of freedom (dof) at temperature $T$. The evolution of the comoving number densities then reads  \cite{Binder:2021vfo}
\begin{subequations}
\label{eq:SystemBoltzmann}
\label[pluralequation]{eqs:SystemBoltzmann}
\begin{align}
\frac{d Y_X}{d x} 
=& -\frac{\lambda}{x^2} \left[ 
\langle\sigma^{\rm ann} v_{\rm rel}\rangle 
\left(Y_X^{2}-\left(Y_X^{\rm eq}\right)^{2} \right)  
+ \sum_{n\ell}  \langle \sigma_{n\ell}^{\rm BSF} v_{\rm rel}\rangle
\left(Y_X^{2}-\frac{Y_{n\ell}}{Y_{n\ell}^{\rm eq}} \left(Y_X^{\rm eq}\right)^{2}\right) 
\right], 
\label{eq:SystemBoltzmann_chi}
\\
\frac{d Y_{n\ell}}{d x} 
=& -\Lambda x \left[ 
  \langle \Gamma_{n\ell}^{\rm dec} \rangle \left(Y_{n\ell} - Y_{n\ell}^{\rm eq}\right) 
+ \langle \Gamma_{n\ell}^{\rm ion} \rangle  \left(Y_{n\ell} - \left(\frac{Y_X}{Y_X^{\rm eq}} \right)^{2} Y_{n\ell}^{\rm eq}\right) \right],
\end{align}
\end{subequations}
where 
\begin{align}
\lambda \equiv \sqrt{\frac{\pi}{45}} m_{{\rm Pl}} \, m_X \cdot g_*^{1 / 2}, 
\qquad
\Lambda  \equiv \frac{\lambda}{s x^3} =
\sqrt{\frac{45}{4 \pi^3}} \frac{m_{{\rm Pl}}}{m_X^2} \cdot \frac{g_*^{1/2}}{g_{* S}},
\label{eq:Lambda}
\end{align}
with $m_{\rm Pl} \approx 1.2 \cdot 10^{19} \, {\rm GeV}$ being the Planck mass, and
\begin{align}
g_*^{1/2} &\equiv \frac{g_{* S}}{\sqrt{g_{* \rho}}} 
\left(1 - \frac{x}{3 g_{* S}} \frac{d g_{* S}}{dx} \right),
\label{eq:gstar}
\end{align}
where $g_{*\rho}$ denotes the energy dof. 
In the non-relativistic regime, the equilibrium comoving number densities of the scattering and bound states are \cite{Binder:2021vfo}
\begin{subequations}
\label{eq:EquilibriumDensities}
\label[pluralequation]{eqs:EquilibriumDensities}
\begin{align}
Y_X^{\rm eq}  &\simeq 
\frac{90}{(2 \pi)^{7/2}} \frac{g_X}{g_{* S}} x^{3/2} e^{-x},
\label{eq:EquilibriumDensities_YX}
\\
Y_{n\ell}^{\rm eq}  &\simeq 
\frac{90}{(2\pi)^{7/2}} \frac{g_{n\ell}}{g_{* S}}  
(2x-\left|E_n\right| / T)^{3/2}
\, e^{-2 x}
\, e^{\left|E_n\right| / T},
\label{eq:EquilibriumDensities_YB}
\end{align}
\end{subequations}
where $g_X = 2$ and $g_{n\ell}= 2\ell +1$. The thermally-averaged cross sections for $X$ annihilations directly into radiation and BSF, $\langle \sigma^{\rm ann} v_{\rm rel} \rangle$ and $\langle \sigma^{\rm BSF}_{n \ell} v_{\rm rel} \rangle$ respectively, along with the bound-state decay and ionisation rates, $\langle \Gamma_{n\ell}^{\rm dec} \rangle, \langle \Gamma_{n\ell}^{\rm ion} \rangle$, are introduced in the following section.

\smallskip

For simplicity and concreteness, we assume that the dark sector plasma is at the same temperature as the SM plasma throughout DM decoupling. This raises the potential concern of extra radiation from dark relativistic dof, possibly in tension with CMB constraints. However, heavy DM decouples early, and the subsequent decoupling of SM dof lowers the dark radiation temperature relative to the SM, compared to the temperature at DM decoupling. This is sufficient to satisfy constraints. For DM masses that decouple later, a somewhat lower initial temperature for the dark plasma must be assumed. While this can affect numerical results, in ways that have been explored in the literature, it does not substantially impact the important features and conclusions of our calculations. To maintain a unified and simple description, and avoid complications tangential to the goals of this work, we assume the SM temperature throughout.

\subsubsection*{Effective DM depletion \label{sec: Effective DM depletion}} 

The system of coupled Boltzmann equations \eqref{eq:SystemBoltzmann} can be reduced to a single effective equation under the assumption that the bound levels reach a quasi-steady state~\cite{Ellis:2015vaa}. This approximation is justified as long as either the formation or decay rates of the bound states are sufficiently large. If this is not the case, their impact on DM decoupling, as well as their contribution to the effective depletion rate are negligible, still allowing the use of the effective description~\cite{Binder:2021vfo}. The effective equation allows to include a very large number of bound levels rather efficiently, as is needed here. It reads~\cite{Ellis:2015vaa,Binder:2021vfo}
\begin{equation}
\label{eq:EffectiveBoltzmann}
\frac{d Y_X}{d x} = 
- \frac{\lambda}{x^2} \langle \sigma^{\mathrm{eff}} v_{\rm rel} \rangle 
\left[Y_X^2-\left(Y_X^{\mathrm{eq}}\right)^2\right] ,
\end{equation}
i.e.~$Y_X^{\rm eq}$ remains the attractor solution of the system~\cite{Binder:2021vfo}, with the effective depletion cross section being
\begin{equation}
\label{eq:EffectiveCrossSection}
\langle \sigma^{\mathrm{eff}} v_{\rm rel} \rangle  = 
\sum_\ell 
\left( 
\langle \sigma^{\rm ann}_{\ell} v_{\rm rel} \rangle + \sum_{n=\ell+1}^{n_{\rm max}} 
\epsilon_{n\ell} \langle \sigma^{\rm BSF}_{n \ell} v_{\rm rel} \rangle 
\right).
\end{equation}
The factors $\epsilon_{n\ell} \in [0,1]$ quantify the efficiency of a bound state $n\ell$ to deplete DM, encapsulating the interplay between ionisation, decay, and transitions processes. 
They can be expressed in closed form in terms of the corresponding rates of the entire bound-state network, upon inversion of a system of algebraic equations~\cite{Binder:2021vfo}. In the absence of bound-to-bound transitions, they take the very simple form~\cite{Ellis:2015vaa,Binder:2021vfo}
\begin{align}
\label{eq:rB_NoTrans}
\epsilon_{n\ell} = \dfrac{\langle \Gamma_{n\ell}^{\rm dec} \rangle}
{\langle \Gamma_{n\ell}^{\rm dec} \rangle + \langle \Gamma_{n\ell}^{\rm ion}\rangle}. 
\end{align}
We sum the bound levels $n$ up to $n_{\rm max}$ that achieves the desired precision, as we discuss in \cref{Sec:Decoupling_Rates_BSF}.

\subsection{Interaction rates \label{Sec:Decoupling_Rates}}

\subsubsection{Annihilation \label{Sec:Decoupling_Rates_Annihilation}}

\begin{figure}[h!]
\centering
\begin{tikzpicture}[line width=1pt, scale=0.9]
\begin{scope}[shift={(0,2)}]

\begin{scope}[shift={(-2,0)}]
\node at (-1.4, 1) {$X$};
\node at (-1.4, 0) {$X^{\dagger}$};
\draw[fermion]    (-1,1) -- (0,1);
\draw[fermion]    ( 0,0) -- (0,1);
\draw[fermionbar] (-1,0) -- (0,0);
\draw[scalarbar] (0,0) -- (1,0);
\draw[scalar] (0,1) -- (1,1);
\node at (1.3,1) {$\Phi$};
\node at (1.35,0) {$\Phi^\dagger$};
\end{scope}
\begin{scope}[shift={(2,0)}]
\node at (-1.4, 1) {$X$};
\node at (-1.4, 0) {$X^{\dagger}$};
\draw[fermion]    ( -1,1) -- (-0.5,0.5);
\draw[fermionbar] ( -1,0) -- (-0.5,0.5);
\draw[vector] (-0.5,0.5) -- (1,0.5);
\draw[scalar] (1,0.5)   -- (1.5,1);
\draw[scalarbar] (1,0.5)   -- (1.5,0);
\node at (1.8,1) {$\Phi$};
\node at (1.85,0) {$\Phi^\dagger$};
\end{scope}

\begin{scope}[shift={(5.6,0)}]
\node at (-0.45, 1) {$X$};
\node at (-0.45, 0) {$X^{\dagger}$};
\draw[fermion]    ( 0,1) -- (0.5,0.5);
\draw[fermionbar] ( 0,0) -- (0.5,0.5);
\draw[scalar] (0.5,0.5)   -- (1,1);
\draw[scalarbar] (0.5,0.5)   -- (1,0);
\node at (1.3,1) {$\Phi$};
\node at (1.35,0) {$\Phi^\dagger$};
\end{scope}

\end{scope}

\end{tikzpicture}
\caption{Tree-level diagrams contributing to the $X X^\dagger \rightarrow \Phi \Phi^\dagger$ annihilation. 
\label{fig:XXPhiPhiabelian}}
\end{figure}
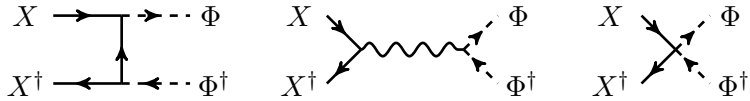

The $X$ species can annihilate into $\Phi$ bosons, via
$X X^\dagger \rightarrow \Phi \Phi^\dagger$. The tree-level diagrams contributing to this process are shown in \cref{fig:XXPhiPhiabelian}.
Due to the attractive long-range forces in $XX^\dagger$ pairs, the annihilation rate is Sommerfeld enhanced. Diagrammatically, this effect corresponds to the resummation of one-boson-exchange diagrams shown in \cref{fig:2PIdiagrams}.

The $\ell$-wave velocity-weighted cross section can be written as (see \cref{App:Annihilation and Decay Rates})
\begin{equation}
\sigma^{\mathsmaller{X X^\dagger \rightarrow \Phi \Phi^\dagger}}_\ell v_{\rm rel} =  
(\sigma^{\mathsmaller{X X^\dagger \rightarrow \Phi \Phi^\dagger}}_{\ell, {\rm tree}} 
v_{\rm rel})
S_{\ell}(\zeta_\Phi),
\end{equation}
where $\zeta_\Phi \equiv \alpha_\Phi / v_{\rm rel}$, and the $\ell$-wave Sommerfeld factor, $S_{\ell}(\zeta_\Phi)$, has been defined in the Coulomb limit in \cref{eq:SommerfeldFactor-l-wave}. Neglecting the $\Phi$ mass, the  tree-level partial-wave $X X^\dagger \rightarrow \Phi \Phi^\dagger$ cross section is
\begin{equation}
\label{eq:PerturbativeAnnihilationCrossSection}
\sigma^{\mathsmaller{X X^\dagger \rightarrow \Phi \Phi^\dagger}}_{\ell, {\rm tree}} v_{\rm rel} = 
\dfrac{4 \pi(2\ell +1)}{m_X^2} 
\dfrac{(\ell!)^4 }{[(2 \ell+1)!]^2} 
\left( \alpha_\Phi+ \alpha_V \cdot \delta_{1,\ell} -\frac{\lambda_{X\Phi}}{8\pi } \cdot \delta _{0,\ell} \right)^2 
\ v_{\rm rel}^{2 \ell}.
\end{equation}
The quartic vertex and the $s$-channel $V_\mu$ exchange diagrams contribute only to the $s$- and $p$-wave annihilation, respectively. In what follows, we take $\alpha_V \to 0$, and also assume that the $\lambda_{X \Phi}$ contribution is subdominant, as this coupling is not essential in the dynamics we aim to explore. For \(\lambda_{X\Phi} \lesssim 0.1\, \alpha_\Phi\), its effect on the $s$-wave annihilation and decay processes is less than 1\%.

As is standard, \cref{eq:PerturbativeAnnihilationCrossSection} shows that the contribution from higher partial waves to the perturbative cross section is velocity-suppressed,  $\sigma^{\mathsmaller{X X^\dagger \rightarrow \Phi \Phi^\dagger}}_{\ell, {\rm tree}} v_{\rm rel} \propto v_{\rm rel}^{2\ell}$. Once the Sommerfeld enhancement is taken into account, at low velocities being 
$S_\ell \simeq [2\pi / (\ell!)^2] (\alpha_\Phi/v_{\rm rel} )^{2\ell +1}$, 
the velocity suppression morphs into a suppression by powers of the coupling, such that the full cross section scales as 
$\sigma^{\mathsmaller{X X^\dagger \rightarrow \Phi \Phi^\dagger}}_\ell v_{\rm rel} \propto  \alpha_\Phi^{2 \ell+2} (\alpha_\Phi/v_{\rm rel})$~\cite{Baldes:2017gzw}. Higher partial waves in annihilation processes remain thus subdominant.

The $XX^\dagger$ annihilation processes are not subject to regularisation by the large $XX$ BSF cross sections, since the initial states are different. They independently approach their unitarity limit at $2\pi \alpha_{\Phi}^{2\ell+3} \sim [(2\ell+1)!/\ell!]^2$, or $\alpha_{\Phi} \sim 0.54$ for $\ell=0$, where regularisation becomes important. While we regularise them for consistency, using the prescription of Ref.~\cite{Flores:2024sfy}, this is only marginally important near the upper edge of the parameter range we consider.

In the $\alpha_V \to 0$ limit, we neglect $XX^\dagger \to V^\mu V^\mu$ and $XX \to \Phi V^\mu$ annihilations, thus setting $\sigma^{\rm ann} = \sigma^{\mathsmaller{XX^\dagger \to \Phi\Phi^\dagger}}$. Assuming Maxwell-Boltzmann statistics, the thermally-averaged annihilation cross section is
\begin{equation} 
\label{eq:ThermalAver_Annihilations}  
\left\langle\sigma^{\rm ann}_{\ell} v_{\rm rel}\right\rangle
=\frac{x^{3 / 2}}{2 \sqrt{\pi}} \int_0^{\infty} 
d v_{\mathrm {rel}} v_{\mathrm {rel}}^2 
\left(\sigma^{\rm ann}_\ell v_{\rm rel}\right) 
e^{-x v_{\mathrm{rel }}^2 / 4}.
\end{equation}

\subsubsection{Bound-state formation \label{Sec:Decoupling_Rates_BSF}}

\begin{figure}[t!]
\centering
\includegraphics[width=1.0\textwidth]{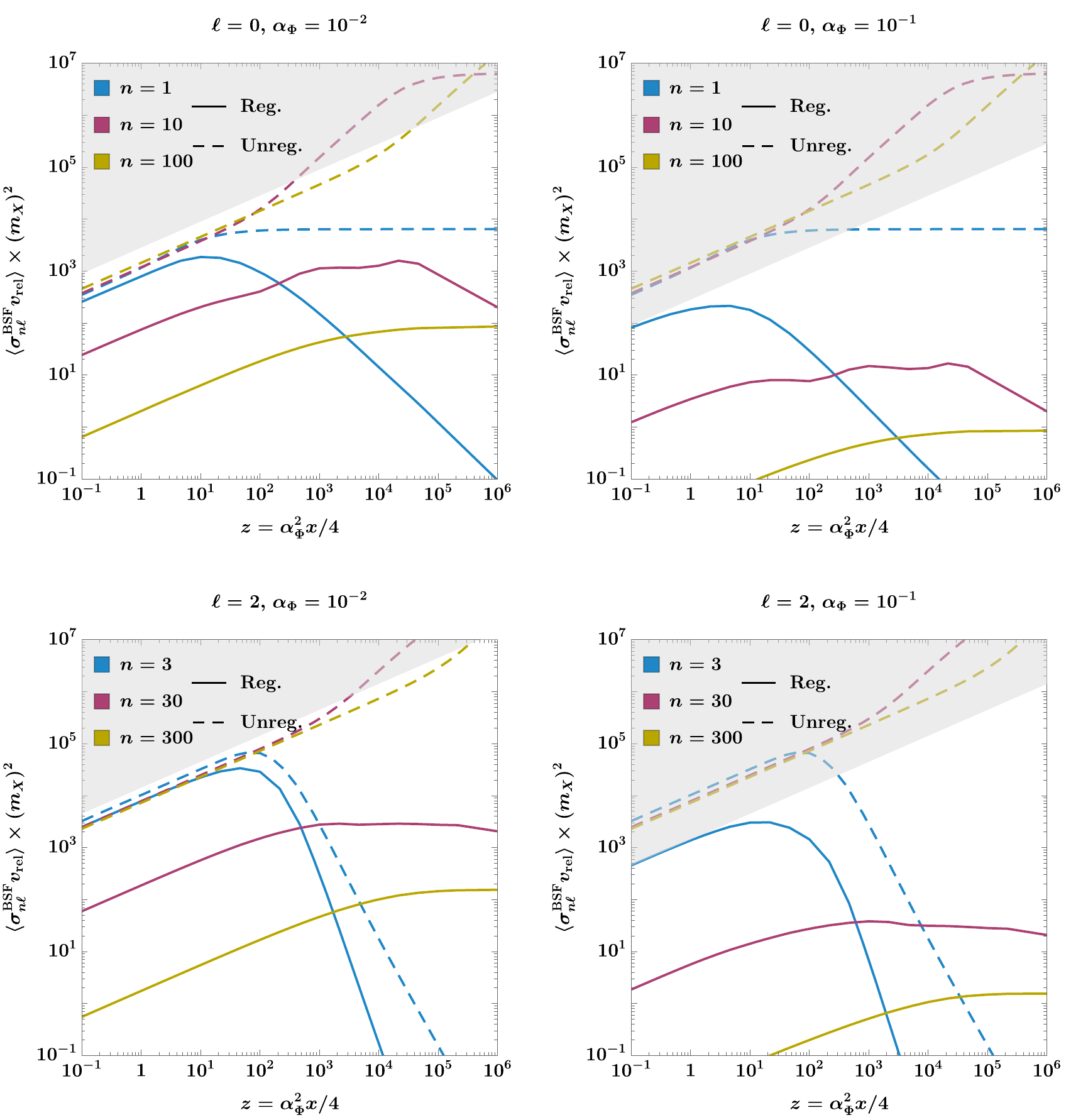}
\caption{Thermally-averaged BSF cross sections weighted by the relative velocity as a function of $z = |E_{n=1}|/T$ for $n = (\ell+1),\ 10(\ell+1),\ 100(\ell+1)$. The gray-shaded region indicates where the cross section exceeds the unitarity bound, i.e., $\langle \sigma_\ell^\mathrm{BSF} v_\mathrm{rel} \rangle > \langle \sigma_\ell^\mathrm{uni} v_\mathrm{rel} \rangle / 4$.}
\label{fig:ThermallyAveragedCrossSectionPlots}
\end{figure}

The thermally-averaged BSF cross sections are
\begin{align}
\label{eq:ThermalAver_BSF}
&\left\langle\sigma^{\rm BSF}_{n\ell} v_{\rm rel}\right\rangle =
\frac{x^{3 / 2}}{2 \sqrt{\pi}} 
\int_0^{\infty} d v_{\rm rel} \, v_{\rm rel}^2 
\left(\sigma^{\rm BSF}_{n\ell}v_{\rm rel}\right) 
e^{-x v_{\rm rel}^2 / 4} 
\times  {\rm BE}_n (v_{\rm rel}, T), 
\end{align}   
where ${\rm BE}_n (v_{\rm rel},T)$ is the Bose enhancement factor, due to the ultra-soft emitted boson,
\begin{align}
{\rm BE}_n(v_{\rm rel}, T) \equiv 
1 + \frac{1}{e^{\omega_n/T} -1},
\label{eq:BoseEnhancement_T}
\end{align}
with $\omega_n = \mu v_{\rm rel}^2 / 2 + |E_{n}|$ being the energy of the radiated $\Phi$. The Bose enhancement becomes particularly significant at high temperatures, where it is essential to maintain detailed balance~\cite{vonHarling:2014kha}. We discuss its impact in more detail below. Having taken the limit $\alpha_V \to 0$, the BSF processes depend solely on one coupling, $\alpha_\Phi$, and the relevant scale is set by the binding energy. It is then convenient to recast \cref{eq:ThermalAver_BSF} in terms of $z = |E_{n=1}| / T$~\cite{vonHarling:2014kha}, 
\begin{align}
\label{eq:ThermalAver_BSF_z}
&\left\langle\sigma^{\rm BSF}_{n\ell} v_{\rm rel}\right\rangle =
\frac{4z^{3/2}}{\sqrt{\pi}} 
\int_0^{\infty} \dfrac{d \zeta_{\cal B}}{\zeta_{\cal B}^4}
\left(\sigma^{\rm BSF}_{n\ell}v_{\rm rel}\right) 
e^{-z /\zeta_{\cal B}^2} 
\times  \tilde{\rm BE}_n (\zeta_{\cal B}, z), 
\end{align} 
where
\begin{equation}
\tilde{{\rm BE}}_n (\zeta_{\cal B},z) = 
1+ \left[ e^{z \left( \frac{1}{\zeta_{\cal B}^2 } + \frac{1}{n^2}\right)}-1 \right]^{-1}.
\label{eq:BoseEnhancement_z }
\end{equation}
\Cref{eq:ThermalAver_BSF_z} holds both for the regulated and unregulated BSF cross sections. Considering their respective expressions, it takes the form 
\begin{subequations}
\label{eq:ThermalAver_BSF_UnregAndReg}
\label[pluralequation]{eqs:ThermalAver_BSF_UnregAndReg}    
\begin{align}
\label{eq:ThermalAver_BSF_unreg}        
\left\langle\sigma^{\rm BSF}_{n\ell}  v_{\rm rel}\right\rangle_{\rm unreg} 
&= N_\ell \, z^{3/2} \int_0^{\infty} 
\dfrac{d \zeta_{\cal B}}{\zeta_{\cal B}^3} 
\, R_{n \ell} (\zeta_{\cal B}) 
\, e^{-z/\zeta_{\cal B}^2}  
\times \tilde{{\rm BE}}_n (\zeta_{\cal B}, z),
\\
\label{eq:ThermalAver_BSF_reg}
\left\langle\sigma^{\rm BSF}_{n\ell} v_{\rm rel}\right\rangle_{\rm reg} 
&=  N_\ell \, z^{3/2} \int_0^{\infty}  
\dfrac{d\zeta_{\cal B}}{\zeta_{\cal B}^3} 
\, \dfrac{R_{n \ell} (\zeta_{\cal B})}
{[1 +  \alpha_\Phi \, R_{\ell}(\zeta_{\cal B})]^2} 
\, e^{-z / \zeta_{\cal B}^2} 
\times \tilde{{\rm BE}}_n (\zeta_{\cal B},z) ,
\end{align} 
\end{subequations}
with 
\begin{equation}
\label{eq:NormalisingFactorBSFcross section}
N_{\ell} \equiv \frac{2^{7}\sqrt{\pi}}{m_X^2} (2 \ell+1) \delta_{\ell,\mathrm{even}}.
\end{equation}
In our numerical calculations, we integrate \cref{eqs:ThermalAver_BSF_UnregAndReg} using the analytic approximation for $R_{n\ell}$ given by \cref{eq:rnlAnalyticalApproximation}. For $R_\ell$, we sum \cref{eq:rnlAnalyticalApproximation} over $n \in [1+\ell, 10\zeta_{\cal B}]$, following \cref{eq:nmaxforRl}, to ensure better than 1\% precision. The regulated and unregulated thermally-averaged BSF cross sections are shown in \cref{fig:ThermallyAveragedCrossSectionPlots}. We now briefly discuss their features, noting beforehand for comparison, that the thermally-averaged velocity-weighted unitarity cross section \eqref{eq:Unitarity-cross section} is
\begin{align}
\langle \sigma_\ell^{\rm uni} v_{\rm rel} \rangle 
= 2^\delta \ \dfrac{4 (2\ell+1)}{\mu^2}
\ \sqrt{\pi x}
\ \to \ (N_\ell/4) \sqrt{x}, 
\label{eq:ThermalAver_Unitarity}
\end{align}
where we used \cref{eq:ThermalAver_Annihilations} for the thermal average, i.e. no Bose enhancement or Fermi suppression factor due to the final-state particles has been included. The arrow denotes the value for identical particles in the initial state, setting $\delta=1$ and $\mu = m_X/2$.

For the unregulated BSF cross sections, it is possible to obtain analytic approximations for their thermal averages (cf.~\cref{App:IndividualCrossSectionsApproximation}):
\begin{align}
\label{eq:BSFsvbarAnalyticalExpressionsFinal}
& \left\langle \sigma^{\rm BSF}_{n\ell} v_{\rm rel} \right\rangle_{\rm unreg}
\simeq N_{\ell} \times
\\[1em]
\nonumber 
&\times
\left\{
\begin{array}{ll}
2^{4}\sqrt{z} \bigg[  
\dfrac{e^{-z/n^2}}{3} - \dfrac{c_\ell}{n} + \dfrac{17}{3} z^{3/2} 
+ 2\sqrt{\pi} z^{3/2} \, \mathrm{Erf}\left(\dfrac{\sqrt{z}}{n}\right) 
+ \dfrac{\sqrt{\pi} n}{2 z^{1/2}} \, \mathrm{Erf}\left(\dfrac{\sqrt{z}}{n}\right) \bigg],
& z < n^2,
\\[1em]
\dfrac{4 \sqrt{\pi} z}{n} 
\left[ \mathrm{Erf}\left(\dfrac{\sqrt{z}}{n}\right) 
- \mathrm{Erf}\left(\dfrac{c_\ell \sqrt{z}}{n^2}\right)\right] 
+ \dfrac{4 n^{4 \ell+3}}{c_\ell^{2\ell+4} z^{\ell}} 
\bigg[\Gamma\left(\ell + \dfrac{3}{2}\right) - \Gamma \left(\ell+\dfrac{3}{2}, \dfrac{c_\ell^2 z}{n^4}\right)\bigg],
& z > n^2. 
\end{array}
\right.
\end{align}
Note that all dependence of $\left\langle \sigma^{\rm BSF}_{n\ell} v_{\rm rel} \right\rangle_{\rm unreg}$ on the coupling is absorbed in the parameter $z$. At high temperatures ($T \gg |E_n|$), a (sizeable) part of the $XX$ distribution corresponds to relative velocities with $\mu v_{\rm rel}^2/2 \lesssim |E_n| \ll T$, which yields $\omega_n/T \ll 1$. This results in a significant Bose enhancement, ${\rm BE}_n(v_{\rm rel}, T) \sim T/|E_n| \gg 1$, which partially offsets the suppression of the unaveraged BSF cross section at large $v_{\rm rel}$ and ensures that BSF and ionisation processes reach equilibrium~\cite{Harz:2018csl,Harz:2019rro}. In this high-temperature regime, corresponding to $z \ll n^2$, 
$\left\langle \sigma^{\rm BSF}_{n\ell} v_{\rm rel} \right\rangle_{\rm unreg}$
grow as $z^{1/2}$ and exhibit only a mild dependence on $n$ for $n \gg \ell$, which results from the near-cancellation between the $n$-scaling of the unaveraged cross section $\sigma^{\rm BSF}_{n\ell, \rm unreg}$ and the Bose enhancement factor. At lower temperatures, $T \ll |E_n| \lesssim \omega_n$, the Bose enhancement becomes negligible, ${\rm BE}_n \simeq 1$. When $z > n^2$, two distinct scaling regimes appear. In the intermediate regime, where $n^2 < z < n^4 / c_\ell^2$, $\left\langle \sigma^{\rm BSF}_{n\ell} v_{\rm rel} \right\rangle_{\rm unreg}$ scales approximately as $z$. However, at even lower temperatures, corresponding to $z \gg n^4 / c_\ell^2$, the scaling transitions to $z^{-\ell}$, reflecting the standard suppression of higher partial waves in the absence of Sommerfeld enhancement.

In contrast, the regulated thermally-averaged cross sections exhibit very different behaviour. At high temperatures, they scale similarly to the unregulated case, 
$\left\langle \sigma^{\rm BSF}_{n\ell} v_{\rm rel} \right\rangle_{\rm reg} \propto z^{1/2}$. However, the cancellation that led to the mild $n$-dependence without regularisation is no longer present, because the low-velocity particles responsible for the Bose enhancement now have strongly suppressed cross sections. At low temperatures, the suppression of the thermally-averaged regulated cross sections directly reflects the suppression of the regulated cross sections at low velocities, given by \cref{eq:BSFreg}, i.e., $\left\langle \sigma^{\rm BSF}_{n\ell} v_{\rm rel} \right\rangle_{\rm reg} \propto z^{-\ell-1}$, which represents a faster decline than that of the unregulated BSF cross sections in the same regime. With respect to $n$, the regulated cross sections scale as $1/n^2$ in the low temperature limit.

\subsubsection*{Number of bound levels for the effective cross section}

Using the above, we now determine how many bound levels must be included in the effective cross section \eqref{eq:EffectiveCrossSection} to achieve a given precision. For the regulating factor, the number of bound states is given by \cref{eq:nmaxforRl} as a function of $\zeta_{\cal B}$. Upon thermal averaging, \cref{eq:nmaxforRl} suggests that for $\sim 1\%$ precision, we need
\begin{align}
n_{\rm max} = 10 \sqrt{z} .
\label{eq:nmax}
\end{align} 
This ensures that at a given temperature $T$, levels with binding energy as low as $|E_{\cal B}| = T/100$ are included. Considering that states with binding energy much lower than the temperature get rapidly ionised and are not efficient in depleting DM (cf.~\cref{Sec:Decoupling_Stages}), this choice is more than sufficient for our purposes. We verify numerically that \cref{eq:nmax} provides adequate convergence for the unregulated cross sections, and adopt this value in our computations. 
The impact of different choices of $n_{\rm max}$ is explored in \cref{Sec:Decoupling_Results}.

Regularisation suppresses the BSF cross sections, particularly at large $\zeta_{\cal B}$, decreasing the number of bound levels that contribute significantly to the sum of \cref{eq:EffectiveCrossSection}. Considering that the thermally-averaged regulated BSF cross sections scale as $1/n^2$ at large $z$, it follows that, for a given partial wave $\ell$, the relative error due to neglecting levels $n > n_{\rm max}^{\rm reg}$ is $\sim (1+\ell)/n_{\rm max}^{\rm reg}$. To achieve $\sim 1\%$ precision for the regulated BSF cross sections, we need, at large $z$,
\begin{align}
n_{\rm max}^{\rm reg} (\ell) = 100 \, (1+\ell) .
\label{eq:nconv_reg}
\end{align} 

\Cref{eq:nmax} supersedes \cref{eq:nconv_reg}, but we use the latter to determine the total integration time in \cref{Sec:Decoupling_Stages}.

\subsubsection{Bound-state ionisation \label{Sec:Decoupling_Rates_Ion}}

The ionisation rates of the bound states due to the radiation of the thermal bath can be determined via detailed balance, 
\begin{equation}
\label{eq:IonisationRate_DetailedBalance}
\langle \Gamma^{\mathrm{ion}}_{n\ell} \rangle  =  s(T) \,  \langle  \sigma^{\rm BSF}_{n\ell} v_{\rm rel} \rangle  
\frac{(Y_X^{\mathrm{eq} })^2}{Y^{\mathrm{eq}}_{n\ell}}.
\end{equation}
Substituting the equilibrium densities \eqref{eq:EquilibriumDensities} into \cref{eq:IonisationRate_DetailedBalance}, one finds
\begin{equation}
\label{eq:IonisationRate}
\langle\Gamma^{\mathrm{ion}}_{n \ell} \rangle =   \langle  \sigma^{\rm BSF}_{n\ell} v_{\rm rel} \rangle \left( 
\frac{m_X T}{4 \pi}\right)^{\frac{3}{2}} \frac{g_X^2}{g_{n\ell}} e^{-|E_{n}|/T}.
\end{equation}
We note that, unlike in hydrogen recombination in the early universe, the resonant radiation produced in BSF during DM freeze-out does not contribute significantly to the ionisation of the bound states~\cite{Vasilaki:2024fph}, due to the suppressed densities of the metastable bound states~\cite{Binder:2021vfo}.

\subsubsection{Bound-state decay \label{Sec:Decoupling_Rates_BSDecay}}

The decay rates for the process ${\cal B}(X X^{\dagger}) \rightarrow\Phi \Phi^\dagger$ are (see \cref{App:Annihilation and Decay Rates})
\begin{equation}
\label{eq:DecayRateXXPhiPhidagger}
 \Gamma^{\rm dec}_{n \ell} = \Gamma^{\mathsmaller{{\cal B}(X X^\dagger) \to \Phi \Phi^\dagger}}_{n \ell} = \frac{m_X}{2} \frac{ \alpha_\Phi^{2 \ell+5}}{n^{4+2\ell}}
 \frac{(\ell!)^2}{[(2\ell+1)!]^2} \frac{\Gamma (\ell+n+1)}{\Gamma (n-\ell)}.
\end{equation}
The above rates, computed in the bound-state rest frame, serve as a good approximation for the thermally-averaged decay widths, since the bound states are non-relativistic.

\subsubsection{Bound-to-bound transitions}
Bound-to-bound transitions enhance the effective DM depletion cross sections, potentially making highly excited states relevant during thermal decoupling~\cite{Binder:2021vfo}. However, in the present model, such transitions are suppressed. 
In the $\alpha_V \rightarrow 0$ limit, dipole transitions via $V^\mu$ emission vanish. Since $XX^\dagger$ is the only configuration that supports bound states, the dominant bound-to-bound transitions are $\mathcal{B}(XX^\dagger) \to \mathcal{B}'(XX^\dagger) + \Phi\Phi^\dagger$. These processes can arise through two Yukawa $y$ vertices, or via quartic couplings in the scalar potential~\cite{Oncala:2018bvl}. They are dominated by $\Delta \ell \neq 0$ modes, as monopole contributions vanish due to the orthogonality of the initial and final bound-state wavefunctions. Additionally, they are suppressed by the three-body phase space of the final state. We shall therefore neglect them in the following.

\subsubsection{Effective cross section \label{Sec:Decoupling_Rates_EffectiveCrossSection}}

\begin{figure}[t!]
\centering   
\includegraphics[width=.98\linewidth]{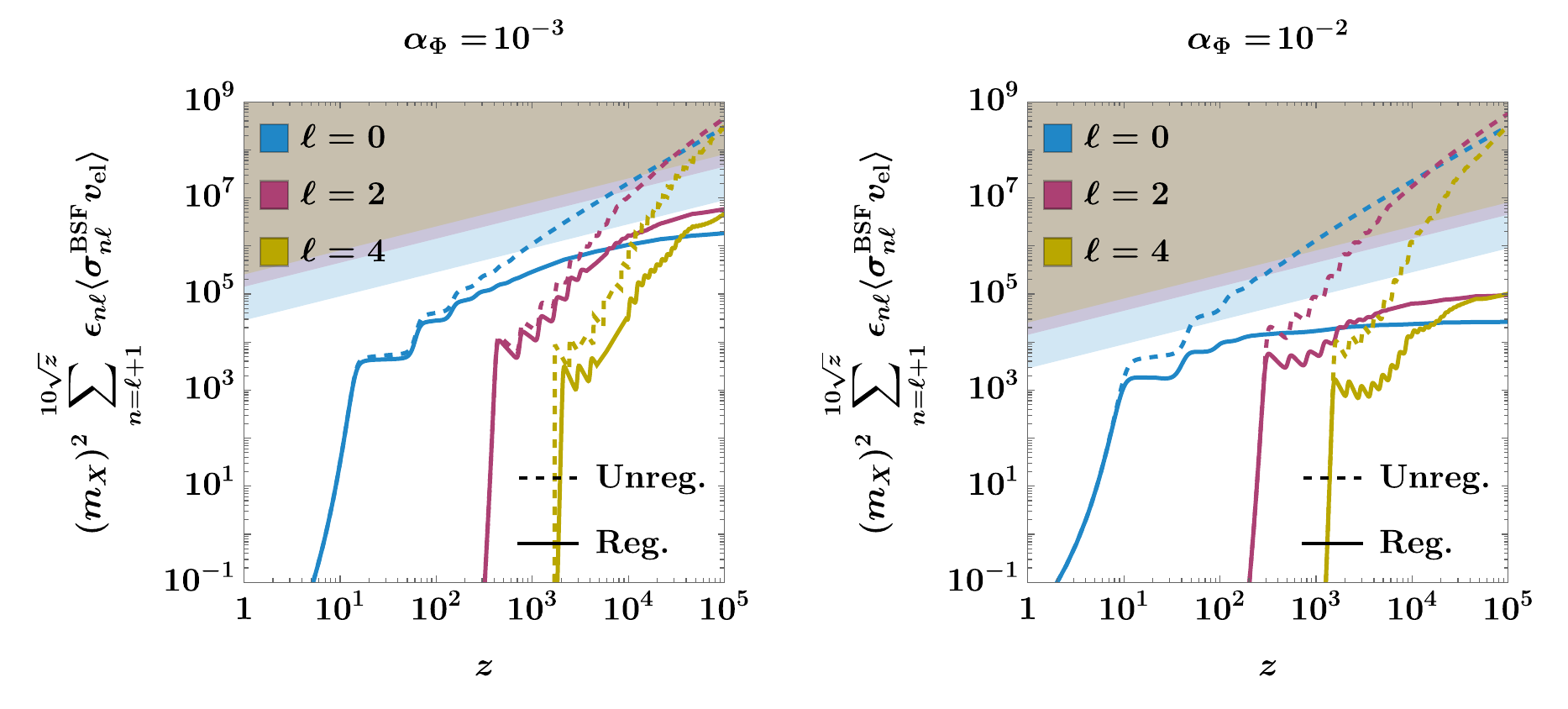}
\caption{The thermally-averaged velocity-weighted BSF cross sections, multiplied by the efficiency factors $\epsilon_{n\ell}$, and summed over bound levels $n$ (cf.~\cref{eq:EffectiveCrossSection}), for different partial waves. The peaks correspond to consecutive levels exiting ionisation equilibrium. The coloured shaded regions correspond to the regions where the thermally-averaged cross section exceeds the unitarity bound, i.e., $\langle \sigma_\ell^\mathrm{BSF} v_\mathrm{rel} \rangle > \langle \sigma_\ell^\mathrm{uni} v_\mathrm{rel} \rangle / 4$ for each partial wave. 
\label{fig:EffectiveBSFCrossSectionSummedOverN}}
\centering
\includegraphics[width=0.98\linewidth]{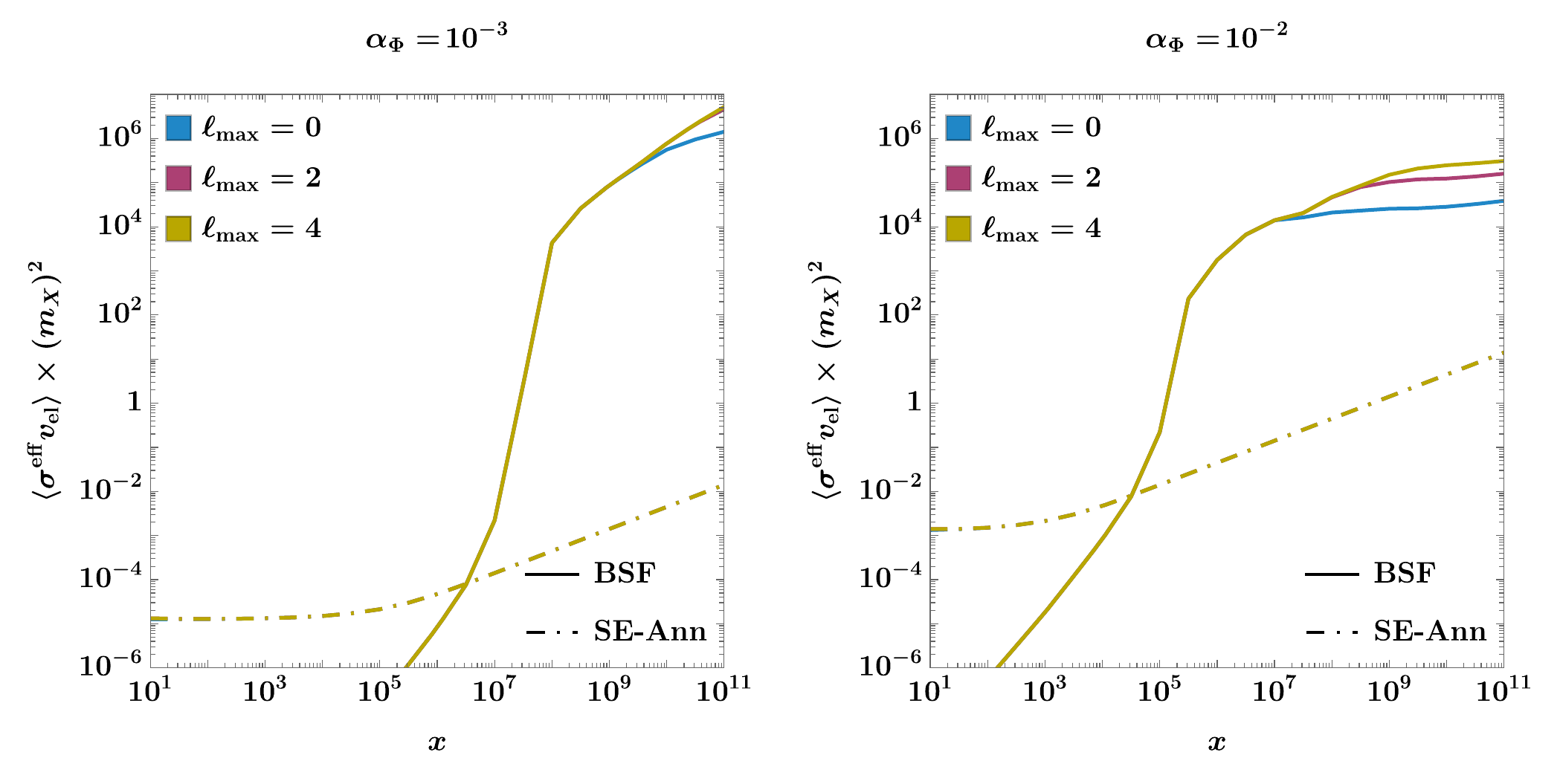}
\caption{
Thermally-averaged velocity-weighted cross sections as functions of $x=m_X/T$, for two indicative values of the coupling $\alpha_\Phi$: annihilation (dot-dashed lines) and effective regulated BSF, summed over partial waves up to $\ell_{\max}$ as indicated in the legend (solid lines). 
\label{fig:PlotTotalEffectiveCrossSection}}
\end{figure}

Using the rates from the previous sections, we compute the effective DM depletion cross section, defined in \cref{eq:EffectiveCrossSection}. \Cref{fig:EffectiveBSFCrossSectionSummedOverN} shows the BSF contributions to the effective cross sections, summed over $n$, for different partial waves.  
\Cref{fig:PlotTotalEffectiveCrossSection} compares the effective BSF cross section summed over partial waves, with the annihilation cross section. Although the BSF cross sections exceed the annihilation cross section by several orders of magnitude, DM depletion via BSF becomes efficient only at lower temperatures due to ionisation, for both the regulated and unregulated cases.

We can identify two regimes during DM thermal decoupling for each bound level~\cite{vonHarling:2014kha}:
\begin{enumerate}
\item[(i)]  
At $T \gg |E_n|$, ionisation dominates over decay, keeping the bound states in ionisation equilibrium~\cite{Binder:2018znk}. In this regime, the contribution to the effective DM depletion rate is independent of the BSF cross section for that level. Due to the large BSF cross sections, ionisation equilibrium persists down to temperatures well below the binding energy, approximately $T \sim 10 |E_n|$ for $\ell=0$. Excited states remain in equilibrium even longer, as their ionisation rates are larger and their decay rates are smaller. In fact, the decay rates are suppressed both by higher $n$ and $\ell$ values; besides the numerical factors, the decay rates of higher $\ell$ bound states are suppressed by larger powers of the coupling, $\alpha_\Phi^{2\ell}$ (cf.~\cref{eq:DecayRateXXPhiPhidagger}). Nevertheless, even if the contribution of excited states to the  effective cross section is not yet maximal, it can be very sizeable, potentially exceeding direct annihilations by orders of magnitude.

\item[(ii)]  
At $T \ll |E_{n}|$,  the ionisation rate drops below the decay rate, yielding an efficiency factor of order one, and maximising the contribution of the bound level to the effective DM depletion rate. The multi-peak structure in the effective cross section seen in \cref{fig:EffectiveBSFCrossSectionSummedOverN} reflects the sequential departure of higher-lying bound levels from ionisation equilibrium.
Summing over all bound levels, and considering the low-velocity regime as given by \cref{eq:RlSum}, as well as the regularization formula \eqref{eq:Rnlregulated},
we find that at low temperatures, 
\begin{subequations}
\label{eq:SumOverLevels_ThermalAver}
\label[pluralequation]{eqs:SumOverLevels_ThermalAver}
\begin{multline}
\sigma_{\ell, {\rm unreg}}^{\rm BSF} v_{\rm rel} 
\approx  \dfrac{2^7\pi (2\ell+1)}{m_X^2} 
\zeta_{\cal B}^2 \ln (\zeta_{\cal B}/c_\ell)   
\Rightarrow
\\
\langle\sigma_{\ell, {\rm unreg}}^{\rm BSF} v_{\rm rel} \rangle
\approx   \dfrac{2^7\pi (2\ell+1)}{m_X^2} \, z [\ln z + \ln (4/c_\ell^2) +\gamma_{\rm E}] ,
\label{eq:SumOverLevels_ThermalAver_unreg}
\end{multline}
where $\gamma_{\rm E}\simeq 0.577$ is Euler's constant, and 
$\sigma_{\ell, {\rm reg}}^{\rm BSF} \approx
(\sigma_{\ell}^{\rm uni})^2 / \sigma_{\ell, {\rm unreg}}^{\rm BSF}$, or
\begin{align}
\sigma_{\ell, {\rm reg}}^{\rm BSF} v_{\rm rel} 
\approx  \dfrac{2^3\pi (2\ell+1)}{m_X^2} 
\dfrac{1}{\alpha_{\cal B}^2 \ln (\zeta_{\cal B}/c_\ell)} 
\quad &\Rightarrow \quad
\langle\sigma_{\ell, {\rm reg}}^{\rm BSF} v_{\rm rel} \rangle
\sim   \dfrac{2^3\pi (2\ell+1)}{m_X^2 \alpha_{\cal B}^2} ,
\label{eq:SumOverLevels_ThermalAver_reg}
\end{align}
\end{subequations}
i.e.~$\langle\sigma_{\ell, {\rm reg}}^{\rm BSF} v_{\rm rel} \rangle$ becomes almost constant. More detailed analytical calculations for the unregulated case are given in \cref{App:Thermally Averaged BSF summed over n}. Notably, the thermally-averaged unregulated BSF cross sections \eqref{eq:SumOverLevels_ThermalAver_unreg} increase faster than linearly with $z$ (or $x$). This \emph{super-critical} behaviour prevents DM from freezing out~\cite{Binder:2023ckj}. Regularisation curbs the growth of the total cross section at low temperatures, ensuring DM freeze-out. 
We estimate the necessary integration time of the Boltzmann equations in \cref{Sec:Decoupling_Stages}, after discussing the stages of thermal decoupling.
\end{enumerate}

\subsection{Stages of dark-matter decoupling \label{Sec:Decoupling_Stages}}

\begin{figure}
\centering
\vspace{-0.5cm}
\includegraphics[width=0.90\textwidth]{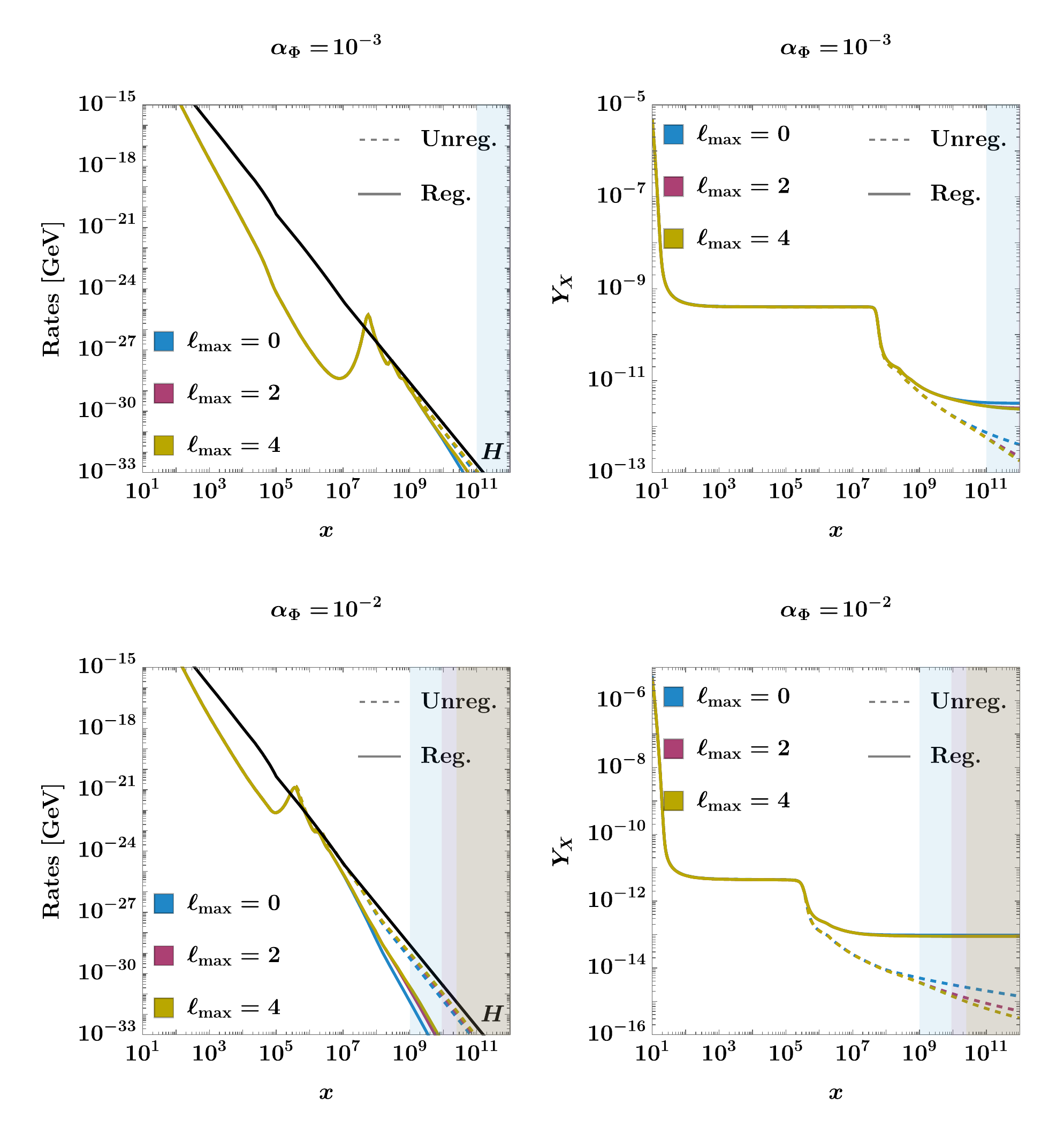}
\vspace{-0.5cm} 
\caption{\textit{Left:}
The evolution of the Hubble rate $H$, (black solid line) and  depletion rates, $\Gamma_X$, obtained for the unregulated (dashed) and regulated (solid) effective BSF cross sections, summed up to $\ell_{\mathrm{max}} \in \{0,2,4 \}$ partial waves for three indicative values of $\alpha_\Phi \in \{ 10^{-3}, 10^{-2}\}$. \textit{Right: }The corresponding particle yield, $Y_X(x)$. The shaded coloured regions marks the maximum $\ell$- dependent $x$ (see \cref{eq:xmax}) up to which we integrate the effective Boltzmann equation. Here we fixed the mass of the DM particle to be $m_X = 10$ TeV.} 
\label{fig:PlotDMDeplitionRates}
\end{figure}

In \cref{fig:PlotDMDeplitionRates}, we compare the DM depletion rate, $\Gamma_X \equiv n_X \langle \sigma^{\mathrm{eff}} v_{\rm rel} \rangle$, with the Hubble expansion rate, $H = \left(4 \pi^3 g_{*\rho}/ 45\right)^{1/2} T^2/m_{\mathrm{Pl}}$, and present the particle yields, $Y_X$, obtained by integrating the effective Boltzmann \cref{eq:EffectiveBoltzmann}. The $Y_X$ evolution reveals three distinct stages~\cite{Oncala:2019yvj}.

\subsubsection*{I. First freeze-out}
At very high temperatures ($x \ll x_{\mathrm{FO1}} \approx 30$), the DM depletion rate exceeds the Hubble expansion rate. All bound levels are in ionisation equilibrium, bound states form rapidly but are ionised by the ambient radiation before decaying, and the depletion rate is dominated by direct annihilations into radiation. Around $x \sim x_{\mathrm{FO1}}$, $\Gamma_X$ falls below $H$, causing the DM comoving number density to freeze out, mimicking the standard scenario.

\subsubsection*{II. Recoupling of DM depletion}
 
Ionisation processes impede the DM depletion via BSF until late, when the DM density is low. However, the large BSF cross sections compensate for the reduced density, leading to recoupling of DM depletion. Recoupling occurs only if and when $s Y^{FO1}_X \langle \sigma^{\mathrm{eff}} v_{\rm rel} \rangle \gtrsim H$~\cite{Oncala:2019yvj}, as shown in \cref{fig:PlotDMDeplitionRates}. Dark matter depletion via BSF is initially driven by the lowest-lying bound level, $n=1+\ell$, for a given partial wave $\ell$. As the temperature decreases further, higher $n,\ell$ bound levels decay into radiation before getting ionised, thereby beginning to contribute to the DM depletion. Their cumulative effect leads to an increase in the total effective cross section at lower temperatures, as illustrated in \cref{fig:EffectiveBSFCrossSectionSummedOverN,fig:PlotTotalEffectiveCrossSection}. For each partial wave, the onset of this stage is set by when the lowest-lying bound level approaches the exit from ionisation equilibrium, with its contribution to the effective cross section saturating the actual value of the corresponding BSF cross section ($\epsilon_{n\ell} \to 1$). 
This occurs at temperatures somewhat lower than the binding energy, around $z \sim 10 (1+\ell)^2$, or equivalently $x \sim 40 (1+\ell)^2 /\alpha_\Phi^2$, due to the large BSF cross sections. We emphasise however that the DM depletion via BSF can be very significant even while $\epsilon_{n\ell} < 1$.

\subsubsection*{III.A~~Unregulated BSF cross sections: eternal DM depletion}

As shown in \cref{fig:PlotDMDeplitionRates}, once DM annihilations have recoupled, DM continues to deplete indefinitely in the absence of proper regularisation. To understand this, we consider the Boltzmann equation governing the particle yield. After the first freeze-out, the comoving DM density has significantly deviated from its equilibrium value ($Y_X \gg Y_X^{\mathrm{eq}}$), which allow us to neglect the second term in the right-hand-side of the effective Boltzmann equation \eqref{eq:EffectiveBoltzmann}, obtaining 
\begin{equation}\label{eq:Eternal_Depletion}
Y_{X}(x) \simeq \lambda^{-1} \left(
\int^{x}_{x_{\mathrm{FO1}}} x^{-2} 
\langle \sigma^{\mathrm{eff}} v_{\rm rel} \rangle 
\, dx 
\right)^{-1}.
\end{equation}
As noted in \cite{Binder:2023ckj}, the scaling of the effective cross section with $x$ determines whether the DM density eventually freezes out. For a power-law dependence, $\langle \sigma^{\mathrm{eff}} v_{\rm rel} \rangle \propto x^{\gamma}$, the integral in \cref{eq:Eternal_Depletion} converges only if $\gamma < 1$, while for $\gamma \geqslant 1$, it becomes divergent. Since the convergence of this integral is a necessary condition for freeze-out, the DM density is depleted indefinitely if $\gamma \geqslant 1$. The critical case of $\gamma = 1$, corresponds to logarithmic depletion and marks the onset of \emph{super-critical behaviour}, where DM never fully freezes out. In this case, the DM depletion rate, $\Gamma_X$, remains proportional to the Hubble rate, preventing the final freeze-out. 

In our model, the unregulated thermally-averaged velocity-weighted BSF cross section for a given partial wave, summed over $n$, scales approximately as $z \ln z$ at low temperatures (cf.~\cref{eq:SumOverLevels_ThermalAver_unreg}). The growth of $\langle \sigma^{\mathrm{eff}} v_{\rm rel} \rangle$ with $z$ being steeper than linear, leads to indefinite DM depletion. This is clearly shown in \cref{fig:PlotDMDeplitionRates}, where the DM depletion rate, $\Gamma_X$, remains proportional to the Hubble rate $H$ at late times, indicating the absence of a final freeze-out.

\subsubsection*{III.B~~Regulated BSF cross sections: final decoupling}

When proper regularisation is applied, the DM density undergoes a final freeze-out, as seen in \cref{fig:PlotDMDeplitionRates}.
Importantly, regularisation reduces the contribution of higher partial waves, which become efficient at later times due to suppressed decay rates. In particular, as the coupling strength increases, the cross section is more strongly regulated, shortening the recoupling period and limiting the impact of higher partial waves.

To compute the relic density, we integrate the effective Boltzmann equation down to sufficiently low temperatures. As discussed, BSF-driven DM depletion is most efficient at the earliest time ionisation becomes suppressed, typically within the temperature range $|E_\mathcal{B}|/T \sim 0.1$–$10$ for a bound level ${\cal B}$. At higher temperatures, bound states are ionised before decaying, while at lower temperatures, the DM dilution due to the expansion of the universe suppresses the effect. To accurately capture the BSF effects, we thus integrate down to $T \sim |E_\mathcal{B}|_{\min}/ 10$, where $|E_\mathcal{B}|_{\min}$ is the binding energy of the highest-lying state that contributes significantly. Since the unregulated BSF cross sections do not allow for freeze-out, we determine this level considering the regulated case. Using \cref{eq:nconv_reg}, we arrive at
\begin{align}
z_{\max} = 10 \, [n_{\rm max}^{\rm reg} (\ell_{\max})]^2 = 10^5 (1+\ell_{\max})^2
\quad \text{or} \quad
x_{\max} = 4 \cdot 10^5 (1+\ell_{\max})^2 / \alpha_{\cal B}^2,
\label{eq:xmax}
\end{align}
where $\ell_{\max}$ is the maximum number of partial waves considered.
 
\subsection{Results and discussion \label{Sec:Decoupling_Results}}

To compute the relic abundance of the $X$ species, $Y_X(\infty)$, we integrate the Boltzmann \cref{eq:EffectiveBoltzmann} with the effective cross sections defined in \cref{eq:EffectiveCrossSection}, using both the regulated and unregulated BSF cross sections. The $X$ fractional energy density is
\begin{equation}
\Omega_X = \frac{2 m_{X} Y_{X}(\infty) s_{0}}{\rho_{c}},
\end{equation}
where the factor of 2 accounts for both $X$ and $X^\dagger$, while $s_0 \simeq 2839.5~\text{cm}^{-3}$, and $\rho_c = 4.78 \times 10^{-6}$ GeV cm$^{-3}$ are the present-day entropy and the critical energy \cite{Aghanim:2018eyx}.

\begin{figure}[h!]
\centering
\includegraphics[width=0.97\linewidth]{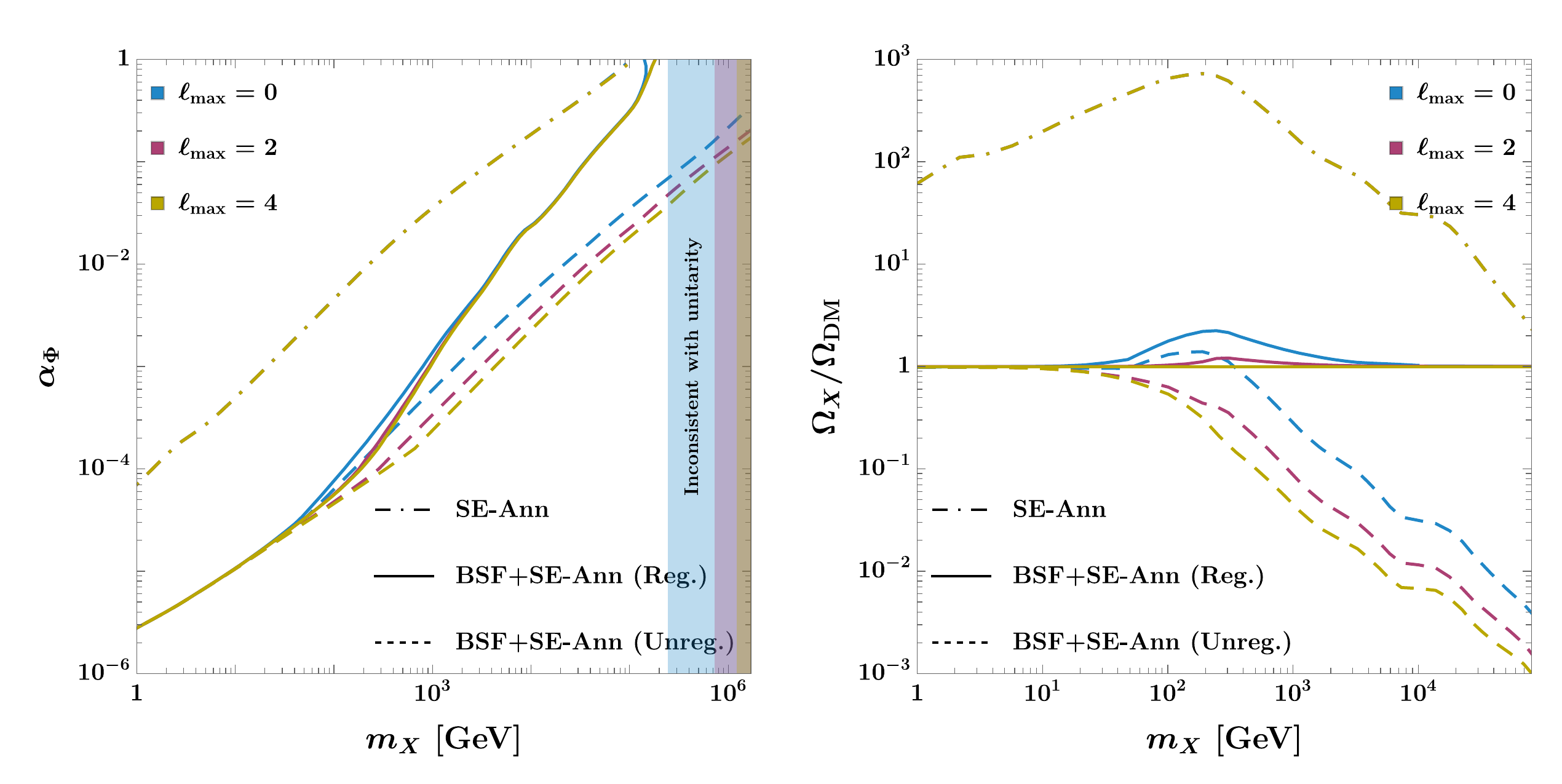}
\caption{\textit{Left}: The mass-coupling relation required to obtain the observed DM abundance. We show results considering 
(i) only SE direct annihilation \eqref{eq:ThermalAver_Annihilations} (dotted-dashed line), 
(ii) including BSF without regularisation (dashed lines), and 
(iii) including regularisation (solid lines). 
For each curve, we consider up to $\ell_{\mathrm{max}}$ partial waves, as indicated in the labels. 
\textit{Right}: The ratio $\Omega_{X}/\Omega_{\rm DM}$, using the $m_X - \alpha_\Phi$ values that reproduce the observed relic abundance in case (iii) with $\ell_{\rm max} =4$, and considering all three scenarios described above, as indicated by the labels. 
We emphasise that for the unregulated BSF cross sections, DM does \emph{not} freeze-out. In these cases, we terminate the integration of the effective Boltzmann equation at the same time as for the regulated cross sections, given by \cref{eq:xmax}, and take the corresponding yield as an estimate of the present abundance. The dashed lines are thus indicatively only; the yield would in fact be vanishingly small, with no mass-coupling relation reproducing the observed density. 
\label{fig:RequiredCouplingPlot}}
\end{figure}

\begin{figure}[h!]
\centering
\includegraphics[width=0.5\linewidth]{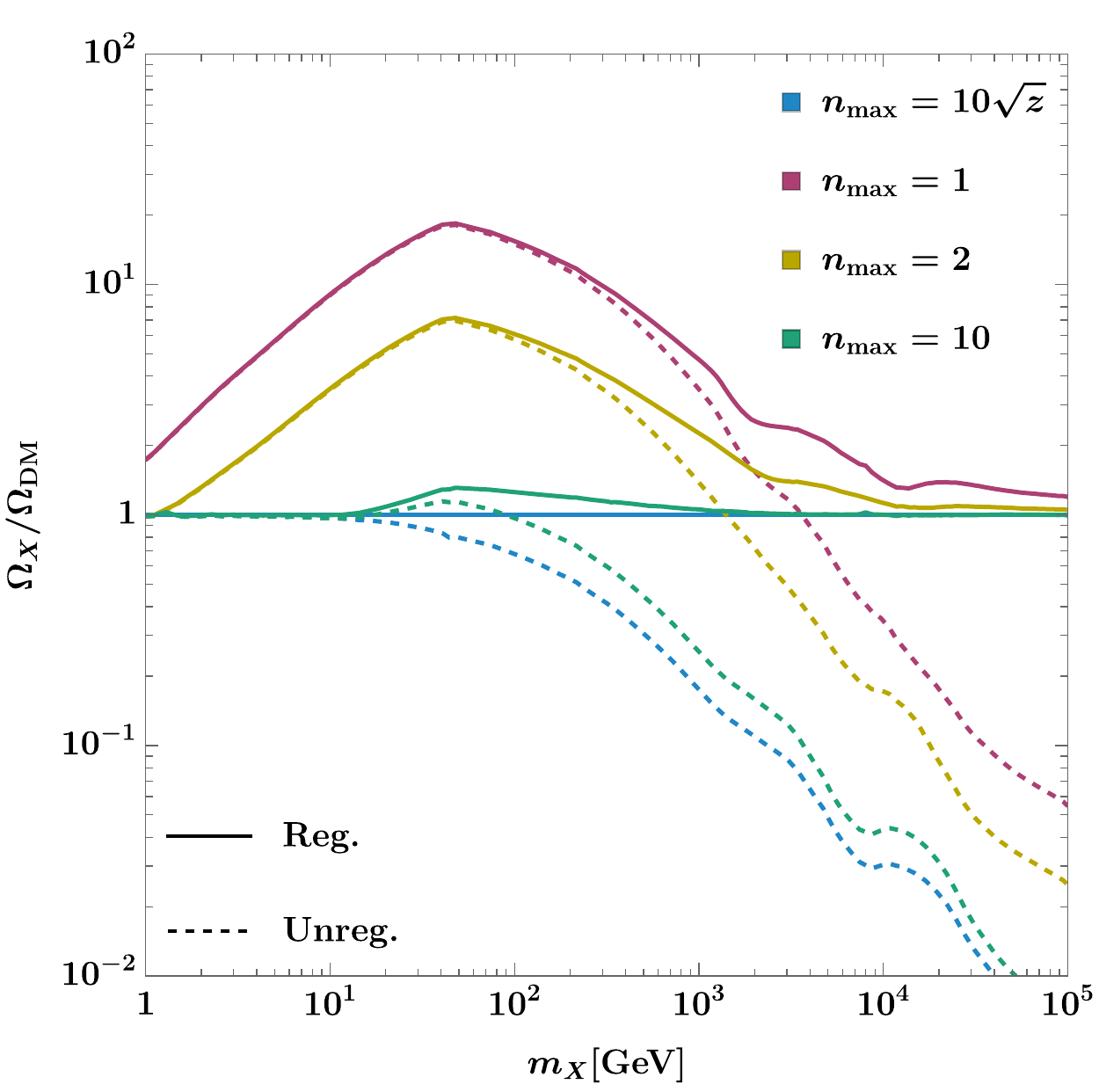}
\caption{The effect of higher-$n$ bound states on the relic density for $\ell_{\mathrm{max}} = 0$. Shown is the ratio $\Omega_X / \Omega_{\rm DM}$, computed using the (un)regulated  effective cross section with $\ell_{\mathrm{max}} = 0$, and summing up to $n_{\mathrm{max}}$ as indicated by the labels (solid lines: regulated , dashed lines: unregulated). For each value of $m_X$, the coupling $\alpha_\Phi$ is chosen to reproduce the observed relic abundance with regulated BSF cross sections and summing up to $n_{\mathrm{max}} = 10 \sqrt{z}$. 
\label{fig:RelicDensityNmaxContributing} }
\end{figure}

Our goal is to determine the relation between the coupling strength, $\alpha_\Phi$, and the mass, $m_X$, for which the predicted relic density of $X$ particles, $\Omega_{X}$, matches the observed DM abundance $\Omega_{\rm DM} \simeq 0.26$. We consider three cases: (i) Sommerfeld-enhanced (SE) direct annihilation, (ii) SE annihilation plus BSF via charged scalar emission with unregulated cross sections summed over relevant bound levels, and (iii) SE annihilation plus regulated BSF. Our results are shown in \cref{fig:RequiredCouplingPlot,fig:RelicDensityNmaxContributing}. The left panel of \cref{fig:RequiredCouplingPlot} shows the mass-coupling relation that reproduces the observed relic density, while the right panel illustrates the effects on the relic density of neglecting BSF, regularisation or higher partial waves. \Cref{fig:RelicDensityNmaxContributing} examines the effect of excited bound levels, with and without regularisation, on the relic density. We now discuss their main features of these results. 


\begin{description}

\item[BSF effect on DM density and mass-coupling relation.] 
Our results reaffirm that BSF via emission of a charged scalar significantly enhances DM depletion compared to SE direct annihilations alone~\cite{Oncala:2019yvj}. Due to the very efficient monopole transitions, this occurs down to much smaller values of the coupling and mass in comparison to models with dipole transitions (see e.g.~\cite{vonHarling:2014kha,Harz:2018csl,Bottaro:2021snn,Bollig:2021psb}). In our most accurate computation --- which includes regulated BSF and partial-wave contributions up to $\ell_{\max}=4$ --- neglecting BSF leads to a substantial overestimate of the DM relic density, by factors ranging from $10^2$ to $10^3$, depending on the value of $m_X$, as illustrated in the right plot of \cref{fig:RequiredCouplingPlot}. The BSF effect on the predicted mass-coupling relation, shown in the left plot of \cref{fig:RequiredCouplingPlot}, is also very significant, with potential implications for many observables, such as DM indirect detection and self-interactions.

\item[Unregulated BSF and eternal depletion.] 
Without regularisation, increasing the coupling $\alpha_\Phi$ leads to a prolonged recoupling period of DM depletion which in turn enhances the impact of excited bound states and higher partial waves in the depletion process. In fact, as already discussed in \cref{Sec:Decoupling_Stages}, DM does not freeze out.  Nevertheless, for comparison, in \cref{fig:RequiredCouplingPlot} we show a mass-coupling relation for the unregulated cross sections, obtained by integrating the Boltzmann equations down to the same temperature as for the regulated cross sections (which allow for freeze-out). We reiterate that the actual yield corresponding to these parameters for the unregulated cross sections would in fact be vanishingly small, due to eternal depletion. This becomes evident when considering the effect on the relic density, as shown in the left panel of  \cref{fig:RequiredCouplingPlot} and in \cref{fig:RelicDensityNmaxContributing}.

\item[Regularisation.] 
To interpret the mass-coupling curves for the regulated cross sections, we must examine how regularisation modifies the scaling of the BSF cross sections. For small values of the coupling, $\alpha_\Phi \lesssim 10^{-5}$, regularisation has negligible impact. As the coupling increases, however, the effects of regularisation become more pronounced: BSF cross sections are increasingly suppressed, leading to a shorter recoupling period and thus reduced efficiency of DM depletion via BSF. Interestingly, for an intermediate coupling range, roughly $10^{-5} \lesssim \alpha_\Phi \lesssim 10^{-3}$, higher partial waves, up to $\ell = 4$, still contribute in the DM depletion, even after regularisation is applied. Neglecting higher partial waves can over-predict the DM density by up to a factor of $\sim 3$ for masses of a few 100 GeV. For $\alpha_\Phi \gtrsim 10^{-3}$, decoupling occurs earlier, before higher partial waves can substantially contribute. Combined with the suppression of higher-$\ell$ decay rates, contributions from $\ell_\mathrm{max} > 4$ become negligible. 

As the coupling grows, even the regulated cross sections saturate the unitarity limit, but decrease with increasing $\alpha_\Phi$ beyond that point~\cite{Flores:2024sfy}. This explains why the effect of regulated BSF on the relic density becomes smaller for $\alpha_\Phi \gtrsim 10^{-3}$ or $m_X \gtrsim 10^3$, even though it remains very significant all the way to the highest mass attained.

\item[Excited bound levels.]
The number of bound levels that significantly contribute to the DM effective depletion cross section \eqref{eq:EffectiveCrossSection} depends on the temperature. Considering the estimation \eqref{eq:nmaxforRl} for the unaveraged unregulated BSF cross sections, we set $n_{\rm max} = 10\sqrt{z}$ for the thermally-averaged ones, to attain better than 1\% precision. Regulated cross sections necessitate a lower number. In \cref{fig:RelicDensityNmaxContributing}, we illustrate how the relic density changes when the sum is truncated at smaller $n$. The effective cross section remains sensitive to the inclusion of higher-$n$ levels, even when contributions from higher partial waves have become negligible. We observe that, for $\ell = 0$, convergence in the computation of the DM density is achieved with $n_{\max} \sim {\cal O} (10)$, which we have checked numerically that suffices to accurately estimate the regulated effective cross section at least up until $z \sim 10^2$.  While an increasingly larger $n_{\max}$ is required for the accurate determination of the effective cross section at larger $z$, particularly in the unregulated case, the depletion of the DM density that occurs at $z\sim 1-10$, once the recoupling takes place, along with the continued expansion of the universe, render the further DM depletion less significant at later times. 

It is therefore reasonable to conclude that the number of bound levels required to accurately estimate the DM density is determined by the sensitivity of the effective cross section to $n$ up until $z\sim \text{few}\times(1+\ell)^2$, with $\ell$ spanning the relevant angular modes, that we now discuss. We emphasise though that estimating the regulating factor may necessitate a much larger number of bound levels, as suggested by \cref{eq:nmaxforRl}.

\item[Higher partial waves.]
In the present model, DM depletion due to higher partial waves is doubly suppressed, first by the regularisation of the BSF cross sections, and second by the smaller decay rates,  $\Gamma_{n\ell}^{\rm dec} \propto \alpha_{\cal B}^{2\ell}$ (cf.~\cref{eq:DecayRateXXPhiPhidagger}). While the regularisation affects all partial waves similarly, the $1/\alpha_{\cal B}^2$ suppression it introduces in the BSF cross sections (cf.~\cref{eq:SumOverLevels_ThermalAver_reg}) makes BSF less significant at large values of the coupling, where the suppression of the higher-$\ell$ bound-state decay rates is eased. This interplay limits the overall impact of higher partial waves.

However, this conclusion may not be generic. If $\ell$-changing transitions are allowed (e.g., in this model, by considering $\alpha_V \neq 0$), then bound states of higher $\ell$ can de-excite into lower-$n$ and lower-$\ell$ states that decay faster, potentially enhancing their impact very significantly. Indeed, the effective cross section increases monotonically with the bound-to-bound transition rates~\cite{Binder:2021vfo}. Moreover, higher partial waves can be important, even dominant, in models in which the BSF cross sections remain large at large coupling, where the higher-$\ell$ decay rates are large~\cite{Baldes:2017gzw}. 

These considerations underscore the model dependence of the higher-$\ell$ effect on DM thermal decoupling, and consequently the absence of a model-independent unitarity bound on the mass of thermal-relic DM~\cite{Flores:2024sfy,Baldes:2017gzw}. 

\item[Unitarity limits on the DM mass.]
The unitarity bound \eqref{eq:UniLimit_Inelastic} on partial-wave inelastic cross sections sets upper limits on the mass of thermal-relic DM~\cite{Griest:1989wd}, provided that a finite number of partial waves are relevant~\cite{Flores:2024sfy,Baldes:2017gzw}. Which and how many partial waves are important depends on the model, there is thus no model-independent upper bound on the mass of thermal-relic DM~\cite{Flores:2024sfy,Baldes:2017gzw}. In the left panel of \cref{fig:RequiredCouplingPlot}, the coloured bands mark the unitarity limits on the DM mass for the partial waves considered. 
Considering that non-self-conjugate DM can be depleted both by 
particle-antiparticle and particle-particle inelastic processes, this is estimated to be $m_X \lesssim 140 \times \sqrt{3(\ell + 1)}$ TeV for a single partial wave~\cite{Flores:2024sfy}. The present model features indeed both types of processes, however $XX$ BSF processes are significantly more efficient in depleting DM than $XX^\dagger$ annihilation, and $\ell = 0$ dominates upon regularisation. The highest mass attained is thus $m_X \sim \sqrt{2} \times 140$ TeV $\sim 197$ TeV, below the theoretical upper limit of $\sqrt{3} \times 140~{\rm TeV} \simeq 242$~TeV for $s$-wave annihilation, marked by the first coloured band.

\begin{figure}[t!]
\centering
\includegraphics[width=0.5\linewidth]{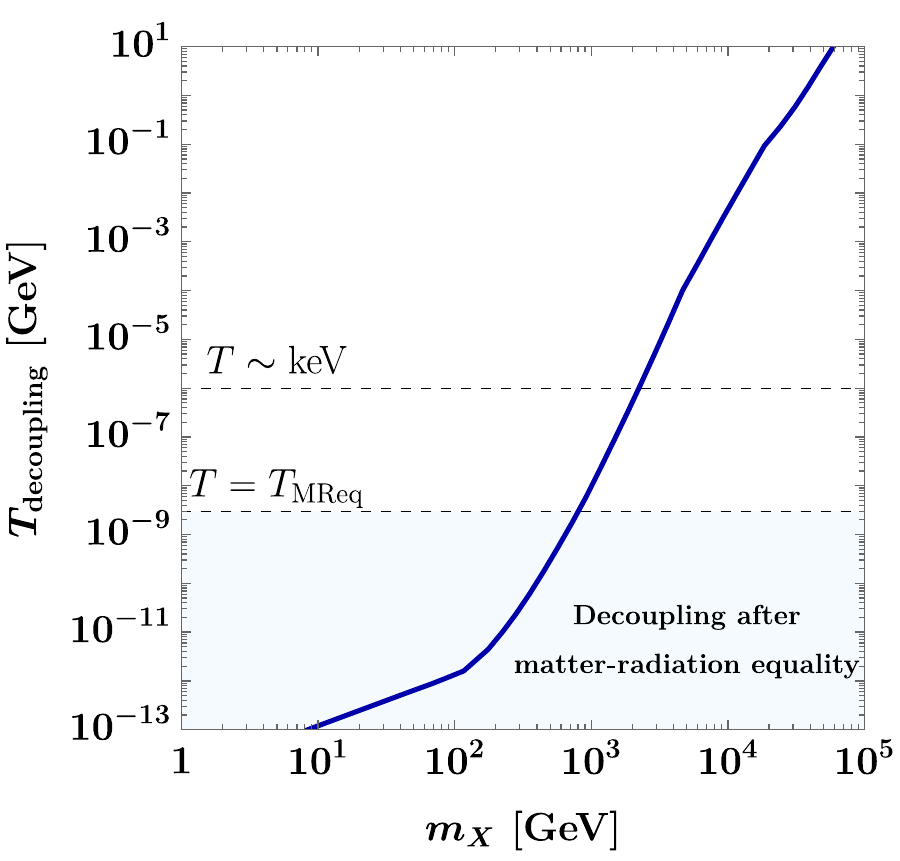}
\caption{The temperature of chemical decoupling as a function of DM mass, defined as the point at which the DM density reaches within 10\% of its final value. We have used the coupling determined by the computation of the relic density, as a function of the DM mass. The large BSF cross sections imply that DM may remain coupled until very late, even for large DM masses. While decoupling after matter-radiation equality is excluded, decoupling around keV temperatures can affect structure formation at scales that are being probed by cosmological observations.}
\label{fig:ThermalDecoupling}
\end{figure}

\item[Very late decoupling.] 
The very high efficiency of monopole BSF processes implies that DM can stay chemically coupled until rather late, which can have implications for structure formation. In \cref{fig:ThermalDecoupling}, we show the temperature of final decoupling for the coupling-mass correlation predicted by the observed relic density considering the most complete calculation that includes annihilation and regularised BSF with $\ell_{\max} =4$. We define the decoupling temperature, $T_{\rm decoupling}$, at the point when the DM density reaches within 10\% of its final value. Because of BSF, $T_{\rm decoupling}$ is much lower than the first freeze-out temperature. We find that for $m_X \leqslant 2$ TeV and $m_X \leqslant 800$ GeV, the final decoupling occurs after $T \lesssim$~keV and matter-radiation equality (MReq), respectively. While the latter possibility is observationally excluded, decoupling around keV temperatures raises the interesting possibility of affecting galactic structure at small scales (see e.g.~\cite{vandenAarssen:2012ag,Villasenor:2022aiy}).
We reiterate that the computations in the present paper have been carried out neglecting the mediator mass. A non-zero mediator mass introduces cutoffs that can affect the results in various ways. We leave this for future work.

\end{description}

\clearpage
\section{Conclusions \label{Sec:Conclusions}}

The formation and decay of bound states play a major role in DM thermal decoupling, adding new annihilation channels that strongly affect the predicted relic abundance, especially for TeV-scale DM~\cite{vonHarling:2014kha}. In models where bound states form from scattering states with different potentials --- such as models with charged scalars or non-Abelian forces --- standard computations of BSF cross sections violate unitarity even at very small couplings~\cite{Oncala:2019yvj,Oncala:2021tkz,Oncala:2021swy,Binder:2023ckj,Beneke:2024nxh,Petraki:2025}. Besides being theoretically inconsistent, this can lead to the incorrect prediction of continuous DM depletion with no freeze-out.

In this work, we refine the treatment of BSF in DM freeze-out calculations to ensure consistency with unitarity. Using the regularisation scheme of Ref.~\cite{Flores:2024sfy}, we study a simple gauged $U(1)$ extension of the SM, which includes a complex scalar DM candidate $X$, singly charged under the gauge symmetry, a massless gauge boson $V_\mu$, and a light doubly charged scalar $\Phi$~\cite{Oncala:2019yvj}. In this setup, DM bound states form dominantly via emission of the charged scalar, via monopole transitions between different Coulomb potentials~\cite{Oncala:2019yvj,Ko:2019wxq}. Besides this model being a viable possibility, it emulates the dynamics of Higgs portal scenarios involving near-degenerate multiplets~\cite{Oncala:2021tkz,Oncala:2021swy}, as well as of dipole transitions in non-Abelian theories where initial and final states are subject to different potentials~\cite{Harz:2018csl,Harz:2019rro,Oncala:2021tkz,Oncala:2021swy,Binder:2023ckj,Bottaro:2021snn}.

We extend the analysis of Ref.~\cite{Oncala:2019yvj} by including all relevant bound levels. We find that, without regularisation, including excited bound levels leads to unitarity violation at low velocities, causing the thermally-averaged cross section to scale as $\langle \sigma^{\rm eff} v_{\rm rel} \rangle \propto x \ln x$, where $x=m_X/T$, and preventing DM from freezing out. However, once regularisation is applied, the model becomes viable. $X$ undergoes two (re-)coupling and decoupling stages: an initial one driven by direct annihilations, and a subsequent one governed by regulated BSF processes.

Our results reaffirm that BSF can significantly affect the DM decoupling. Even under high ionisation rates, capture into excited states remains efficient and alters the relic abundance. However, in the absence of $\ell$-changing bound-to-bound transitions, regularisation suppresses the influence of higher partial waves. This is due to the interplay between regularisation and higher-$\ell$ bound-state decay rates: the former suppresses cross sections at large couplings, while the latter require large couplings to be significant.

The prolonged DM decoupling due to BSF raises an intriguing possibility: DM may remain coupled to relativistic particles until very low temperatures. Remarkably, this is possible even for heavy DM. Decoupling at $T\sim$~keV can influence the formation of galactic structures on small scales (see e.g.~\cite{vandenAarssen:2012ag,Villasenor:2022aiy}) that are being probed by Lyman-$\alpha$ forest observations of quasars~\cite{Boera:2018vzq,eBOSS:2018qyj}, including recently by the DESI survey~\cite{Karacayli:2025svi}. This adds a novel element to the interplay between early-universe DM decoupling and late-time self-interactions within halos.

\smallskip

These findings suggest several avenues for further research, motivated by many existing theories, including WIMPs, coloured coannihilation scenarios and hidden sector models.
 
The role of unitarisation in models featuring transitions between bound states warrants close examination. Bound-to-bound transitions can only enhance DM depletion through BSF~\cite{Binder:2021vfo}. After unitarisation, $\ell$-changing transitions can substantially increase the contribution of higher partial waves.

Different long-range potentials, in Abelian or non-Abelian frameworks, can lead to qualitatively and quantitatively distinct outcomes. This is because the velocity scaling, before and after regularisation, depends on the long-range potentials governing the initial and final states, particularly on the ratio $\alpha_{\cal S}/\alpha_{\cal B}$~\cite{Beneke:2024nxh,Petraki:2025}. Unitarisation modifies the velocity scaling of the cross sections, unless the cross sections, including their sum over a given partial wave, already scale as the unitarity limit~\cite{Flores:2024sfy}. In this work, we have focused on the case $\alpha_{\cal S}$=0, but exploring other possibilities is an important direction for the future.

In the present model, the large BSF cross sections regulate only the BSF processes themselves. However, in other models, they may also regulate conventional annihilation channels, when the initial states for annihilation and BSF coincide~\cite{Oncala:2021tkz,Oncala:2021swy}. This can significantly affect the DM relic abundance, even if DM depletion via BSF is suppressed due to high ionisation rates or low bound-state decay rates.

\acknowledgments
This work was supported by the European Union’s Horizon 2020 research and innovation programme under grant agreement No 101002846, ERC CoG CosmoChart.

\clearpage
\appendix
\section*{Appendices}

\section{Numerical validity of the approximations\label{App:Approximations}}

\begin{figure}[h!]
\centering
\includegraphics[width=1\textwidth]{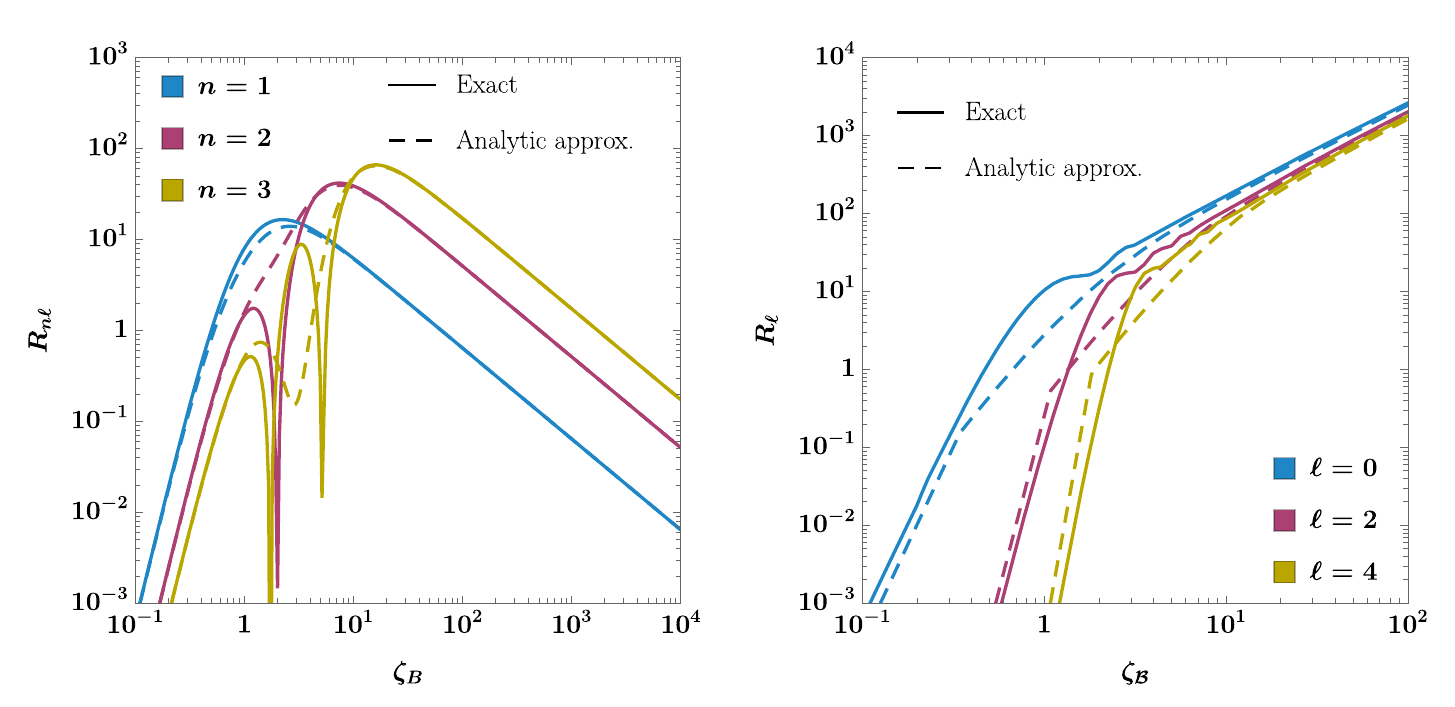}
\caption{\textit{Left:} The factor $R_{n \ell} \propto \sigma^{\rm BSF}_{n \ell}/ \sigma^{\rm uni}_{\ell}$ as a function of  $\zeta_{\cal B}$ for $\ell =0$ and $n$ as denoted in the labels. Solid (dashed) lines correspond to the exact expression \cref{eq:rnlExact} (approximation \eqref{eq:rnlAnalyticalApproximation}). \textit{Right:} $R_\ell = \sum_n R_{n\ell}$ vs $\zeta_{\cal B}$ for different values of $\ell$. We compare the $R_\ell$ calculated using the exact expression \cref{eq:rnlExact} for $R_{n\ell}$ considering up to $n= 10 \zeta_{\cal B}$ for each value of $n$,  and \cref{eq:RlSum}.} 
\label{fig:App- ApproximationCheck1}
\end{figure}

\begin{figure}[t!]
\centering
\includegraphics[width=0.5\textwidth]{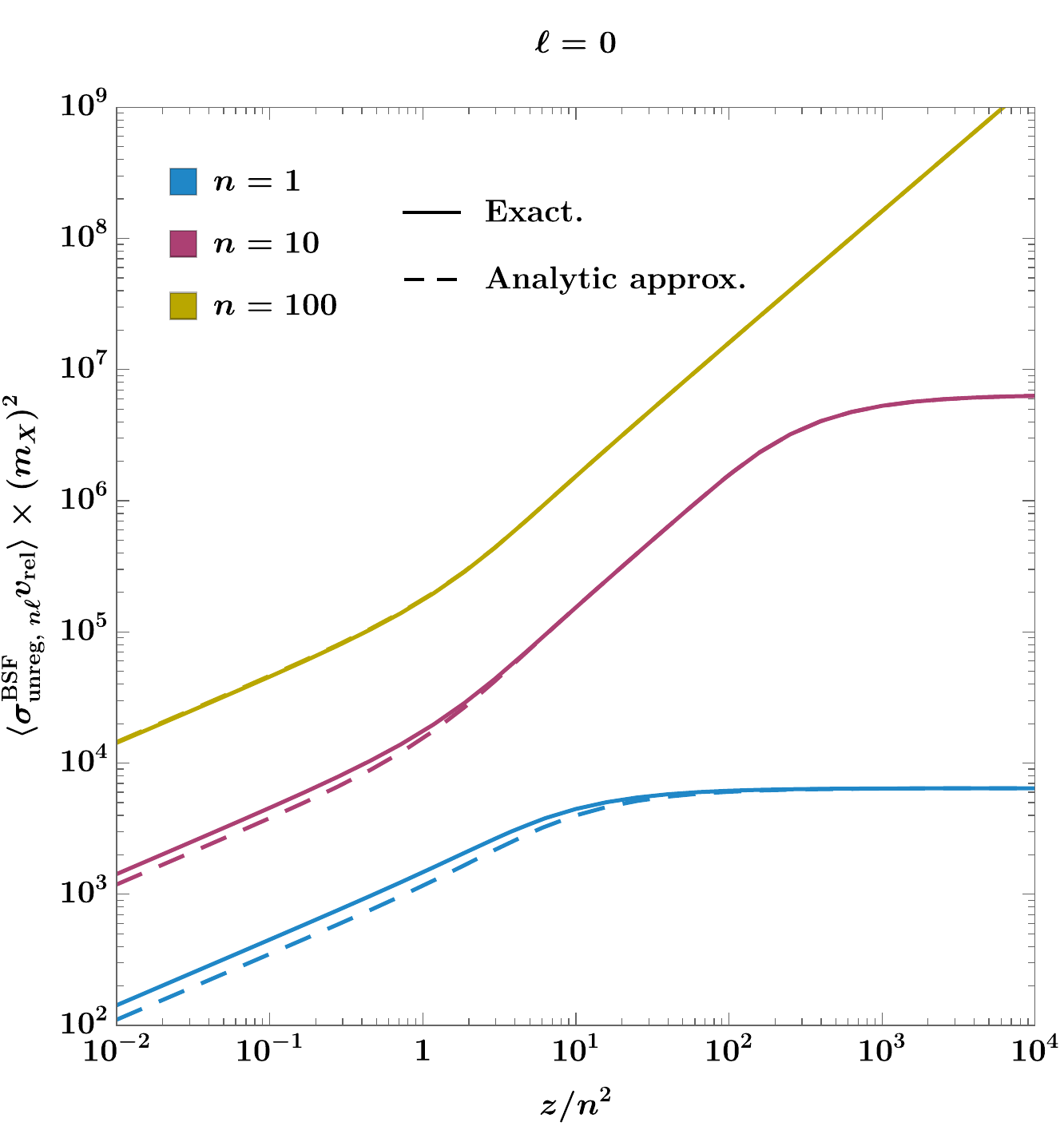}
\caption{The unregulated thermally-averaged velocity-weighted BSF cross sections vs $z/n^2$, where $z = m_X \alpha_{\cal B}^2/(4T)$ considering (i) the exact expression \eqref{eq:rnlExact} and (ii) the analytic approximation \eqref{eq:rnlAnalyticalApproximation}.}
\label{fig:AppExactEnvelopComparisonThermalAverage}
\end{figure}

We examine the approximations introduced in \cref{Sec:MonopoleTransitions_UniViolation}. Figure \eqref{fig:App- ApproximationCheck1} compares the analytical formula for $R_{n\ell}(\zeta_{\cal B}, 0)$ (cf.~\cref{eq:rnlExact}) with its analytical approximation $R_{n\ell}^{\rm A}(\zeta_{\cal B})$ (cf.~\cref{eq:rnlAnalyticalApproximation}, derived in \cite{Petraki:2025}). 
Although the exact and approximate expressions differ in the phase of their oscillations, they are enveloped by the same function. In addition, the approximation accurately predicts the position and magnitude of the peak. Upon thermal averaging, as prescribed by \cref{eq:ThermalAver_BSF}, the oscillations are smoothed out, and the approximation improves further, as seen in \cref{fig:AppExactEnvelopComparisonThermalAverage}. 

To regularise the BSF cross sections, $R_{n\ell}$ should be summed over $n$ for a given partial wave $\ell$. Approximating the sum $R_\ell = \sum_{n = \ell +1}^\infty R_{n \ell}$ by an integral over $n$, and using the asymptotic expansions of the spherical Bessel function leads to \cref{eq:RlSum} for $R_\ell$. \Cref{fig:App- ApproximationCheck1} compares this approximation with $R_\ell$ obtained using the exact formula \eqref{eq:rnlExact} summed up to $n_\mathrm{max}= 10 \zeta_{\cal B}$. While at large velocities, $\zeta_{\cal B} \ll 1$, there is some noticeable deviation, at low relative velocities, $\zeta_{\cal B} \gg 1$, the approximation becomes increasingly accurate. Given that the exact formula (\cref{eq:rnlExact}) is computationally demanding, we use the asymptotic form of \cref{eq:RlSum} for large $\zeta_{\cal B}$, in numerical calculations.

\newpage
\section{Thermally-averaged bound-state-formation cross sections \label{App:Crosssection}}

We find analytical approximations for the unregulated thermally-averaged BSF cross sections. We first derive formulas for the individual bound levels, and then sum over $n$ to obtain the total cross sections for a given partial wave.

\subsection{Individual bound levels \label{App:IndividualCrossSectionsApproximation}}
We calculate the integral \eqref{eq:ThermalAver_BSF_unreg} for the in the four intervals where the $R_{n \ell}$ scaling with $\zeta_{\cal B}$ differs, as seen in \cref{eq:rnlAnalyticalApproximationFull}.

\subsubsection*{Large temperatures, $ z \leq c_\ell^2 $}

\begin{itemize}
\item \textbf{\( \zeta_{\mathcal{B}} \leq c_\ell \):} \\
In this region, the BE factor can be approximated as 
\begin{align}
{\rm BE}(z,\zeta_{\cal B}) \simeq
\begin{cases}
1,& \zeta_{\cal B} \in (0,  \sqrt{z}), \\
\zeta_{\cal B}^2/z, &  \zeta_{\cal B} \in (\sqrt{z}, c_\ell),
\end{cases}
\end{align}
as a result the integral splits into two parts
\begin{align}
\label{eq: ThermallyAveragedCrossSection_High_Temp_Interval_1}
\frac{1}{N_\ell z^{\frac{3}{2}}}
\left\langle\sigma^{\rm BSF}_{n\ell} v_{\rm rel}\right\rangle_{\rm unreg}  
& \supsetsim
\frac{8}{c_\ell^{2\ell +2} n^3} \Bigg[\int_{0}^{\sqrt{z}} d \zeta_{\cal B} \,\zeta_{\cal B}^{2\ell +2} e^{ - z/ \zeta_{\cal B}^2} + \int_{\sqrt{z}}^{c_\ell} d \zeta_{\cal B} \, \zeta_{\cal B}^{2\ell +2} e^{ - z/ \zeta_{\cal B}^2}  \left( \zeta_{\cal B}^2 /z \right) \Bigg] 
\nonumber  \\
& =  \frac{4 c_\ell}{n^3}\left( \frac{z}{c_\ell^2}\right)^{\ell + \frac{3}{2}}  
\bigg[\Gamma \left(-\ell-\frac{5}{2},\frac{z}{c_\ell^2}\right)+E_{\ell+\frac{5}{2}}(1)-E_{\ell+\frac{7}{2}}(1)\bigg]\nonumber 
\nonumber \\
& \approx \frac{4}{z}\left( \frac{c_\ell}{n}\right)^{3}  E_{\ell +\frac{
7}{2}} \left( \frac{z}{c_\ell^2}\right)
\end{align}
where $E_{n} (z)$ denotes the generalised exponential integral function.

\item \textbf{\( c_\ell < \zeta_{\mathcal{B}} \leq n \):} \\
In this region, ${\rm BE}(z, \zeta_{\cal B}) \approx \zeta_{\cal B}^2/z$, which implies 
\begin{align}
\label{eq: ThermallyAveragedCrossSection_High_Temp_Interval_2}
\frac{1}{N_\ell z^{\frac{3}{2}}}\left\langle\sigma^{\rm BSF}_{n\ell} v_{\rm rel}\right\rangle_{\rm unreg}  & \supsetsim    \frac{2^4}{z n^3} \int_{c_\ell}^n d \zeta_{\cal B} \; \zeta_{\cal B}^2 \, e^{-z/\zeta_{\cal B}^2 } = \frac{16}{3 n^3 z}  \bigg\{ 2 \sqrt{\pi} z^{3/2} \left[ \mathrm{Erf}\left(\frac{\sqrt{z}}{c_\ell} \right) - \mathrm{Erf}\left(
\frac{\sqrt{z}}{n} \right) \right] 
\nonumber \\
&- c_\ell (c_\ell^2 - 2z) e^{-z/c_\ell^2} + n (n^2 - 2z) e^{-z/n^2} \bigg\} 
\nonumber \\
& \overset{c_{\ell} \ll n}{\approx}\frac{16}{z} \bigg\{ \frac{1}{3} -\frac{z}{n^2} + \frac{2 \sqrt{\pi }}{3} z^{3/2}\text{Erf}\left(\frac{\sqrt{z}}{c_\ell}\right) \bigg\}    \end{align}

\item \textbf{\( n < \zeta_{\mathcal{B}} \leq n^2 / c_\ell \):}  \\
Here, the BE factor becomes ${\rm BE}(z, \zeta_{\cal B}) \approx n^2/z$, and the cross section can be approximated by the following expression: 
\begin{align}
\label{eq: ThermallyAveragedCrossSection_High_Temp_Interval_3}
      \frac{1}{N_\ell z^{\frac{3}{2}}}\left\langle\sigma^{\rm BSF}_{n\ell} v_{\rm rel}\right\rangle_{\rm unreg} & \supsetsim 
     \frac{2^4 n}{z} \int_{n}^{n^2/c_\ell}\frac{d \zeta_{\cal B}}{\zeta_{\cal B}^2} e^{ -z /\zeta_{\cal B}^2}   
      =  \frac{2^3 \sqrt{\pi } n}{z^{\frac{3}{2}}}\left[\mathrm{Erf}\left(\sqrt{\frac{z}{n^2}}\right)-\mathrm{Erf}\left(\frac{c_\ell}{n} \sqrt{\frac{z}{n^2}}\right)\right] \nonumber \\& \approx  \frac{16}{z} \left( 1- \frac{c_{\ell}}{n}\right)
\end{align}
    \item \textbf{\( \zeta_{\mathcal{B}} > n^2 / c_\ell \):}   \\
    Analogously, we find In this region ${\rm BE} (z, \zeta_{\cal B}) \approx n^2/z$, which implies 
 \begin{align}
\label{eq: ThermallyAveragedCrossSection_High_Temp_Interval_4}
      \frac{1}{N_\ell z^{\frac{3}{2}}}\left\langle\sigma^{\rm BSF}_{n\ell} v_{\rm rel}\right\rangle_{\rm unreg} & \supsetsim 
     \frac{2^3 n^{4 \ell +5 }}{c_\ell^{2\ell +2} z} \int_{n^2/c_\ell}^{\infty} \frac{d \zeta_{\cal B}}{\zeta_{\cal B}^{2\ell+4}} e^{-z/\zeta_{\cal B}^2}   \\ 
     & = \frac{4 n^{4 \ell+5} } {c_\ell^{2\ell+2}  z^{\ell+\frac{5}{2}}} \left[\Gamma \left(\ell+\frac{3}{2}\right)-\Gamma \left(\ell+\frac{3}{2},\frac{c_\ell^2 z}{n^4}\right)\right]
     \nonumber \\
     & \approx \frac{8}{n z} \frac{c_\ell}{2\ell +3}  \nonumber
     \end{align}
\end{itemize}

Approximate scaling of the individual unregulated cross sections for \( z < c_\ell^2 \)
\begin{align}
\label{eq:ThermallyAveragedCrossSection_High_Temp_Total}
    \frac{1}{N_\ell z^{3/2}} \langle \sigma_{n\ell}^{\text{BSF}} v_{\text{rel}} \rangle_{\text{unreg}} 
    &\approx \frac{16}{z} \bigg\{ \frac{4}{3} - \frac{c_\ell}{n} \left( 1 - \frac{1}{2(2\ell +3)}\right)-\frac{z}{n^2} + \frac{2 \sqrt{\pi }}{3} z^{3/2}\text{Erf}\left(\frac{\sqrt{z}}{c_\ell}\right) \bigg\} 
\end{align}

\subsubsection*{Intermediate temperatures, $c_\ell^2 < z \leq n^2$}
Here, the behaviour changes as follows:
\begin{itemize}
    \item \textbf{\( \zeta_{\mathcal{B}} \leq c_\ell \):} \\
 In this subregion, the BE factor is $ \approx 1$, giving:
    \begin{align}
\label{eq: ThermallyAveragedCrossSection_Intermediate_Temp_Interval_1}
        \frac{1}{N_\ell z^{\frac{3}{2}}}\left\langle\sigma^{\rm BSF}_{n\ell} v_{\rm rel}\right\rangle_{\rm unreg} 
        &\supset   \frac{8}{c_\ell^{2\ell +2} n^3} \int_{0}^{c_\ell} d \zeta_\mathcal{B} \zeta_\mathcal{B}^{2\ell +2} e^{-z/\zeta_\mathcal{B}^2} =\frac{4 c_\ell E_{\ell+5/2}(z / c_\ell^2)}{n^3}. 
    \end{align}

    \item \textbf{\( c_\ell < \zeta_{\mathcal{B}} \leq n \):} \\
   For $\zeta_{\cal B} \in (c_\ell, \sqrt{z})$, the BE factor is negligible (${\rm BE}(z, \zeta_{\cal B}) \approx 1$), while for $\zeta_{\cal B} \in (\sqrt{z}, n)$ it grows as ${\rm BE}(z, \zeta_{\cal B}) \approx \zeta_{\cal B}^2/z$. Therefore, 
\begin{align}
\label{eq: ThermallyAveragedCrossSection_Intermediate_Temp_Interval_2}
 \frac{1}{N_\ell z^{\frac{3}{2}}}\left\langle\sigma^{\rm BSF}_{n\ell} v_{\rm rel}\right\rangle_{\rm unreg} & \supsetsim  
\frac{2^4}{n^3} \left(\int_{c_\ell}^{\sqrt{z}} d \zeta_{\cal B} \; e^{ - z/ \zeta_{\cal B}^2}   +\frac{1}{z} \int_{\sqrt{z}}^{n} d \zeta_{\cal B} \;  \zeta_{\cal B}^2  \;  e^{ - z/ \zeta_{\cal B}^2}   \right) \nonumber   \\ 
 & = \frac{16}{3 n^3} \bigg\{ - 3 c_\ell e^{-z/c_\ell^2} + n \left( \frac{n^2}{z} - 2\right)e^{-z/n^2} + 4 \frac{\sqrt{z}}{e} + 5 \sqrt{\pi z} \nonumber \\
 &+\sqrt{\pi z } \left[3 \, \mathrm{Erf}\left(\frac{\sqrt{z}}{c_\ell}\right)+2 \, \mathrm{Erf}\left(\frac{\sqrt{z}}{n}\right)+5 \, \mathrm{Erfc}(1)\right] \bigg\}\nonumber \\
 &\approx \frac{16}{z} \bigg \{ \left(1- \frac{2z}{n^2}\right) \frac{e^{-z/n^2}}{3} + \frac{17}{3} z^{\frac{3}{2}} + 2\sqrt{\pi} z^{\frac{3}{2}} \text{Erf}\left(  \frac{\sqrt{z}}{n}\right)  \bigg\}
\end{align}
where $\mathrm{Erfc}(x)$ denotes the complementary error function. 
\item \textbf{\( n < \zeta_{\mathcal{B}} \leq n^2 / c_\ell \):} \\
In this subregion, following \cref{eq: ThermallyAveragedCrossSection_High_Temp_Interval_3}, the result is given by 
\begin{align}
\label{eq: ThermallyAveragedCrossSection_Intermediate_Temp_Interval_3}
      \frac{1}{N_\ell z^{\frac{3}{2}}}\left\langle\sigma^{\rm BSF}_{n\ell} v_{\rm rel}\right\rangle_{\rm unreg} & \supsetsim \frac{2^3 \sqrt{\pi } n}{z^{\frac{3}{2}}}\left[\mathrm{Erf}\left(\sqrt{\frac{z}{n^2}}\right)-\mathrm{Erf}\left(\frac{c_\ell}{n} \sqrt{\frac{z}{n^2}}\right)\right] \\
      & \approx   \frac{2^3 \sqrt{\pi } n}{z^{\frac{3}{2}}}\mathrm{Erf}\left(\sqrt{\frac{z}{n^2}}\right)-  \frac{2^4 }{z} \frac{c_\ell}{n}. 
\end{align}
    \item \textbf{\( \zeta_{\mathcal{B}} > n^2 / c_\ell \):} \\
Similarly, the result is given by \cref{eq: ThermallyAveragedCrossSection_High_Temp_Interval_4}.
\end{itemize}

Approximate total for \( c_\ell^2 < z \leq n^2 \):
\begin{align}
\label{eq: ThermallyAveragedCrossSection_Intermediate_Temp_Total}
    \frac{1}{N_\ell z^{3/2}} \langle \sigma_{n\ell}^{\text{BSF}} v_{\text{rel}} \rangle_{\text{unreg}} 
    &\approx    \frac{16}{z} \bigg \{ \frac{e^{-z/n^2}}{3} - \frac{c_\ell}{n}+ \frac{17}{3} z^{\frac{3}{2}} \nonumber\\
    & + 2\sqrt{\pi} z^{\frac{3}{2}} \text{Erf}\left(  \frac{\sqrt{z}}{n}\right)   + \frac{\sqrt{\pi} n}{2 z^{1/2}}  \text{Erf}\left(  \frac{\sqrt{z}}{n}\right) \bigg\}
\end{align}

\subsubsection*{Low temperatures, $ z > n^2 $}
In the high-temperature regime, the Bose enhancement factor is constant, i.e., ${\rm BE}(z, \zeta_{\cal B}) \approx 1$, and we find
\begin{itemize}
    \item \textbf{\( \zeta_{\mathcal{B}} \leq c_\ell \):} 
\begin{align}
\label{eq: ThermallyAveragedCrossSection_Low_Temp_Interval_1}
      \frac{1}{N_\ell z^{\frac{3}{2}}}\left\langle\sigma^{\rm BSF}_{n\ell} v_{\rm rel}\right\rangle_{\rm unreg}  &\supsetsim 
         \frac{8}{c_\ell^{2\ell +2} n^3} \int_{0}^{c_\ell} d \zeta_{\cal B} \zeta_{\cal B}^{2\ell +2} e^{ - z/ \zeta_{\cal B}^2}   = \frac{4 c_\ell E_{\ell+\frac{5}{2}}\left(\frac{z}{c_\ell^2}\right)}{n^3}  \nonumber \\
      &\approx\frac{4 c_\ell^3 \; e^{- z/c_\ell^2}}{n^3 z}.
\end{align}

   \item \textbf{\( c_\ell < \zeta_{\mathcal{B}} \leq n \):}
\begin{align}
\label{eq: ThermallyAveragedCrossSection_Low_Temp_Interval_2}
     \frac{1}{N_\ell z^{\frac{3}{2}}}\left\langle\sigma^{\rm BSF}_{n\ell} v_{\rm rel}\right\rangle_{\rm unreg} & \supsetsim   
     \frac{2^4}{n^3} \int_{c_\ell}^{n} d \zeta_{\cal B} \; e^{-z/\zeta_{\cal B}^2} 
      =   \frac{2^4}{n^3} \bigg\{ n e^{-z/n^2} - c_\ell e^{-z/c_\ell^2} \nonumber \\
      &+\sqrt{\pi z} \left[\mathrm{Erf}\left( \frac{\sqrt{z}}{n}\right) - \mathrm{Erf}\left( \frac{\sqrt{z}}{c_\ell} \right)\right] \bigg\} \nonumber \\
      &\overset{c_\ell^2 \ll n^2 \ll z}{\approx} \frac{8}{z}  e^{-z/n^2}.
\end{align}
\item \textbf{\( n < \zeta_{\mathcal{B}} \leq n^2 / c_\ell \):} 
\begin{align}
\label{eq: ThermallyAveragedCrossSection_Low_Temp_Interval_3}
      \frac{1}{N_\ell z^{\frac{3}{2}}}\left\langle\sigma^{\rm BSF}_{n\ell} v_{\rm rel}\right\rangle_{\rm unreg}  &\supsetsim
     2^3 \int_{n}^{\frac{n^2}{c_\ell}}\frac{d \zeta_{\cal B}}{\zeta_{\cal B}^3} \left(\frac{\zeta_{\cal B}}{n}\right) e^{-z/\zeta_{\cal B}^2} 
   \nonumber  \\
     & = \frac{4\sqrt{\pi }}{n \sqrt{z}}\left[\mathrm{Erf}\left(\frac{\sqrt{z}}{n} \right)-\mathrm{Erf}\left(\frac{c_\ell}{n} \frac{\sqrt{z}}{n} \right)\right] 
\end{align}
    \item \textbf{\( \zeta_{\mathcal{B}} > n^2 / c_\ell \):}
\begin{align}
\label{eq: ThermallyAveragedCrossSection_Low_Temp_Interval_4}
       \frac{1}{N_\ell z^{\frac{3}{2}}}\left\langle\sigma^{\rm BSF}_{n\ell} v_{\rm rel}\right\rangle_{\rm unreg}  &\supsetsim  
 \frac{2^3 n^{4 \ell +3 }}{c_\ell^{2\ell +2}} \int_{\frac{n^2}{c_\ell}}^{\infty} \frac{d \zeta_{\cal B}}{\zeta_{\cal B}^{2\ell+4}} e^{\frac{-z}{\zeta_{\cal B}^2}} \nonumber \\
      &= \frac{4  n^{4 \ell+3}}{ c_\ell^{2 (\ell+2)} z^{\ell+\frac{3}{2}} }\bigg[\Gamma\left(\ell + \frac{3}{2} \right)-\Gamma \left(\ell+\frac{3}{2},\frac{c_\ell^2 z}{n^4}\right)\bigg] 
\end{align}

\end{itemize}
Approximate total for \( z >   n^2 \):
\begin{align}
\label{eq: ThermallyAveragedCrossSection_Low_Temp_Total}
    \frac{1}{N_\ell z^{3/2}} \langle \sigma_{n\ell}^{\text{BSF}} v_{\text{rel}} \rangle_{\text{unreg}} 
    &\approx \frac{4 \sqrt{\pi }}{n \sqrt{z}}\left[\mathrm{Erf}\left(\frac{\sqrt{z}}{n} \right)-\mathrm{Erf}\left(\frac{c_\ell}{n} \frac{\sqrt{z}}{n} \right)\right] \nonumber\\
    & + \frac{4  n^{4 \ell+3}}{ c_\ell^{2 (\ell+2)} z^{\ell+\frac{3}{2}} }\bigg[\Gamma\left(\ell + \frac{3}{2} \right)-\Gamma \left(\ell+\frac{3}{2},\frac{c_\ell^2 z}{n^4}\right)\bigg] .
\end{align}

\subsection{Sum over bound levels \label{App:Thermally Averaged BSF summed over n}}

At late times, when $z$ becomes large, one can neglect the Bose enhancement, and set $\epsilon_{n\ell} \approx 1$ in the effective cross section \eqref{eq:EffectiveCrossSection}. In this regime,  the total thermally-averaged BSF cross section for a given partial wave $\ell$ reads 
\begin{align}
\label{eq:sumRnlThermalAverage}
\left\langle\sigma^{\rm BSF}_{\ell}v_{\rm rel}\right\rangle_{\rm unreg} = \sum_{n=\ell+1}^{\infty} \left\langle\sigma^{\rm BSF}_{n\ell} v_{\rm rel}\right\rangle_{\rm unreg}
& \simeq N_\ell z^{\frac{3}{2}} \int_0^{\infty} \frac{d \zeta_{\cal B}}{\zeta_{\cal B}^3} R_{\ell} (\zeta_{\cal B}) e^{ -z/\zeta_{\cal B}^2}.
\end{align}
Using \cref{eq:RlSum} we can evaluate this integral in the three relevant regions.

\subsubsection*{In the interval $\zeta_{\cal B} < c_\ell$:}
For small $\zeta_{\cal{B}}$, one finds
\begin{align}
\label{eq:rlThermal-Average-First-Interval}
\frac{1}{N_\ell z^{\frac{3}{2}}}\left\langle\sigma^{\rm BSF}_{\ell} v_{\rm rel}\right\rangle_{\rm unreg} &\supsetsim
\frac{4}{c_{\ell}^{2 \ell+2}(1+\ell)^2} \int_0^{c_\ell} d \zeta_{\cal B} \; \zeta_{\cal B}^{2 \ell+2} e^{-z/\zeta_{\cal B}^2}  =\frac{2 c_\ell}{(\ell+1)^2}  E_{\ell + \frac{5}{2}}\left(\frac{z}{c_\ell^2}\right) \nonumber \\
&\overset{z \gg c_\ell^2}{\simeq} \frac{2 c_\ell^5}{(\ell+1)^2 z^2} \bigg[ \frac{z}{c_\ell^2}-\left(\ell + \frac{5}{2} \right) \bigg]  e^{- z/c_\ell^2}.  
\end{align}

\subsubsection*{In the interval $ c_\ell< \zeta_{\cal B} < (\ell+1)^2/ c_\ell$:}
In the intermediate region, we get
\begin{align}
\label{eq:rlThermal-Average-Second-Interval}
 \frac{1}{N_\ell z^{\frac{3}{2}}}\left\langle\sigma^{\rm BSF}_{\ell} v_{\rm rel}\right\rangle_{\rm unreg} &\supsetsim  4 \int_{c_\ell}^{\frac{(1+\ell)^2}{c_\ell}} \frac{d \zeta_{\cal B}}{\zeta_{\cal B}^3} \zeta_{\cal B} \ln \left[1+\frac{\zeta_{\cal B}^2}{(1+\ell)^2}\right] e^{-z/\zeta_{\cal B}^2}  \nonumber  \\
&=\frac{2}{\ell+1} \left[ \Gamma \left(0,\frac{c_\ell^2 z}{(\ell+1)^4}\right)-\Gamma \left(0,\frac{z}{c_\ell^2}\right)\right]  \nonumber  \quad \\
& \quad + 2(\ln 2 -1) \sqrt{\frac{\pi}{z}} \left[ \mathrm{Erf}\left( \frac{z}{c_\ell}\right)- \mathrm{Erf}\left(\frac{z c_\ell}{(1+\ell)^2}\right) \right] \nonumber  \\ 
& \overset{ z \gg e (1+\ell)^2}{\simeq}
\frac{2 c_\ell^2}{ z(\ell+1)} \left(  e^{1- \frac{c_\ell^2 z}{(\ell+1)^4}} - e^{-\frac{z}{c_\ell}^2}\right),
\end{align}
where we have adopted the following approximation: $\ln\left[1 + \frac{\zeta_{\cal B}^2}{(1+\ell)^2}\right] \approx \frac{\zeta_{\cal B}}{\ell+1} + (\ln 2 - 1)$.

\subsubsection*{In the interval $  \zeta_{\cal B}  >   (\ell+1)^2/ c_\ell$}
In this regime, the integral splits into three parts:
\begin{subequations}
\label{eq:rlThermal-Average-Third-Interval}
\begin{align}
  \frac{1}{N_\ell z^{\frac{3}{2}}}\left\langle\sigma^{\rm BSF}_{\ell} v_{\rm rel}\right\rangle_{\rm unreg}^{(1)} & \supsetsim  \left(\frac{2}{ (1+ \ell)}- 4 \ln c_\ell \right) 
\int_{\frac{(1 + \ell)^2}{c_\ell}}^{\infty}  \frac{d \zeta_{\cal B}}{\zeta_{\cal B}^2}  e^{-z/\zeta_{\cal B}^2}  
\nonumber \\
&= 2 \sqrt{\frac{\pi}{z}} \mathrm{Erf}\left( \frac{c_\ell \sqrt{z}}{(1+\ell)^2}\right)  \left(\frac{1}{2 (1+ \ell)}- \ln c_\ell \right), 
\label{eq:rlThermal-Average-Third-Interval-1} 
\\
 \frac{1}{N_\ell z^{\frac{3}{2}}}\left\langle\sigma^{\rm BSF}_{\ell} v_{\rm rel}\right\rangle_{\rm unreg}^{(2)} & \supsetsim -\frac{2}{(\ell+1)}\int_{\frac{(1 + \ell)^2}{c_\ell}}^{\infty} \frac{d \zeta_{\cal B}}{\zeta_{\cal B}^{2\ell +4}}
\left( \frac{(1+\ell)^2}{c_\ell}\right)^{2\ell +2} e^{-z/\zeta_{\cal B}^2} 
\nonumber   \\   
& = \frac{(\ell+1)^{4 \ell+3}}{ c_\ell^{2 (\ell+1)} z^{\ell+\frac{3}{2}}} \left[\Gamma \left(\ell+\frac{3}{2},\frac{c_\ell^2 z}{(\ell+1)^4}\right) -
\Gamma\left(\ell + \frac{3}{2} \right) \right],
\label{eq:rlThermal-Average-Third-Interval-2} 
\\
\frac{1}{N_\ell z^{\frac{3}{2}}}\left\langle\sigma^{\rm BSF}_{\ell} v_{\rm rel}\right\rangle_{\rm unreg}^{(3)} & \supsetsim 4 \int_{\frac{(1 + \ell)^2}{c_\ell}}^{\infty}  \frac{d \zeta_{\cal B}}{\zeta_{\cal B}^2} \ln\zeta_{\cal B} \,  e^{-z/\zeta_{\cal B}^2} = \int^{ \frac{z c_\ell^2}{(1+\ell)^4} }_{0} \frac{du}{\sqrt{z u}} \ln\left(\frac{z}{u}\right) e^{-u} 
\nonumber \\
&= \sqrt{\frac{\pi}{z} }\ln z \mathrm{Erf}\left(\frac{z c_\ell^2}{(1+ \ell)^4}\right) - \frac{1}{\sqrt{z}}\frac{\partial \gamma\left(k,\frac{z c_\ell^2}{(1+\ell)^4}\right)}{\partial k}\bigg|_{k= 1/2},
\label{eq:rlThermal-Average-Third-Interval-3}
\end{align}
\end{subequations}
where, in \cref{eq:rlThermal-Average-Third-Interval-3}, we have used the definition of the derivative of the lower incomplete gamma function $\gamma(k,a)$ given by
\begin{equation}
\frac{\partial \gamma(k, a)}{\partial k}
=\int_0^{a}(\ln x) \cdot x^{k-1} e^{-x} d x.
\end{equation}
In the limit $z \gg (\ell+1)^2 e$, the total contribution becomes
\begin{equation}
    \begin{aligned}
    \label{eq:rlThermal-Average-Third-Interval-Large-z-limit}
        \sum_{i=1}^3 \frac{1}{N_\ell z^{\frac{3}{2}}}\left\langle\sigma^{\rm BSF}_{\ell} v_{\rm rel}\right\rangle_{\rm unreg}^{(i)} 
       &  \supsetsim \sqrt{\frac{\pi}{z}} \left[\ln \left(\frac{z}{c_\ell^2} \right) + \frac{1}{ (1+ \ell)}
     - \psi^{(0)}\left(\frac{1}{2} \right)\right],
    \end{aligned}
\end{equation}
where $\psi^{(0)}$ denotes the logarithmic derivative of the gamma function.

The above result indicates that, in the low-temperature regime, the unregulated thermally-averaged BSF cross section grows approximately linearly with \( z \), up to a logarithmic correction, and exhibits only a mild dependence on $\ell$.

\clearpage
\section{Scattering-state annihilation and bound-state decay \label{App:Annihilation and Decay Rates}}

We derive the partial-wave cross section for the $X X^\dagger \rightarrow \Phi \Phi^\dagger$ annihilation, and the decay rates of bound states,  ${\cal B}(X X^\dagger) \to \Phi \Phi^\dagger$,  incorporating the corresponding Sommerfeld factors. 

\subsection*{Useful identities}

The completeness relation of spherical harmonics, and the partial-wave expansion of the exponential read
\begin{align}
P_\ell ({\bf \hat{x} \cdot \hat{y}}) &= 
\dfrac{4\pi}{2\ell+1} 
\sum_{m=-\ell}^\ell  
Y_{\ell m}^* (\Omega_{\bf\hat{x}})
Y_{\ell m} (\Omega_{\bf\hat{y}}),
\label{eq:SphericalHarmonics_Completeness}
\\
e^{\pm \mathbb{i} {\bf \hat{q} \cdot \hat{r}}} &= 
\sum_{\ell=0}^\infty (2\ell+1) (\pm\mathbb{i})^\ell
j_\ell (qr) P_\ell ({\bf \hat{q} \cdot \hat{r}}),
\label{eq:Exponential_Expansion}
\end{align}
with the following orthonormality relations,
\begin{align}
\int d\Omega \, Y_{\ell m} (\Omega) \, Y_{\ell' m'}^* (\Omega) 
&= \delta_{\ell\ell'} \, \delta_{mm'}  ,
\label{eq:SphericalHarmonics_Orthonormality}
\\
\int_{-1}^1 dz \, P_\ell (z) \, P_{\ell'} (z) 
&= \dfrac{2}{2\ell+1} \ \delta_{\ell\ell'}.
\label{eq:Legendre_Orthonormality}
\end{align}
We will also need the following derivatives of the spherical Bessel function and the generalised Laguerre polynomials,
\begin{align}
\left.\dfrac{d^\ell j_\ell (x)}{dx^\ell} \right|_{x=0} 
&= \dfrac{\ell!}{(2\ell+1)!!}
= \dfrac{2^\ell \, (\ell!)^2}{(2\ell+1)!},
\label{eq:BesselDerivative}    
\\
\left(
\dfrac{d^\ell}{dx^\ell} \left[
e^{-x/n} x^\ell L_{n-\ell-1}^{2\ell+1} (2x/n)
\right]\right)_{x\to 0} &= 
\dfrac{\ell!}{(2\ell+1)!}
\dfrac{(n+\ell)!}{(n-\ell-1)!} .
\label{eq:LaguerreDerivative}    
\end{align}
The Fourier transforms of the wavefunctions are
\begin{align}
\psi ({\bf r}) = \int \dfrac{d^3q}{(2\pi)^3} 
\, e^{\mathbb{i} {\bf q \cdot r}}
\, \tilde{\psi} ({\bf q}),
\qquad
\tilde{\psi} ({\bf q}) = \int d^3r 
\, e^{-\mathbb{i} {\bf q \cdot r}}
\, \psi ({\bf r}),
\label{eq:Wavefunctions_Fourier}
\end{align}
both for the scattering-state and bound-state wavefunctions, which can be expanded in terms of spherical harmonics as follows,
\begin{subequations}
\label{eq:Wavefunctions}
\label[pluralequation]{eqs:Wavefunctions}
\begin{align}
\psi_{\bf k} ({\bf r}) &= 
\sum_{\ell=0}^\infty (2\ell+1)
\mathfrak{R}_{|{\bf k}|,\ell} (r) \, 
P_{\ell} ({\bf \hat{k} \cdot \hat{r}})    
= 
\sum_{\ell=0}^\infty \sum_{m=-\ell}^{\ell}
4\pi \, 
\mathfrak{R}_{|{\bf k}|,\ell} (r) \, 
Y_{\ell m}  (\Omega_{\bf r})    
Y_{\ell m}^*(\Omega_{\bf k})   ,
\label{eq:Wavefunction_Scattering}
\\
\psi_{n\ell m} ({\bf r}) &= 
\mathfrak{R}_{n\ell} (r) \, Y_{\ell m} (\Omega_{\bf r})   . 
\label{eq:Wavefunction_Bound}
\end{align}
\end{subequations}

\subsection*{Perturbative amplitude and annihilation cross section}

The diagrams contributing to the annihilation process $X X^\dagger \rightarrow \Phi \Phi^\dagger$ are shown in Fig.~\ref{fig:XXPhiPhiabelian}. Let ${\bf k}$ and ${\bf p}$ be the momenta of each of the incoming and outgoing particles, respectively, in the CM frame. For non-relativistic $XX^\dagger$, 
$|{\bf k}| = (m_X/2) v_{\rm rel}$ and 
$|{\bf p}| = \sqrt{m_X^2+{\bf k}^2}$, 
where we have neglected the $\Phi$ mass. 
The tree-level amplitude is
\begin{equation}
{\cal A}^{\rm ann} ( {\bf k}, {\bf p}) = 
\dfrac{8\pi\,\alpha_\Phi m_X^2}{|{\bf p}| (|{\bf p}| - |{\bf k}| \cos\theta_{{\bf k, p}})} 
+ 8\pi \, \alpha_V \dfrac{|{\bf k}|}{|{\bf p}|} 
\cos\theta_{{\bf k,p}} 
- \lambda_{X\Phi}.
\label{eq:Aann}
\end{equation}
We expand ${\cal A}^{\rm ann} ( {\bf k}, {\bf p})$ in partial waves as follows,
\begin{subequations}
\label{eq:Aann_PW}
\label[pluralequation]{eqs:Aann_PW}
\begin{align}
{\cal A}^{\rm ann} ({\bf k,p}) &= 
16 \pi \sum_{\ell =0}^\infty (2 \ell +1) 
{\cal A}_{\ell}^{\rm ann} (|{\bf k}|, |{\bf p}|) 
P_{\ell} (\cos{\theta_{{\bf k}, {\bf p}}}),    
\label{eq:Aann_PW_expansion}
\\
{\cal A}_{\ell}^{\rm ann} (|{\bf k}|, |{\bf p}|) &=
\frac{1}{64 \pi^2} \int d \Omega_{{\bf k}, {\bf p}} 
\,{\cal A}^{\rm ann} ({\bf k}, \mathbf{p}) P_\ell (\cos{\theta_{{\bf k}, {\bf p}}}).
\label{eq:Aann_PW_inverse}
\end{align}
\end{subequations}
From \cref{eq:Aann,eq:Aann_PW}, we find
\begin{align}
{\cal A}_{\ell}^{\rm ann} (|{\bf k}|, |{\bf p}|)  = 
\frac{ \alpha_\Phi m_X^2}{4 |{\bf p}|^2} 
\sum_{j=0}^\infty \left(\dfrac{|{\bf k}|}{|{\bf p}|}\right)^{\ell+2j} {\cal I}_{2j}   
+ \frac{\alpha_V}{6} \frac{|{\bf k}|}{|{\bf p}|} \delta_{\ell, 1}
- \frac{ \lambda_{X \Phi}}{16 \pi} \delta_{\ell, 0},
\label{eq:Aann_ell}
\end{align}
with 
\begin{align}
{\cal I}_{2j} \equiv 
\dfrac{2^{\ell+1} (\ell+j)! (\ell+2j)!}{j! (2\ell+2j+1)!}.
\label{eq:Aann_PW_ExpansionCoeff_I}
\end{align}
Keeping the leading order term in $v_{\rm rel}$ for every partial wave,
\begin{align}
{\cal A}_\ell^{\rm ann}
\simeq a_\ell^{} \left(\dfrac{2|{\bf k}|}{m_X}\right)^\ell    
\simeq a_\ell^{} v_{\rm rel}^\ell  ,  
\label{eq:Aann_ell_LO}
\end{align}
where \cref{eq:Aann_ell} implies
\begin{align}
a_\ell^{} = 
\dfrac{1}{2} 
\frac{(\ell!)^2}{(2 \ell+1)!} 
\left( \alpha_\Phi  + \alpha_V \, \delta_{1, \ell} - \frac{\lambda_{X \Phi}}{8 \pi} \delta_{0, \ell} \right).
\label{eq:a_ell}
\end{align}
The tree-level annihilation cross section times relative velocity is
\begin{align}
\sigma^{\rm ann, tree} v_{\rm rel} &= 
\frac{1}{64 \pi^2 m_X^2} 
\int d \Omega_{\bf p} 
|{\cal A}^{\rm ann} ({\bf k,p})|^2 
= \frac{16 \pi}{m_X^2} \sum_{\ell =0}^ \infty (2 \ell +1) |{\cal A}_\ell^{\rm ann}|^2,    
\label{eq:sigma_ann_tree}
\end{align}
where we used the expansion \eqref{eq:Aann_PW_expansion} and the completeness relation \eqref{eq:SphericalHarmonics_Completeness}. Considering \cref{eq:Aann_ell_LO}, this yields
\begin{align}
\sigma^{\rm ann, tree} v_{\rm rel} = \sum_\ell \sigma_\ell^{\rm tree} v_{\rm rel},
\qquad
\sigma_\ell^{\rm tree} v_{\rm rel} = h_\ell \, v_{\rm rel}^{2\ell},
\qquad
h_\ell^{} = \dfrac{16\pi (2\ell+1)}{m_X^2} |a_\ell^{}|^2.
\label{eq:sigma_ann_tree_modes}
\end{align}

\subsection*{Sommerfeld factors}

The full annihilation and bound-state decay amplitudes read~\cite{Cassel:2009wt,Petraki:2015hla}
\begin{subequations}
\label{eq:FullAmplitudes_def}
\label[pluralequation]{eqs:FullAmplitudes_def}
\begin{align}
\label{eq:FullAmplitude_def_Annihilation}
{\cal M}_{\bf k}^{\rm ann} ({\bf p}) &\simeq 
\int \frac{d^3 q}{(2 \pi)^3} 
\,\tilde{\psi}_{\mathbf{k}}({\bf q}) \,{\cal A}^{\rm ann} ({\bf q,p}),
\\
\label{eq:FullAmplitude_def_Decay}
{\cal M}^{\rm dec}_{n\ell m}  ({\bf p}) &\simeq 
\frac{1}{\sqrt{m_X}} \int \frac{d^3 q}{(2 \pi)^3} 
\, \tilde{\psi}_{n\ell m}({\bf q}) \,{\cal A}^{\rm ann} ({\bf q,p}),
\end{align}
\end{subequations}
where $\tilde{\psi}_{\mathbf{k}}({\bf q})$ and $\tilde{\psi}_{n\ell m}({\bf q})$ are the momentum-space wavefunctions of the scattering and bound states respectively. Considering the partial-wave expansion \eqref{eq:Aann_PW_expansion} of the tree-level amplitude, and the form of the leading order contributions for each partial wave as given by \cref{eq:Aann_ell_LO}, as well as \cref{eq:SphericalHarmonics_Completeness,eq:Exponential_Expansion,eq:BesselDerivative,eq:Wavefunctions_Fourier}, it can be shown that~\cite{Cassel:2009wt} (see also \cite[Appendix~D]{Petraki:2015hla})
\begin{align}
\int \frac{d^3 q}{(2 \pi)^3} 
&
\,\tilde{\psi} ({\bf q}) 
\,{\cal A}^{\rm ann} ({\bf q,p}) =
\nonumber \\
&=2^6\pi^2 \sum_{\ell,m}
\dfrac{a_\ell}{(m_X/2)^\ell}
Y_{\ell m} (\Omega_{\bf p})
\int \dfrac{d^3q}{(2\pi)^3} q^\ell 
\, \tilde{\psi} ({\bf q})
\, Y_{\ell m}^* (\Omega_{\bf q})
\nonumber \\
&=2^6\pi^2 \sum_{\ell,m}
\dfrac{a_\ell}{(m_X/2)^\ell}
Y_{\ell m} (\Omega_{\bf p})
\ \dfrac{(2\ell+1)!}{4\pi \mathbb{i}^\ell \, 2^\ell \, (\ell!)^2}
\left[
\dfrac{d^\ell}{d r^\ell}
\int d\Omega_{\bf r}
\, \psi ({\bf r})
\, Y_{\ell m}^* (\Omega_{\bf r})
\right]_{r=0}.
\label{eq:FullAmplitude_MomentumToSpace}
\end{align}
With this, considering the general forms \eqref{eq:Wavefunctions} of the wavefunctions, the amplitudes \eqref{eq:FullAmplitudes_def} become
\begin{subequations}
\label{eq:FullAmplitudes}
\label[pluralequation]{eqs:FullAmplitudes}
\begin{align}
{\cal M}_{\bf k}^{\rm ann} ({\bf p}) &=   
2^6\pi^2 \sum_{\ell, m}
\dfrac{a_\ell}{(m_X/2)^\ell}
\dfrac{(2\ell+1)!}{4\pi \mathbb{i}^\ell \, 2^\ell \, (\ell!)^2}
\left[
\dfrac{d^\ell}{d r^\ell}
\, \mathfrak{R}_{|{\bf k}|,\ell} (r)
\right]_{r=0}
\, 4\pi
\, Y_{\ell m} (\Omega_{\bf p})
\, Y_{\ell m}^* (\Omega_{\bf k}) ,  
\\
{\cal M}_{n\ell m}^{\rm dec} ({\bf p}) &=   
\dfrac{2^6\pi^2}{\sqrt{m_X}} 
\dfrac{a_\ell}{(m_X/2)^\ell}
\dfrac{(2\ell+1)!}{4\pi \mathbb{i}^\ell \, 2^\ell \, (\ell!)^2}
\left[
\dfrac{d^\ell}{d r^\ell}
\, \mathfrak{R}_{n\ell} (r)
\right]_{r=0}
\, Y_{\ell m} (\Omega_{\bf p}).  
\end{align}
\end{subequations}
The full velocity-weighted annihilation cross section and the bound-state decay rates are
\begin{subequations}
\label{eq:FullRates}
\label[pluralequation]{eqs:FullRates}
\begin{align}
\sigma^{\rm ann} v_{\rm rel} &= 
\frac{1}{64 \pi^2 m_X^2} \int d \Omega_{\bf p} 
|{\cal M}^{\rm ann}_{\bf k} ({\bf p})|^2 ,
\\ 
\Gamma^{\rm dec}_{n\ell m} 
&= \frac{1}{64 \pi^2 m_X} \int d \Omega_{\bf p} 
|{\cal M}^{\rm dec}_{n\ell m} ({\bf p}) |^2 .
\end{align}
\end{subequations}
Combining \cref{eq:FullAmplitudes,eq:FullRates}, we obtain
\begin{subequations}
\begin{gather}
\sigma^{\rm ann} v_{\rm rel} = 
\sum_\ell \sigma_\ell^{} v_{\rm rel},
\qquad
\sigma_\ell^{} v_{\rm rel} = 
\big(h_\ell^{} \, v_{\rm rel}^{2\ell} \big) \, S_\ell^{\rm ann},
\\
\Gamma_{n\ell m}^{\rm dec} = h_\ell \, S_{n\ell m}^{\rm dec} ,
\end{gather}
\end{subequations}
where $h_\ell$ is given in \cref{eq:sigma_ann_tree_modes}, and the Sommerfeld factors are \cite{Petraki:2015hla}
\begin{subequations}
\begin{align}
v_{\rm rel}^{2 \ell} S_\ell^{\rm ann} &\equiv 
\left[\dfrac{(2 \ell +1)!}{(\ell !)^2}\right]^2
\dfrac{1}{m_X^{2 \ell}}
\left| \dfrac{d^\ell}{d r^\ell} 
\, \mathfrak{R}_{|\bf k|,\ell} (r) 
\right|^2_{r=0}, 
\\
S_{n\ell m}^{\rm dec} &\equiv 
\dfrac{1}{4\pi (2\ell+1)}
\left[\dfrac{(2 \ell +1)!}{(\ell !)^2}\right]^2
\dfrac{1}{m_X^{2 \ell}}
\left| \dfrac{d^\ell}{d r^\ell} 
\, \mathfrak{R}_{n\ell} (r) 
\right|^2_{r=0}. 
\end{align}
\end{subequations}
Note that $S_{n\ell m}^{\rm dec}$ is independent of the quantum number $m$.

\subsection*{Coulomb wavefunctions}

For a Coulomb potential, $V(r) = -\alpha/r$, the radial wavefunctions, defined via \cref{eqs:Wavefunctions}, are
\begin{subequations}
\begin{align}
\mathfrak{R}_{|{\bf k}|,\ell} (r) &= 
e^{\pi \zeta/2} 
\, \dfrac{\Gamma(1 + \ell - \mathbb{i} \zeta)}{(2\ell+1)!} 
\, (2\mathbb{i} k r)^\ell
\, e^{-\mathbb{i} k r}
\, {}_1 F_{1} (1+\ell + \mathbb{i} \zeta;~2\ell+2;~2\mathbb{i} kr) , 
\\
\mathfrak{R}_{n \ell} (r) &=  
\left( \frac{2 \kappa}{n}\right)^{3/2} \left[\frac{(n- \ell -1)!}{2n(n + \ell)!} \right]^{1/2} e^{- \kappa r/n} \left( \frac{2 \kappa r}{n}\right)^\ell L_{n - \ell -1}^{(2 \ell +1)} \left( \frac{2 \kappa r}{n}\right),
\end{align}
\end{subequations}
where here
$\zeta \equiv \mu \alpha / k \to  \alpha_{\Phi} / v_{\rm rel}$
and $\kappa \equiv \mu \alpha \to (m_X/2)\alpha_{\Phi}$. 
The factor $S_\ell^{\rm ann}$ is given by \cref{eq:SommerfeldFactor-l-wave}~\cite{Cassel:2009wt}, whereas for bound-state decays, we find, using \cref{eq:LaguerreDerivative},
\begin{align}
S_{n\ell m}^{\rm dec} = 
\dfrac{\kappa^3}{\pi} 
\left(\dfrac{\alpha}{n}\right)^{2\ell}
\dfrac{1}{(2 \ell +1) (\ell!)^2}
\dfrac{(n+\ell)!}{n^4 (n - \ell -1)!}.
\end{align}

\subsection*{Final result}

Considering \cref{eq:a_ell,eq:sigma_ann_tree_modes} that give the tree-level annihilation cross sections, and the Sommerfeld factors derived above, the full annihilation cross sections read
\begin{align}
\sigma^{\rm ann}_\ell v_{\rm rel} &\simeq 
4 \pi (2 \ell +1) \left[ \frac{(\ell!)^2}{(2 \ell +1)!} \right]^2 \frac{\alpha_\Phi^2}{m_X^2} v_{\rm rel}^{2 \ell} 
\, S_\ell^{\rm ann} (\alpha_{\Phi}/v_{\rm rel}),
\label{eq:AnnihilationCrossSection_final}
\end{align}
and the bound-state decay rates are
\begin{align}
\Gamma^{\rm dec}_{n\ell m} &= \frac{m_X}{2}  \frac{\alpha_\Phi^{2\ell+5}}{n^{2\ell +4}}   
\dfrac{(n+\ell)!}{(n-\ell-1)!} 
\left[ \dfrac{\ell !}{ (2\ell +1)!} \right]^2.
\label{eq:DecayRates_final}
\end{align}

\clearpage
\bibliography{Bibliography}

\end{document}